\documentclass[a4paper, 10pt]{memoir}                
\usepackage{graphicx} 
\usepackage{amsfonts, amsmath,amssymb}      

\usepackage{longtable}

\ifpdf
 \usepackage[pdftex,colorlinks=true,urlcolor=gray,pdfstartview=FitH,plainpages=false,pdfpagelabels,pagebackref=true]{hyperref}
 \usepackage{memhfixc}
 \hypersetup{
   pdftitle={Logics For XML},
   pdfauthor={Pierre Genev\`es},
   pdfsubject={Ph.D. Dissertation with Applications for the Static Analysis of XML-manipulating Programming Languages},
   pdfkeywords={Logic, regular tree languages, XPath, XML, modal logic, mu-calculus, static, analysis, containment, satisfiability}
 }
\else
\usepackage[pdfmark]{hyperref}
\usepackage{memhfixc}
\fi

\usepackage{stmaryrd}
\usepackage{mathpartir}
\usepackage{notations} 

\usepackage{multirow}

\usepackage{tikz} 
\usepackage{pgf}

\newtheorem{thm}{Theorem}[section]

\newtheorem{lem}[thm]{Lemma}
\newtheorem{prop}[thm]{Proposition}
\newtheorem{defn}[thm]{Definition}

\newenvironment{mypfo}{\noindent{\emph{Proof outline:\ }}}{\hfill$\Box$ }

\newcommand{\ntobin}[1]{\mathcal{\beta}({#1})}
\newcommand{\binton}[1]{\mathcal{\beta}^{-1}({#1})}

\newcommand{\tsbin}[1]{\mathcal{B}({#1})}

\newcommand{\setofbintrees}{\mathcal{T}_{\Sigma}^2}
\newcommand{\setofunrankedtrees}{\mathcal{T}_{\Sigma}^n}
\newcommand{\setofhedges}{\mathcal{H}_{\Sigma}}

\renewcommand{\phi}{\varphi}

\newcommand{\automaton}[1]{\mathcal{A} \llbracket {#1} \rrbracket}

\newcommand{\lang}[1]{\mathcal{L}({#1})}

\newcommand{\tuplerep}[1]{\widetilde{{#1}}}
\newcommand{\matricialrep}[1]{\ddddot{{#1}}}
\newcommand{\treerep}[1]{\widehat{{#1}}}

\newcommand{\dom}[1]{\text{\emph{#1}}}

\newcommand{\lxpath}{\mathcal{L}_{\text{XPath}}}

\newcommand{\xpath}[1]{$${#1}$$}
\newcommand{\name}[1]{\text{#1}}
\newcommand{\step}[2]{\text{{#1}::}{#2}}
\newcommand{\axis}[1]{\text{#1}}
\newcommand{\axisvar}{\emph{a}}

\newcommand{\punion}{\shortmid}

\newcommand{\qualif}[2]{{#1}\text{[}{#2}\text{]}}
\newcommand{\op}[1]{\mathbin{\text{\small {#1}}}}

\newcommand{\fun}[2]{\text{\emph{{#1}}}({#2})}
\newcommand{\rbefore}{\ll}
\newcommand{\semanEFunc}{\mathcal{S}_e}
\newcommand{\semanPFunc}{\mathcal{S}_p}
\newcommand{\semanQFunc}{\mathcal{S}_q}
\newcommand{\semanAFunc}{\mathcal{S}_{\axisvar}}

\newcommand{\semanE}[2]{\semanEFunc \llbracket {#1} \rrbracket #2}
\newcommand{\semanP}[2]{\semanPFunc \llbracket {#1} \rrbracket #2}
\newcommand{\semanQ}[2]{\semanQFunc \llbracket {#1} \rrbracket #2}
\newcommand{\semanA}[2]{\semanAFunc \llbracket {\text{{#1}}} \rrbracket #2}

\newcommand{\tefun}{\mathcal{S}_e}
\newcommand{\tpfun}{\mathcal{S}_p}
\newcommand{\tqfun}{\mathcal{S}_q}
\newcommand{\tafun}{\mathcal{S}_{\axisvar}}

\newcommand{\teinterp}[3]{\tefun \llbracket {#1} \rrbracket_{({#2},{#3})}}
\newcommand{\tpinterp}[3]{\tpfun \llbracket {#1} \rrbracket_{({#2},{#3})}}
\newcommand{\tqinterp}[3]{\tqfun \llbracket {#1} \rrbracket_{({#2},{#3})}}
\newcommand{\tainterp}[3]{\tafun \llbracket {\text{{#1}}} \rrbracket_{({#2},{#3})}}

\newcommand{\xaf}[2]{f_{({#1},{#2})}}
\newcommand{\xap}[2]{p_{({#1},{#2})}}
\newcommand{\xaa}[2]{a_{({#1},{#2})}}

\newcommand{\kripkexmltrees}{\mathcal{K}_T}
\newcommand{\kripkexmltreenodes}{\mathcal{W}(\kripkexmltrees)}
\newcommand{\kripkexmltreeroot}[1]{\texttt{root}(#1)}

\newcommand{\xefun}{\mathcal{S}_e}
\newcommand{\xpfun}{\mathcal{S}_p}
\newcommand{\xqfun}{\mathcal{S}_q}
\newcommand{\xafun}{\mathcal{S}_{\axisvar}}

\newcommand{\xeinterp}[2]{\xefun \llbracket {#1} \rrbracket_{{#2}}}
\newcommand{\xpinterp}[2]{\xpfun \llbracket {#1} \rrbracket_{{#2}}}
\newcommand{\xqinterp}[2]{\xqfun \llbracket {#1} \rrbracket_{{#2}}}
\newcommand{\xainterp}[2]{\xafun \llbracket {\text{{#1}}} \rrbracket_{{#2}}}

\newcommand{\logicEFunc}{\mathcal{W}_e}
\newcommand{\logicPFunc}{\mathcal{W}_p}
\newcommand{\logicQFunc}{\mathcal{W}_q}

\newcommand{\logicE}[3]{\logicEFunc \llbracket {#1} \rrbracket ^{#3}_{#2}}
\newcommand{\logicP}[3]{\logicPFunc \llbracket {#1} \rrbracket ^{#3}_{#2}}
\newcommand{\logicQ}[2]{\logicQFunc \llbracket {#1} \rrbracket _{#2}}

\newcommand{\et}{\wedge}
\newcommand{\ou}{\vee}
\newcommand{\equalsdef}{\stackrel{\text{def}}{=}}

\newcommand{\optlogicEFunc}{\mathcal{W}'_e}
\newcommand{\optlogicPFunc}{\mathcal{W}'_p}
\newcommand{\optlogicQFunc}{\mathcal{W}'_q}
\newcommand{\optlogicE}[4]{\optlogicEFunc \llbracket {#1} \rrbracket({#2},{#3}, {#4})}
\newcommand{\optlogicP}[4]{\optlogicPFunc \llbracket {#1} \rrbracket({#2}, {#3},{#4})}
\newcommand{\optlogicQ}[3]{\optlogicQFunc \llbracket {#1} \rrbracket({#2},{#3})}

\newcommand{\restrict}[2]{{#1} \; [{#2}] \;}

\newcommand{\levelEFunc}{L_e}
\newcommand{\levelPFunc}{L_p}

\newcommand{\levelE}[2]{\levelEFunc \llbracket {#1} \rrbracket _{#2}}
\newcommand{\levelP}[2]{\levelPFunc \llbracket {#1} \rrbracket _{#2}}

\newcommand{\N}{\mathbb{N}}

\newcommand{\mucalcEFunc}{E^\rightarrow}
\newcommand{\mucalcPFunc}{P^\rightarrow}
\newcommand{\mucalcAFunc}{A^\rightarrow}
\newcommand{\mucalcE}[2]{\mucalcEFunc \llbracket{#1}  \rrbracket_{#2}} 
\newcommand{\mucalcP}[2]{\mucalcPFunc \llbracket {#1}  \rrbracket_{#2}}
\newcommand{\mucalcA}[2]{\mucalcAFunc \llbracket {#1} \rrbracket_{#2}}

\newcommand{\mucalcPexFunc}{P^\leftarrow}
\newcommand{\mucalcQexFunc}{Q^\leftarrow}
\newcommand{\mucalcAexFunc}{A^\leftarrow}
\newcommand{\mucalcPex}[2]{\mucalcPexFunc \llbracket{#1}  \rrbracket_{#2}}
\newcommand{\mucalcQex}[2]{\mucalcQexFunc \llbracket{#1}  \rrbracket_{#2}}
\newcommand{\mucalcAex}[2]{\mucalcAexFunc \llbracket{#1}  \rrbracket_{#2}}

\newcommand{\ourimplies}{\Rightarrow}
\newcommand{\ourequiv}{\Leftrightarrow}

\newcommand{\contained}{\subset}
\newcommand{\contains}{\supset}

\newcommand{\equivalent}{\equiv}
\newcommand{\norelation}{\not \sim}

\newcommand{\atomprop}{\nodelabel}
\newcommand{\atomprops}{\alphabet}

\newcommand{\umod}[1]{\left[ {#1} \right]}
\newcommand{\emod}[1]{\left< {#1} \right>}

\newcommand{\isleftsucc}{\overline{1}}
\newcommand{\isrightsucc}{\overline{2}}
\newcommand{\hasleftsucc}{1}
\newcommand{\hasrightsucc}{2}

\newcommand{\domFProg}{\{\hasleftsucc, \hasrightsucc\}}
\newcommand{\domBProg}{\{\isleftsucc, \isrightsucc\}}
\newcommand{\domProg}{\{\hasleftsucc, \hasrightsucc, \isleftsucc, 
\isrightsucc\}}

\newcommand{\fullmucalculus}{\mathcal{L}_\mu^{\text{full}}}
\newcommand{\mulogic}{\mathcal{L}_\mu}

\newcommand{\false}{\bot}
\newcommand{\true}{\top}

\newcommand{\signature}{\Xi}

\newcommand{\powerset}[1]{2^{{#1}}}
\newcommand{\setcardinal}[1]{\left| #1 \right|}

\newcommand{\fc}{1}
\newcommand{\ns}{2}

\newcommand{\invfc}{\overline{\fc}}
\newcommand{\invns}{\overline{\ns}}

\newcommand{\fcrel}{\prec_\text{fc}}
\newcommand{\nsrel}{\prec_\text{ns}}

\newcommand{\nsrelplus}{\prec_\text{ns}^+}

\newcommand{\someFCverifies}{\left<\fc\right>}
\newcommand{\someNSverifies}{\left<\ns\right>}

\newcommand{\allFCverify}{\left[\fc\right]}
\newcommand{\allNSverify}{\left[\ns\right]}

\newcommand{\someinvFCverifies}{\left<\invfc\right>}
\newcommand{\someinvNSverifies}{\left<\invns\right>}

\newcommand{\allinvFCverify}{\left[\invfc\right]}
\newcommand{\allinvNSverify}{\left[\invns\right]}

\newcommand{\musem}[3]{\llbracket {#1} \rrbracket^{{#2}}_{{#3}}}
\newcommand{\any}{\cdot}

\newcommand{\kripke}[1]{\left< {#1} \right>}

\newcommand{\phiroot}{\phi_{\text{root}}}
\newcommand{\phift}{\phi_{\text{ft}}}

\newcommand{\instr}[1]{\textbf{{#1}}~}
\newcommand{\func}[1]{{\texttt{{#1}}}}

\newcommand{\startatom}{\circledS}
\newcommand{\startatomopt}{\circ}

\newcommand{\cl}[1]{\text{cl}({#1})}
\newcommand{\extendedcl}[1]{\text{cl}^*({#1})}
\newcommand{\lean}[1]{\text{Lean}({#1})}
\newcommand{\types}[1]{\text{Typ}({#1})}

\newcommand{\expand}[1]{\text{exp}({#1})}

\newcommand{\existsvectnesp}[2]{\exists #1 ~\left[~ #2 ~\right]}
\newcommand{\existsvect}[2]{\existsvectnesp{#1}{#2}~}
\newcommand{\bigexists}[2]{\begin{array}{l}~\begin{huge}~\text{$\exists$}~\end{huge}~ \left[~{#2}~\right]\\\begin{small}\text{${#1}$}\end{small}\end{array}}

\newcommand{\extendedin}{\stackrel{.}{\in}}
\newcommand{\notextendedin}{\stackrel{.}{\notin}}

\newcommand{\inbase}[4]{#1 \extendedin #2 \Longrightarrow (#3,#4)}
\newcommand{\notinbase}[4]{#1 \notextendedin #2 \Longrightarrow (#3,#4)}

\newcommand{\erel}{\rightarrow_e}
\newcommand{\ap}[1]{\Sigma({#1})}

\newcommand{\vect}[1]{\vec{{#1}}}

\newcommand{\charact}[1]{\chi_{{#1}}}

\newcommand{\needwitness}[2]{\text{isparent}_{{#1}}({#2})}
\newcommand{\iswitness}[2]{\text{ischild}_{{#1}}({#2})}

\newcommand{\status}[2]{\text{status}_{{#1}}({#2})}

\newcommand{\implication}{\rightarrow}
\newcommand{\equivalence}{\leftrightarrow}

\newcommand{\typebind}[2]{\text{let } \overline{{#1}_i.{#2}_i} \text{ in }}
\newcommand{\typebindifree}[2]{\text{let } \overline{{#1}.{#2}} \text{ in }}
\newcommand{\singletypebind}[2]{{\text{let } {#1}.{#2}} \text{ in }}
\newcommand{\doubletypebind}[4]{{\text{let } {#1}.{#2}, {#3}.{#4}} \text{ in }}

\newcommand{\lwss}{\mathcal{L}_\text{ws2s}}
\newcommand{\lcft}{\mathcal{L}_\text{cft}}
\newcommand{\lrtt}{\mathcal{L}_\text{rt}}
\newcommand{\lbtt}{\mathcal{L}_\text{bt}}
\newcommand{\ldtd}{\mathcal{L}_\text{dtd}}
\newcommand{\Tsem}[2]{\llbracket {#1} \rrbracket_{{#2}}}
\newcommand{\Ttomu}[1]{\llbracket {#1} \rrbracket}

\newcommand{\tou}{\mid}
\newcommand{\subtaggingrel}{\prec}

\newcommand{\leastfp}[1]{\text{\emph{lfp}}({#1})}

\newcommand{\nullable}[1]{\text{\emph{nullable}}({#1})}

\newcommand{\Tsucc}[2]{\text{\emph{succ}}_{#2}({#1})}

\newcommand{\satisfiabilitypb}{\semanE{e}{x} \neq \emptyset }
\newcommand{\containmentpb}{\semanE{e_1}{x} \subseteq \semanE{e_2}{x}}
\newcommand{\equivalencepb}{\semanE{e_1}{x} = \semanE{e_2}{x}}
\newcommand{\overlappb}{\semanE{e_1}{x} \cap \semanE{e_2}{x} \neq \emptyset}
\newcommand{\coveragepb}{\semanE{e_1}{x} \subseteq \bigcup_{2 \leq i \leq n}\semanE{e_i}{x}}
\newcommand{\setofdocs}{\mathcal{T}}

\newcommand{\phitype}{\phi_\setofdocs}
\newcommand{\phitested}{\phi_{\text{tested}}}

\newcommand{\nodelabel}{\sigma}
\newcommand{\nodelabelother}{\sigma_x}
\newcommand{\alphabet}{\Sigma}

\newcommand{\focusedtrees}{\mathcal{F}}
\newcommand{\focusedtree}[2]{(#1,#2)}
\newcommand{\focusedtreevar}{f}
\newcommand{\focusedtreesetvar}{F}

\newcommand{\starttrue}[1]{#1^{\circledS}}
\newcommand{\startunk}[1]{#1^\circ}

\newcommand{\tree}[2]{#1[#2]}
\newcommand{\treevar}{t}

\newcommand{\treelistnil}{\epsilon}
\newcommand{\treelistcons}[2]{#1 :: #2}
\newcommand{\treelistvar}{\mathit{tl}}

\newcommand{\backtree}[2]{#1[#2]}

\newcommand{\contexttop}[2]{(#1,\mathit{Top},#2)}
\newcommand{\contextnode}[4]{(#1,\backtree{#2}{#3},#4)}
\newcommand{\contextvar}{c}

\newcommand{\fname}[1]{\mathtt{nm}#1}

\newcommand{\godown}[1]{#1\emod{\hasleftsucc}}
\newcommand{\goright}[1]{#1\emod{\hasrightsucc}}
\newcommand{\goup}[1]{#1\emod{\isleftsucc}}
\newcommand{\goleft}[1]{#1\emod{\isrightsucc}}

\newcommand{\revappend}[2]{\mathtt{rev\_a}(#1,#2)}

\newcommand{\parentfocusedtree}[1]{\mathtt{parent}(#1)}
\newcommand{\nsiblingsfocusedtrees}[1]{\mathtt{nsibling}(#1)}
\newcommand{\psiblingsfocusedtrees}[1]{\mathtt{psibling}(#1)}
\newcommand{\fchildrenfocusedtrees}[1]{\mathtt{fchild}(#1)}
\newcommand{\rootfocusedtrees}[1]{\mathtt{root}(#1)}

\newcommand{\cyclefree}[5]{#1 \parallel #2 \vdash_{#3}^{#4} #5}

\newcommand{\varcf}{\Gamma}
\newcommand{\varenv}{\Delta}
\newcommand{\varignore}{I}
\newcommand{\varrecurse}{R}
\newcommand{\cfbot}{\bot}
\newcommand{\cfinject}[2]{#1 \lhd #2}
\newcommand{\cfextend}[2]{#1 + #2}
\newcommand{\cfbind}[2]{#1 : #2}
\newcommand{\cfnil}{\_}

\newcommand{\finterp}[2]{\llbracket #1 \rrbracket_{#2}}
\newcommand{\fpinterp}[3]{\llbracket #1 \rrbracket_{#2}^{#3}}
\newcommand{\setof}[2]{\left\{ #1 \;|\; #2\right\}}
\newcommand{\multilinesetof}[2]{\{ #1 \;|\; #2\}}
\newcommand{\subst}[2]{^{#1} \!/\! _{#2}}
\newcommand{\isdefined}[1]{#1 \text{ defined}}
\newcommand{\isundefined}[1]{#1 \text{ undefined}}

\newcommand{\eqdef}{\stackrel{\text{\tiny def}}{=}}

\newcommand{\syntaxdef}{\mathrel{::=}}

\newcommand{\syntaxtable}[1]{
  \def\entry##1[##2]##3[##4]{
    {##1} & \syntaxdef & \hspace{3cm} & \!\!\!\! \mbox{##2}
    \\    &     & {##3} & \mbox{##4} }
  \def\singleentry##1[]##2[##3]{
  {##1} & \syntaxdef & {##2} & \!\!\!\! \mbox{##3} }
  \def\oris##1[##2]{
    \\    & |   & {##1} & \mbox{##2} }
  \def\orisopt##1[##2]{
    \\ \left(   & |   & {##1} & \mbox{##2} \right) }
  \begin{array}{rcll}
  #1
  \end{array}
  }
\newcommand{\smallsyntax}[1]{\[\syntaxtable{#1}\]}

\newcommand{\finiteunfold}[1]{\mathit{unf}(#1)}

\newcommand{\atomin}[1]{\sigma({#1})}

\newcommand{\typeformula}[1]{\phi_c({#1})}
\newcommand{\descofsatisfies}[2]{\func{dsat}({#1},{#2})}
\newcommand{\tripletype}[1]{\func{type}({#1})}

\newcommand{\emptypath}{\epsilon}
\newcommand{\pathvar}{\rho}
\newcommand{\conspath}[2]{#1 #2}            

\newcommand{\depth}[1]{\mathit{depth}(#1)}

\newcommand{\treetypevar}{\mathcal{T}}
\newcommand{\treetypenode}[1]{#1(\bullet)}
\newcommand{\treetypedown}[1]{#1\emod{\hasleftsucc}}
\newcommand{\treetyperight}[1]{#1\emod{\hasrightsucc}}

\newcommand{\marked}[1]{\starttrue{#1}}
\newcommand{\unmarked}[1]{\startunk{#1}}

\newcommand{\allcases}{\_}

\newcommand{\treeval}[3]{#1 \Vdash_{#2} #3}

\newcommand{\phirootexpr}{\mu Z. \neg \emod{\overline{1}}\true \ou \emod{\overline{2}}Z}

\newcommand{\murec}[3]{\mu \overline{#1.#2} \text{ in } #3}
\newcommand{\nurec}[3]{\nu \overline{#1.#2} \text{ in } #3}

\newcommand{\afletmu}[2]{\text{let}_\mu~{({#1}_i . {#2}_i)}_{1 \leq i \leq m} \text{ in }}
\newcommand{\afletmuifree}[2]{\text{let}_\mu~{({#1} . {#2})}_{1 \leq i \leq m} \text{ in }}

\newcommand{\treetypelabel}[2]{{#1}\texttt{[}{#2}\texttt{]}}

\newcommand{\mychapter}[1]{\chapter{{#1}}}

\newcommand{\treedom}[1]{\text{Dom}({#1})}
\newcommand{\infch}{\prec_\text{ch}}
\newcommand{\infsb}{\prec_\text{sb}}
\newcommand{\infchstar}{\prec_\text{ch}^*}
\newcommand{\infsbstar}{\prec_\text{sb}^*}

\newcommand{\cduce}{\mathbb{C}\text{Duce}}

\newcommand{\hname}[1]{\textsc{#1}}

\makeatletter
\def\thickhrulefill{\leavevmode \leaders \hrule height 1ex \hfill \kern \z@}
\def\@makechapterhead#1{%
  \vspace*{10\p@}%
  {\parindent \z@ \raggedleft \reset@font
            \scshape \@chapapp{} \thechapter
        \par\nobreak
        \interlinepenalty\@M
    \Huge \bfseries #1\par\nobreak 
    \hrulefill
    \par\nobreak
    \vskip 100\p@
  }}
\def\@makeschapterhead#1{%
  \vspace*{10\p@}%
  {\parindent \z@ \raggedleft \reset@font
            \scshape \vphantom{\@chapapp{} \thechapter}
        \par\nobreak
        \interlinepenalty\@M
    \Huge \bfseries #1\par\nobreak
    \hrulefill
    \par\nobreak
    \vskip 100\p@
  }}

\begin{document} 

\pagestyle{ruled}

\frontmatter
\newif\ifinpgcover\inpgcoverfalse
\newif\ifdedication\dedicationfalse

\ifinpgcover
  
\newcommand{\g}[1]{#1} 

\begin{titlingpage}
\pagestyle{empty}
\begin{center}
\textbf{INSTITUT NATIONAL POLYTECHNIQUE DE GRENOBLE}
\\[2\baselineskip]
\hfill
\begin{minipage}{5cm}
\begin{center}
\begin{small}
\textsf{\emph{N$^o$ attribu\'e par la biblioth\`eque}} \\
\begin{tabular}{|l|l|l|l|l|l|l|l|l|l|}
\ &\ &\ &\ &\ &\ &\ &\ &\ &\ \\ \hline
\end{tabular}
\end{small}
\end{center}
\end{minipage}
\\[2\baselineskip]
{\LARGE \textbf{TH\`ESE}}
\\[\baselineskip]
pour obtenir le grade de
\\[0.5\baselineskip]
\textbf{DOCTEUR DE l'INPG}
\\[0.5\baselineskip]
\textbf{Sp\'ecialit\'e: Informatique}
\\[0.5\baselineskip]
pr\'epar\'ee au laboratoire
\\[0.5\baselineskip]
\textbf{\g{INRIA Rh\^one-Alpes, projet WAM}},
\\[0.5\baselineskip]
dans le cadre de l'Ecole Doctorale
\\[0.5\baselineskip]
\textbf{\g{Math\'ematiques Informatique Sciences et Technologies de l'Information}}
\\[\baselineskip]
pr\'esent\'ee et soutenue publiquement par
\\[\baselineskip]
{\Large Pierre \textsc{Genev\`es}}
\\[\baselineskip]
{le \large{4 D\'ecembre 2006}\normalsize} \vfill \hrule
~\\[5mm]
{\huge\textsl{Logiques pour XML}}
~\\[5mm]
\hrule \vfill Directeur de th\`ese:
\\[\baselineskip]
Vincent \textsc{Quint} \\
\vfill JURY
\\[\baselineskip]
\begin{tabular}{l@{\protect\hspace{0.5cm}}ll@{\protect\hspace{1.0cm}}l}
M. & Giorgio & \textsc{Ghelli} & Rapporteur \\
M. & Denis & \textsc{Lugiez} & Rapporteur \\
M. & Makoto & \textsc{Murata} & Rapporteur \\
M. & Vincent & \textsc{Quint} & Directeur de th\`ese \\
Mme. & Christine & \textsc{Collet} & Pr\'esidente \\
M. & Nabil & \textsc{Laya\"ida} & Invit\'e \\
\end{tabular}
\end{center}
\end{titlingpage}


\else  
  
\begin{titlingpage} 
\addcontentsline{toc}{chapter}{Title Page}
\begin{centering} 
Institut National Polytechnique de Grenoble
\vspace{4cm}

\begin{HUGE}
\textbf{Logics for XML} \\
\end{HUGE}
\vspace{1cm}
\begin{large}
Pierre \textsc{Genev\`es}  \\
\end{large} 
\vspace{8cm}
\end{centering}

\noindent
Thesis presented in partial fulfillment of the requirements for the degree of Ph.D. in Computer Science and Software Systems from the Institut National Polytechnique de Grenoble. Dissertation prepared at the Institut National de Recherche en Informatique et Automatique, Montbonnot, France. Thesis defended on the $4^\text{th}$ of December 2006.

\vspace{0.5cm}
\begin{center} 
Board of examiners: \\
\vspace{0.4cm}
\begin{tabular}{rl}
Giorgio \textsc{Ghelli}    & Referee \\
Denis \textsc{Lugiez}      & Referee \\
Makoto \textsc{Murata}     & Referee \\
Christine \textsc{Collet}  & Examiner \\
Vincent \textsc{Quint}     & Ph.D. advisor \\
Nabil \textsc{Laya\"ida}   & Invited Member
\end{tabular}
\end{center} 
\end{titlingpage} 
\fi

\ifdedication
\thispagestyle{empty} 
\pagenumbering{roman} \setcounter{page}{3} 
\newenvironment{dedication} 
  {\cleardoublepage \thispagestyle{empty} \vspace*{\stretch{1}} \begin{center}} 
  {\end{center} \vspace*{\stretch{3}} \clearpage} 
\begin{dedication} 

\emph{To ......} 

\end{dedication} 
\thispagestyle{empty} \cleardoublepage 

  \pagenumbering{roman} \setcounter{page}{5} 
\else 
  \pagenumbering{roman} \setcounter{page}{3}   
\fi
\newif\ifglobalcompil\globalcompiltrue

  \ifglobalcompil
  \else 
    \input{Preamble} 
    \input{Markup}                       
    \newcommand{\chapterabstract}[1]{Chapter Abstract: #1}
    \begin{document} 
  \fi

\chapter*{Abstract\markboth{Abstract}{Abstract}} 
\addcontentsline{toc}{chapter}{Abstract}

This thesis describes the theoretical and practical foundations of a system for the static analysis of XML processing languages. The system relies on a fixpoint temporal logic with converse, derived from the $\mu$-calculus, where models are finite trees. This calculus is expressive enough to capture regular tree types along with multi-directional navigation in trees, while having a single exponential time complexity. Specifically the decidability of the logic is proved in time $2^{O(n)}$ where $n$ is the size of the input formula.

Major XML concepts are linearly translated into the logic: XPath navigation and node selection semantics, and regular tree languages (which include DTDs and XML Schemas). Based on these embeddings, several problems of major importance in XML applications are reduced to satisfiability of the logic. These problems include XPath containment, emptiness, equivalence, overlap, coverage, in the presence or absence of regular tree type constraints, and the static type-checking of an annotated query.

The focus is then given to a sound and complete algorithm for deciding the logic, along with a detailed complexity analysis, and crucial implementation techniques for building an effective solver. Practical experiments using a full implementation of the system are presented. The system appears to be efficient in practice for several realistic scenarios.

The main application of this work is a new class of static analyzers for programming languages using both XPath expressions and XML type annotations (input and output). Such analyzers allow to ensure at compile-time valuable properties such as type-safety and optimizations, for safer and more efficient XML processing.

\chapter*{R\'esum\'e \markboth{R\'esum\'e}{French Abstract}}    

Cette th\`ese pr\'esente les fondements th\'eoriques et pratiques d'un syst\`eme pour l'analyse statique de langages manipulant des documents et donn\'ees XML. Le syst\`eme s'appuie sur une logique temporelle de point fixe avec programmes inverses, d\'eriv\'ee du $\mu$-calcul modal, dans laquelle les mod\`eles sont des arbres finis. Cette logique est suffisamment expressive pour prendre en compte les langages r\'eguliers d'arbres ainsi que la navigation multidirectionnelle dans les arbres, tout en ayant une complexit\'e simplement exponentielle. Plus pr\'ecis\'ement, la d\'ecidabilit\'e de cette logique est prouv\'ee en temps $2^{O(n)}$ o\`u $n$ est la taille de la formule dont le statut de v\'erit\'e est d\'etermin\'e.

Les principaux concepts de XML sont traduits lin\'eairement dans cette logique. Ces concepts incluent la navigation et la s\'emantique de s\'election de noeuds du langage de requ\^etes XPath, ainsi que les langages de sch\'emas (incluant DTD et XML Schema). Gr\^ace \`a ces traductions, les probl\`emes d'importance majeure dans les applications XML sont r\'eduits \`a la satisfaisabilit\'e de la logique. Ces probl\`emes incluent notamment l'inclusion, la satisfaisabilit\'e, l'\'equivalence, l'intersection, le recouvrement des requ\^etes, en pr\'esence ou en l'absence de contraintes r\'eguli\`eres d'arbres, et le typage statique d'une requ\^ete annot\'ee.

Un algorithme correct et complet pour d\'ecider la logique est propos\'e, accompagn\'e d'une analyse d\'etaill\'ee de sa complexit\'e computationnelle, et des techniques d'implantation cruciales pour la r\'ealisation d'un solveur efficace en pratique. Des exp\'erimentations avec l'implantation compl\`ete du syst\`eme sont pr\'esent\'ees. Le syst\`eme appara\^it efficace et utilisable en pratique sur plusieurs sc\'enarios r\'ealistes.

La principale application de ce travail est une nouvelle classe d'analyseurs statiques pour les langages de programmation utilisant des requ\^etes XPath et des types r\'eguliers d'arbres. De tels analyseurs permettent de s'assurer, au moment de la compilation, de propri\'et\'es importantes comme le typage correct des programmes ou leur optimisation, pour un traitement plus s\^ur et plus efficace des donn\'ees XML.

 \ifglobalcompil
  \else 
   \end{document}
  \fi
\newif\ifglobalcompil\globalcompiltrue

  \ifglobalcompil
  \else 
    \input{Preamble} 
    \input{Markup}                       
    \newcommand{\chapterabstract}[1]{Chapter Abstract: #1}
    \begin{document} 
  \fi

\chapter*{Preface\markboth{Preface}{Preface}} 
\addcontentsline{toc}{chapter}{Preface}

This manuscript presents my research work done at the \emph{Ins\-ti\-tut Na\-tio\-nal de Re\-cher\-che en In\-for\-ma\-ti\-que et Automatique} (INRIA Rh\^one-Alpes, France), from November 2003 to September 2006, within the WAM research project. 
The work was supported by a personal Ph.D. grant from the french ministry for research (\emph{Minist\`ere d\'el\'egu\'e \`a la Recherche}).

This manuscript focuses on presenting the main results obtained for the static analysis of XML specifications using logical formalisms. 
The list of the main articles and communications I have authored or co-authored during my Ph.D. thesis follows. Some results are not presented in this manuscript (in particular, during this period I have spent several months at IBM T.J. Watson Research Center, New York, United States, working on scalable runtime XML processing architectures, for which I received an IBM invention achievement award).



\paragraph{Main Publications} 

\begin{itemize}

\bibitem[1]{preface:geneves-pldi07}
Pierre Genev{\`e}s, Nabil Laya\"ida, and Alan Schmitt.
\newblock Efficient Static Analysis of XML Paths and Types.
\newblock To appear in \emph{PLDI'07: Proceedings of the 2007 ACM Conference on Programming Language Design and Implementation}, San Diego, California, USA, June 2007. ACM Press.

\bibitem[2]{preface:geneves-tois06}
Pierre Genev\`es and Nabil Laya\"ida.
\newblock A system for the static analysis of {XPath}.
\newblock {\em ACM Transactions on Information Systems (TOIS)}, 24(4), October
  2006.

\bibitem[3]{preface:geneves-dke06}
Pierre Genev\`es and Nabil Laya\"ida.
\newblock Deciding {XPath} containment with {MSO}.
\newblock {\em  To Appear in Elsevier Data \& Knowledge Engineering (DKE)}, 2007.

\bibitem[4]{preface:geneves-doceng06}
Pierre Genev{\`e}s and Nabil Laya\"ida.
\newblock Comparing {XML} Path Expressions.
\newblock In \emph{DocEng'06: Proceedings of the 2006 ACM Symposium on Document
  Engineering}, pages 65--74,  Amsterdam, The Netherlands, October 2006. ACM Press.


\bibitem[5]{preface:geneves-doceng05}
Pierre Genev\`es and Kristoffer~H{\o}gsbro Rose.
\newblock Compiling {XPath} for streaming access policy.
\newblock In \emph{DocEng '05: Proceedings of the 2005 ACM Symposium on Document
  Engineering}, pages 52--54, Bristol, UK, November 2005. ACM Press.

\bibitem[6]{preface:geneves-doceng04}
Pierre Genev{\`e}s and Jean-Yves Vion-Dury.
\newblock Logic-based {XPath} optimization.
\newblock In \emph{DocEng'04: Proceedings of the 2004 ACM Symposium on Document
  Engineering}, pages 211--219, Milwaukee, Wisconsin, USA, October 2004. ACM Press.

\bibitem[7]{preface:geneves-tphols04}
Pierre Genev\`es and Jean-Yves Vion-Dury.
\newblock {XPath} formal semantics and beyond: A {Coq}-based approach.
\newblock In \emph{TPHOLs '04: Emerging Trends Proceedings of the 17th
  International Conference on Theorem Proving in Higher Order Logics}, pages
  181--198, Salt Lake City, Utah, USA, August 2004. University Of
  Utah.

\bibitem[8]{preface:geneves-patent04}
Kristoffer~H{\o}gsbro Rose and Pierre Genev\`es.
\newblock Optimization of {XPath} expressions for evaluation upon streaming
  {XML} data, \emph{IBM Research Patent Filed}, May 2004. This patent was awarded an invention achievement award, given by Samuel J. Palmisano (chairman of IBM Corporation) in July 2004.

\end{itemize}

 \ifglobalcompil
  \else 
   \bibliographystyle{alpha}
   \bibliography{references}
   \end{document}
  \fi
\chapter*{Acknowledgements\markboth{Acknowledgements}{Acknowledgements}} 
\addcontentsline{toc}{chapter}{Acknowledgements}
I would like to take this opportunity to thank the many people who have contributed either directly or indirectly to the development of this thesis.

My acknowledgements first go to Vincent Quint, who accepted me as a PhD candidate in his team, and provided me with a high quality research environment. I also thank him for always having had confidence in me, and letting me freely choose my research directions and the way of investigating them. In his team I have met my colleague and friend Nabil Laya\"ida with whom I have enjoyed sharing many happy moments doing research. 

Giorgio Ghelli, Denis Lugiez, and Makoto Murata honoured me by accepting the role of referee for this dissertation. I would like to thank them for accepting this task. I also thank Christine Collet for accepting the role of examiner for this dissertation.

I am grateful to Alan Schmitt for his insights in the enjoyable collaboration from which the two final chapters of this dissertation benefitted; and since he is very pleasant to work with. I would like to thank Benjamin C. Pierce who was at the origin of my meeting and subsequent collaboration with Alan, following our discussion during a visit at INRIA.

I would like to thank Bob Schloss for giving me several opportunities to join his research team at IBM Watson. I am grateful to Kristoffer H. Rose for being my mentor during my summers spent there.

I also thank Akihiko Tozawa for fruitful discussions by email, Fr\'ed\'eric Lang for helpful discussions, and Jean-Yves Vion-Dury for kindly introducing me to research during my first year.

Finally, I would like to thank all the people from the B aisle of the INRIA building in Montbonnot, who are definitely responsible for the sympathic and enthousiastic research atmosphere at INRIA Rh\^one Alpes.

On a more personal note, I would like to thank my parents and my brother for their unconditional support. I also thank my friends for their continuing support and everything else. Thank you all.
\maxtocdepth{subsection}
\setsecnumdepth{subsection}
\settocdepth{subsection}
\maxsecnumdepth{subsection}
\cleardoublepage 
\renewcommand{\contentsname}{Table of Contents}
\tableofcontents 
\cleardoublepage 
\listoffigures
\cleardoublepage 
\markboth{List of Notations}{List of Notations}
\pagestyle{plain}
\chapter*{Notations}\addcontentsline{toc}{chapter}{List of Notations}
\pagestyle{ruled}
\printnotations

\cleardoublepage 

\mainmatter
\newif\ifglobalcompil\globalcompiltrue
  \ifglobalcompil
  \else 
    \input{Preamble} 
    \input{Markup}                       
    \newcommand{\chapterabstract}[1]{Chapter Abstract: #1}
    \begin{document} 
  \fi

\mychapter{Introduction}
\label{introduction}
\section{Motivation and Objectives}


This work was initially motivated by the need for efficient static type checkers for XML processing languages.
Such programming languages use schemas \cite{xml-schemas} and XPath \cite{xpath} queries as first class language constructs. Current examples of these languages include the W3C recommendation XSLT \cite{xslt} for the transformation of XML documents, and the forthcoming XQuery \cite{xquery} recommendation for querying XML databases. Providing such languages with decidable and efficient static type systems has been one of the major research challenges over the last decade, notably gathering the programming language, database theory, structured documents, and theoretical computer science communities. This work follows the research effort initiated in \cite{murata-pdp96, tozawa-doceng01, milo-jcss03, hosoya-toit03}.

This work resulted in the design of a new logic of finite trees adapted for XML, and its decision procedure, presented in this dissertation. The logical solver has been implemented as the core of a system for the general static analysis and type-checking of XML specifications. The system can be used as a component of static analyzers for programming languages manipulating both XPath expressions and XML type annotations. 


This dissertation presents the theoretical investigations that led to the foundations of this new logic of finite trees, along with the algorithmic bases and implementation principles on which the logical solver relies. These discoveries are applied to the resolution of XML type-checking problems, which are embedded in the logic. Solved problems include static typing of XPath in the presence of regular tree type constraints.



\subsection{XML Documents and Schemas}

\emph{Extensible Markup Language} (XML) \cite{xml} is a text file format for representing tree structures in a standard form. 
 
The whole structure of an XML document, if we abstract over less important details, is a tree of variable arity, in which nodes (also called \emph{elements} in the XML jargon) are labeled, 
leaves of the tree are text nodes, and the ordering between children of a node is significant.
XML can be seen as a concrete syntax for describing such tree structures using mark-up texts. An example of an XML document is as follows:
\label{introduction:well-formed-sample-xml-doc}
\begin{verbatim}
<plant>
  <category>Vascular</category>
  <tissue>
    <name>Phloem</name>
    <def>The phloem is a living tissue that carries organic
         nutrients to all parts of the plant where needed.</def>
    <note>In trees, the phloem is part of the bark.</note>
  </tissue>
</plant>
\end{verbatim}


An element is described by a pair of an opening tag $<...>$ and an closing tag $</...>$, between which the element content is inserted. In the previous example, ``\texttt{plant}'', ``\texttt{category}'', ``\texttt{tissue}'', ``\texttt{name}'', ``\texttt{def}'', and ``\texttt{note}'' are labels (\emph{tag names} in the XML jargon). 

\begin{figure}[h]
\centering

\begin{tikzpicture}
\tikzstyle{every node}=[ball color=white,circle,text=black]
\draw (1,1) node(root) {\begin{small}plant\end{small}};
\draw (-1,-0.5) node(n1) {\begin{tiny}category\end{tiny}};
\draw (3,-0.5) node(n2) {\begin{small}tissue\end{small}};

\draw (1,-2) node(n3) {\begin{small}name\end{small}};
\draw (3,-2) node(n4) {\begin{small}\;def\;\end{small}};
\draw (5,-2) node(n5) {\begin{small}note\end{small}};

\tikzstyle{every node}=[text=black]

\draw (-1,-2) node(n6) {Vascular};
\draw (1,-3.5) node(n7) {Phloem};
\draw (3,-3.5) node(n8) {The(...)};
\draw (5,-3.5) node(n9) {In trees(...)};

\draw [thick, ->, black] (root) -- (n1);
\draw [thick, ->, black] (root) -- (n2);

\draw [thick, ->, black] (n1) -- (n6);

\draw [thick, ->, black] (n2) -- (n3);
\draw [thick, ->, black] (n2) -- (n4);
\draw [thick, ->, black] (n2) -- (n5);

\draw [thick, ->, black] (n3) -- (n7);
\draw [thick, ->, black] (n4) -- (n8);
\draw [thick, ->, black] (n5) -- (n9);

\end{tikzpicture}
\caption{Sample Tree of a Well-Formed Document.}\label{introduction:fig:tree-sample}
\end{figure}
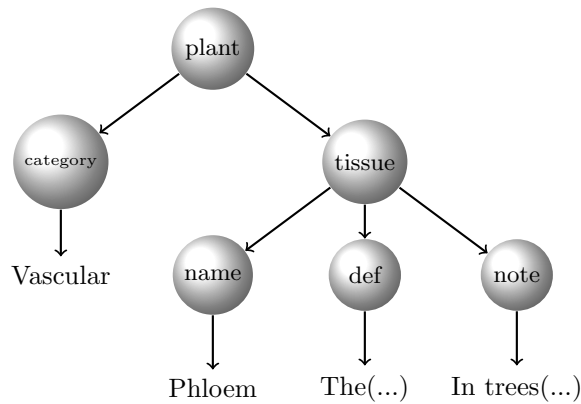

The XML specification does not fix a priori the set of allowed labels in an XML document nor it defines any semantics for labels. Only well-formedness conditions are defined in particular to ensure proper nesting of elements, which allows to consider XML documents as trees. For instance, Figure~\ref{introduction:fig:tree-sample} gives a more visual tree representation of the previous well-formed sample XML document.

The set of labels occurring in an XML document is determined by \emph{schemas} that can freely be defined by users. A \emph{schema} (also called an \emph{XML type}) is a description of constraints on the structure of documents such as allowed labels and their possible nesting structures. A schema thus defines a class of XML documents. Two levels of correctness can therefore be distinguished for XML documents:

\begin{itemize}
\item \emph{well-formedness} which applies to documents that obey the necessary and sufficient syntactic condition (defined by the XML specification) for being interpreted as trees;
\item \emph{validity} which applies to documents that conform to the additional constraints described by a given schema.
\end{itemize}

The validity of a document implies its well-formedness since the schema describes constraints on the tree and not on the text representation of the XML document. 

Each application can define its own data format by defining schemas, at a higher abstract level (tree structures). In that sense, XML is often said to be a metalanguage or a ``format for data formats''. 

Separating the two levels of correctness allows applications to share generic software tools for manipulating well-formed XML documents (parsers, editors, query and transformation tools...). These tools all implement the same syntactic conventions defined by the XML specification (such as the way of including comments, external fragments, special characters...). XML thus allows a first level of processing on an XML document as soon as it is well-formed, without making the additional and much stronger hypothesis that it is valid w.r.t to some schema. This genericity is one of XML strengths. As a consequence, we have seen unprecedented speed and range in the adoption of XML. A large number of schemas have been defined and are actually widely used in practice, for instance: XHTML (the XML version of HTML), SVG (for vector graphics), SMIL (for synchronized multimedia documents), MathML (for mathematical formulas), SOAP (for remote procedure calls), XBRL (for financial information), FIX (for securities transactions), SMD (for music), X3D (for 3D modeling) and CML (for chemical structures).

\subsection{XPath}

\label{introduction:xpath-intro}

XPath \cite{xpath, xpath20} has been introduced by the W3C as the standard query language for addressing and retrieving information in XML documents. It allows to navigate in XML trees and return a set of matching nodes. As such, XPath forms the essence of XML data access.

\begin{figure*}
\centering
  \includegraphics[width=10cm, keepaspectratio=true]{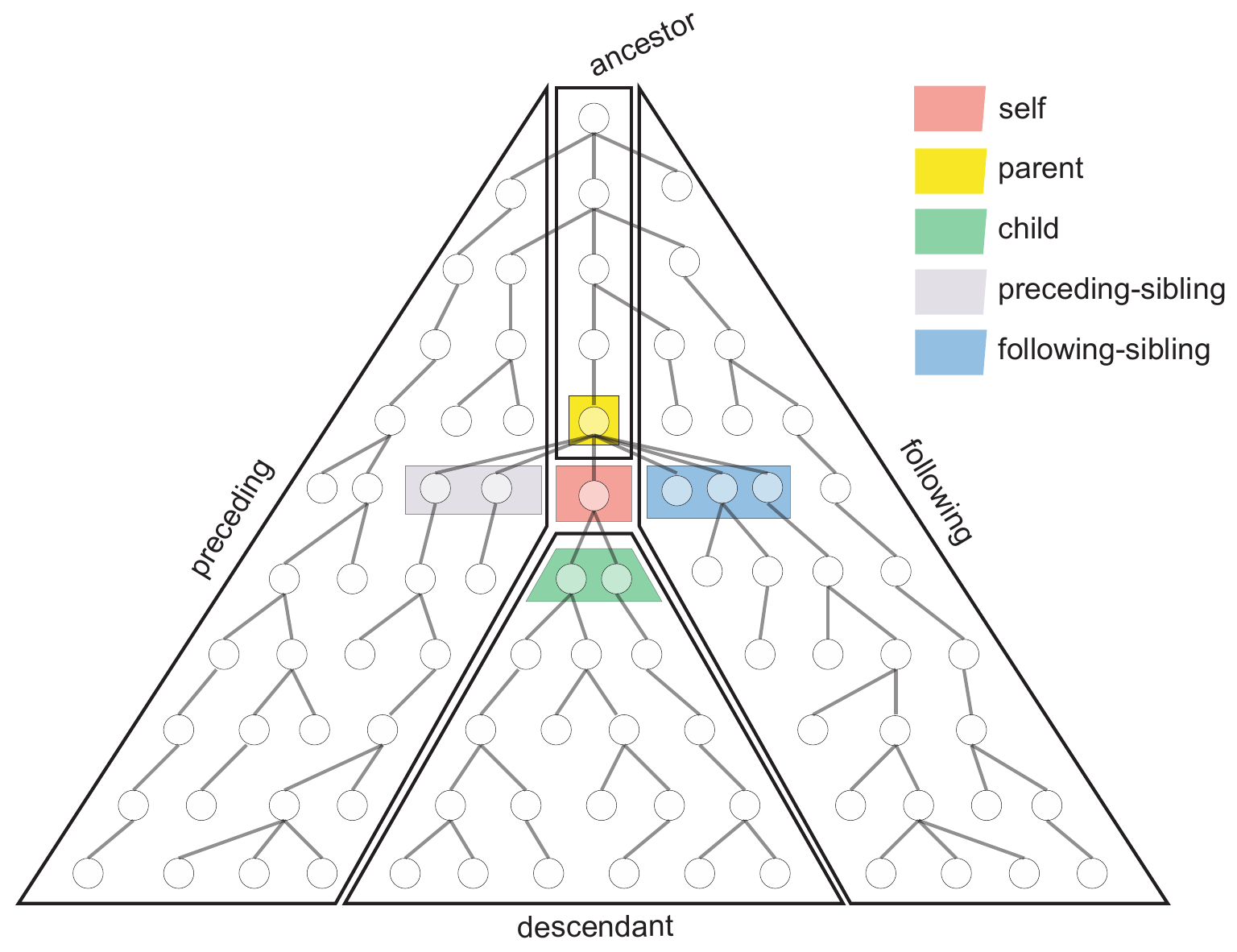}
   \caption{XPath Axes Partition from Context Node.}\label{introduction:fig:xpath-axes}
\end{figure*}

In their simplest form XPath expressions look like ``directory navigation paths''.  For example, the XPath expression
\xpath{/\name{book}/\name{chapter}/\name{section}}
navigates from the root of a document (designated by the leading slash ``/'') through the top-level ``book'' nodes, to their
``chapter'' child nodes, and on to their child nodes named ``section''.  The result of the evaluation of the entire expression is the set of all the ``section'' nodes that can be reached in this manner. Furthermore, at each step in the navigation the selected nodes can be filtered using qualifiers. A qualifier is a boolean expression between brackets that can test the existence or absence of paths. So if we ask for
\xpath{  /\name{book}/\name{chapter}/\qualif{\name{section}}{\name{citation}}}
then the result is \emph{all} ``section'' elements that have a least one child element named ``citation''. The situation becomes more interesting when combined with XPath's capability of searching along ``axes'' other than the shown ``children of'' axis.  Indeed the above XPath is a shorthand for
\xpath{/\axis{child::}\name{book}/\axis{child::}\name{chapter}/\qualif{\axis{child::}\name{section}}{\axis{child::}\name{citation}}}
where it is made explicit that each \emph{path step} is meant to search the ``child''
axis containing all children of the nodes selected at previous step.  If we instead asked for
\xpath{/\axis{child::}\name{book}/\qualif{\axis{descendant::}\name{*}}{\axis{child::}\name{citation}}}
then the last step selects nodes of any kind that are among the descendants of the top element ``book'' and have a ``citation'' sub-element. One may also use other axes such as ``\axis{preceding-sibling}'' for navigating backward through nodes of the same parent, or ``\axis{ancestor}'' for navigating upward recursively (see Figure~\ref{introduction:fig:xpath-axes}). \emph{Document order} is defined as the order in which a depth-first tree traversal visits nodes. Axes that perform navigation in reverse document order are called \emph{reverse axes} (or alternatively \emph{backward} or \emph{upward} axes in the literature).

Previous examples are \emph{absolute} XPath expressions as they start with a ``/'' which refers to the root.
The meaning of a \emph{relative} expression (without the leading ``/'') is defined with respect to a \emph{context node} in the tree. 
The \emph{context node} simply refers to the tree node from which navigation starts. Starting from a particular context node in a tree, every other nodes can easily be reached: XPath axes define a partitioning of a tree from any context node. Figure~\ref{introduction:fig:xpath-axes} illustrates this on a sample tree. More informal details on the complete XPath standard can be found in the W3C specification \cite{xpath}. 
 

XPath is increasingly popular due to its expressive power and its compact syntax. These two advantages have given XPath a central role both in other key XML specifications and XML applications. It is used in XQuery \cite{xquery} as a core query language; in XSLT \cite{xslt} as node selector in the transformations; in XML Schema \cite{xml-schemas} to define keys; in XLink \cite{xlink} and XPointer \cite{xpointer} to reference portions of XML data. XPath is also used in many applications such as update languages \cite{xmlupdates} and access control \cite{xml-access-control}. 




\subsection{Static Type-Checking}


XML applications most commonly use schemas for performing validation (also called \emph{dynamic type-checking}). Validation consists in using a schema validator that analyzes a particular XML document w.r.t a given schema in order to ensure that the document actually conforms to the expectations of the application.

In practice however XML documents are often generated dynamically by some program. Typically, programs that manipulate XML first access data (possibly conforming to an available schema) using XPath expressions, and then build and return an output XML document intended to conform to a given schema.

An ambitious approach is the \emph{static type-checking} of these programs, which consists in ensuring at compile-time that invalid documents can never arise as outputs of XML processing code. A static type checker analyzes a program, possibly in conjunction with schemas that describe its input and output (depending whether such schemas are available). The problem's difficulty is a function of the language in which the program and the schemas are expressed. 

Schema languages have been extensively studied and are now well understood as subsets of regular tree languages \cite{murata-toit05}.
However, although many attempts have been made for better understanding static type-checking techniques, in particular through the design of domain specific languages \cite{hosoya-toit03}, no approach is effectively able to deal with XPath, which nevertheless remains the essence of XML navigation and data access.

\subsection{Research Challenges}

The reason for the limitations of existing approaches is the difficulty of XPath static analysis. It is known that the static analysis of the complete XPath standard is undecidable. Importance and range of applications nevertheless motivate research questions: what is the largest XPath fragment with decidable static analysis? Which fragments can be effectively decided in practice? How to determine if an XPath expression is satisfiable on any of the XML trees defined by a given schema? How to know if two XPath queries will always yield the same result when evaluated on a document valid w.r.t. a given schema? Does the result of an XPath expression over a valid document always conform to another schema? Is there an algorithm able to answer these questions in an efficient way so that it can be used in practice?

One source of difficulty for such an algorithm is that it needs to check properties on a possibly infinite quantification over a set of trees. A variety of factors furthermore contribute to its complexity such as the operators allowed in XPath queries and the combination of them (cf. Chapter~\ref{foundations:queries}). A consequence of these difficulties is that such research questions are still open.






\section{Overview of this Dissertation}


This dissertation starts from the idea that for deciding XML problems, two issues must be addressed. First, identify an appropriate logic with sufficient expressiveness to capture both regular tree types and XPath style navigation and node selection semantics. Second, solve efficiently the satisfiability problem which allows to test if a given formula of the logic admits a satisfying XML document as a model. 


%
%

\subsection{Applications} 

The main application of this work is the static analysis of programs manipulating XML data and documents. This dissertation provides the necessary foundations and system implementations for solving the major XML decision problems that naturally arise from such static analyses.

The most basic decision problem for a query language is the emptiness check \cite{benedikt-pods05}: whether or not an expression yields a non-empty result. XPath emptiness is important for optimization of host languages implementations: for instance, if one can decide at compile time that a query is not satisfiable then subsequent bound computations can be avoided. 

Another basic decision problem is the XPath equivalence problem: whether or not two queries always return the same result. It is important for reformulation and optimization of the query itself \cite{geneves-doceng04}, which aim at enforcing operational properties while preserving semantic equivalence \cite{abiteboul-Jcss99,pierce-dbpl05}. 

The most critical problem for the type-checking of XML transformations is XPath containment: whether or not, for any tree, the result of a particular query is included in the result of another one. It is required for the control-flow analysis of XSLT \cite{moller-rr05}. It is also needed for checking integrity constraints \cite{xml-schemas}, and for checking access control in XML security applications \cite{xml-access-control}.

Other decision problems needed in applications include for example XPath overlap (whether two expressions select common nodes) and coverage (whether nodes selected by an expression are always contained in the union of the results selected by several other expressions). 

This dissertation effectively solves these problems in the presence, or absence, of XML type constraints such as DTDs \cite{xml} or XML Schemas \cite{xml-schemas}. 
This makes possible to ensure valuable properties (such as type-safety and optimizations) at compile-time, toward safer and more efficient runtime XML processing. Results presented in this dissertation thus notably open promising perspectives for the effective static analysis of XML transformations.

\subsection{Outline}

The first part of this dissertation is dedicated to state-of-the-art related tools and techniques. Chapter~\ref{foundations} introduces some known theoretical foundations and formalisms used in the remaining of this dissertation, while progressively introducing related work.

In a second part, Chapter~\ref{containment} and Chapter~\ref{analysis} conduct preliminary investigations with known logics in the context of XML. Specifically, Chapter~\ref{containment} studies to which extent monadic second order logic can be used in practice, despite its high complexity, for solving XML static analysis problems such as XPath containment. Chapter~\ref{analysis} introduces the $\mu$-calculus as a powerful replacement for monadic second order logic, and studies its use for XML reasoning. 

Based on the lessons learned from these investigations, the third part of this dissertation presents the final contribution. Chapter~\ref{xml-calculus:the-logic-for-xml} proposes a logic of finite trees specifically designed for XML. Chapter~\ref{xml-calculus:sec:algo} describes a proposed algorithm for testing the satisfiability of the logic, along with implementation techniques. Finally, Chapter~\ref{conclusion} concludes this dissertation and gives several perspectives.

%
%
%
%
%
%

 \ifglobalcompil
  \else 
   \bibliographystyle{apalike}
   \bibliography{references}
   \end{document}
  \fi

\part*{State of the Art}
\newif\ifglobalcompil\globalcompiltrue

  \ifglobalcompil
  \else 
    \input{Preamble} 
    \input{Markup}                       
    \newcommand{\chapterabstract}[1]{Chapter Abstract: #1}
    \begin{document} 
  \fi

\mychapter{Foundations of XML Processing}
\label{foundations}
In this chapter, some known theoretical foundations and formalisms used in the following chapters of this dissertation are introduced. State of the art related work is presented as underlying concepts are progressively introduced.

\section{Trees and Tree Types}

This section introduces the formal models of XML documents and schemas most often considered in the literature as well as in Chapters~\ref{foundations}, \ref{containment}, and \ref{analysis} of this dissertation~\footnote{Chapter~\ref{xml-calculus} elaborates further on this model by introducing \emph{focused} trees.}.

\subsection{Finite Trees and Hedges}
\label{foundations:unranked-and-binary-trees}
An XML document can be seen as a finite ordered and labeled tree of unbounded depth and arity. Since there is no a priori bound on the number of children of a node; such a tree is therefore \emph{unranked} \cite{neven-sigmod02}. Tree nodes are labeled with symbols taken from a countably infinite alphabet $\notation{\Sigma}{Alphabet of node labels}{n:alphabet}$. 
There is a straightforward isomorphism between sequences of unranked trees and binary trees \cite{hosoya-toit03,neven-sigmod02}. In order to describe it, trees are first formally defined. An unranked tree is defined as $\sigma(h)$ where $\sigma \in \Sigma$ and $h$ is a hedge, i.e. a sequence of unranked trees, defined as follows:
\smallsyntax{
\entry  \notation{\setofhedges}{Labeled hedges}{n:hedges} \ni     h     [hedge]
             \sigma(h), h'               [non-empty sequence of trees]
\oris        ()            [empty sequence]
}
The set of unranked trees is denoted by $\notation{\setofunrankedtrees}{Labeled unranked trees}{n:utrees}$. A binary tree $t$ is either a $\sigma$-labeled root of two subtrees ($\sigma \in \Sigma$) or the empty tree:
\smallsyntax{
\entry   \notation{\setofbintrees}{Labeled binary trees}{n:btrees} \ni     t     [binary tree]
             \sigma(t,t')               [node]
\oris        \epsilon            [empty tree]
}
Unranked trees are translated into binary trees with the following function $\notation{\ntobin{\cdot}}{Mapping from hedges to binary trees}{n:htot}$: 
\begin{align*}
\ntobin{\cdot} &: \setofhedges \rightarrow \setofbintrees\\
\ntobin{\sigma(h), h'} &\eqdef \sigma(\ntobin{h}, \ntobin{h'}) \\
\ntobin{()} &\eqdef \epsilon \\
\end{align*}
 
  
The inverse translation function $\notation{\binton{\cdot}}{Mapping from binary trees to hedges}{n:ttoh}$ converts a binary tree into a sequence of unranked trees: 
\begin{align*}
\binton{\cdot} &: \setofbintrees \rightarrow \setofhedges \\
\binton{\sigma(t,t')} &\eqdef \sigma(\binton{t}), \binton{t'} \\
\binton{\epsilon} &\eqdef () \\
\end{align*}

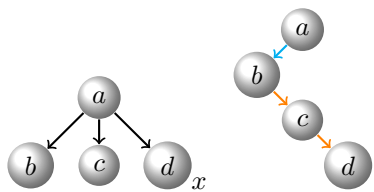
\begin{figure}[h]
\centering
$\begin{array}{ll}
\begin{tikzpicture}[scale=0.6]
\tikzstyle{every node}=[ball color=white,circle,text=black]
\tikzstyle{edge from parent}=[draw,thick,black, ->]
\node {$a$}
 child {node {$b$}}
 child {node {$c$}}
 child {node {$d$}};
\end{tikzpicture}x
&
\begin{tikzpicture}[scale=0.6]
\draw (3,4) node(a)[ball color=white,circle,text=black] {$a$}
      (2,3) node(b)[ball color=white,circle,text=black] {$b$}
      (3,2) node(c)[ball color=white,circle,text=black]{$c$}
      (4,1) node(d)[ball color=white,circle,text=black]{$d$}; 
            
\draw[draw,thick,cyan,->] (a) -- (b);
\draw[draw,thick,orange,->] (b) -- (c);
\draw[draw,thick,orange,->] (c) -- (d);
\end{tikzpicture}
\end{array}$
\caption{Unranked and Binary Tree Representations.}\label{foundations:fig:bijection}
\end{figure}

 For example, Figure~\ref{foundations:fig:bijection} illustrates how the sample tree $a(b,c,d)$ is mapped to its binary representation $a(b(\epsilon,c(\epsilon,d(\epsilon,\epsilon))),\epsilon)$ and vice-versa.

%
%
%
%


Note that the translation of a single unranked tree results in a binary tree of the form $\sigma(t,\epsilon)$. Reciprocally, the inverse translation of such a binary tree always yields a single unranked tree. When modeling XML, it is therefore possible to focus on binary trees of the form $\sigma(t,\epsilon)$, without loss of generality. The following section presents how this isomorphism between binary and unranked trees also extends to tree types. Such binary mappings allow to simplify formal notations used in the remaining.  

\subsection{Schema Languages and Regular Tree Types}

\begin{figure*}
\begin{center}
 \includegraphics[keepaspectratio,width=7cm]{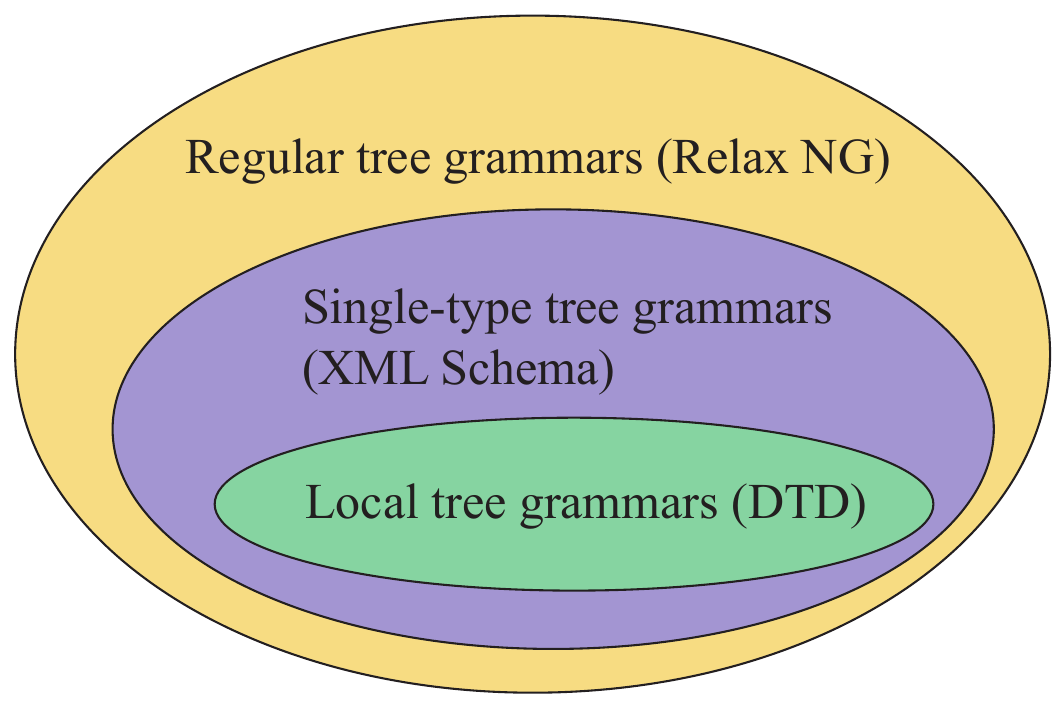}
 \end{center}
 \caption{Relative Expressiveness of Schema Languages.}\label{foundations:fig:schemas}
\end{figure*}

Schemas describe structural constraints for XML documents. There are many formalisms (called \emph{schema languages}) for specifying schemas (or ``types''). For instance: DTD, which is part of the XML specification \cite{xml}, XML Schema (W3C) \cite{xml-schemas}, and RELAX NG (OASIS/ISO) \cite{relax} are actively used by various applications. Each schema language has different constraint mechanisms and different expressivenesses. A detailed characterization of each schema language can be found in \cite{murata-toit05}. No current schema language goes beyond the expressive power of regular tree languages. From an XML point of view, regular tree types form a strict superset of standards such as XML Schemas and DTDs (cf. Figure ~\ref{foundations:fig:schemas}). Therefore, in this dissertation, regular tree languages are considered as the general mechanism for typing XML documents. 

A tree type expression $T$ is syntactically defined as follows:
\smallsyntax{
\entry   \notation{\lcft}{Context-free tree type expressions}{b:lcft} \ni T                 [context-free tree type expression]
             \emptyset                [empty set of trees]
\oris       ()                        [empty sequence]
\oris       X                         [variable]
\oris       \treetypelabel{l}{T}      [label]
\oris       T_1,T_2                   [sequence]
\oris       T_1 \tou T_2              [disjunction]
\oris       \typebind{X}{T} T         [$n$-ary binder]
}
where $l \in \Sigma$ and $X \in \dom{TVar}$ assuming that $\dom{TVar}$ is a countably infinite set of type variables. Abbreviated type expressions can be defined as follows:
\begin{align*}
T?  &\eqdef () \tou T \\
T*  &\eqdef \singletypebind{X}{T} T, X  \tou ()\\
T^+ &\eqdef T, T* \\
\end{align*}
Given an environment $\theta$ of type variable bindings, the semantics of tree types is given by the denotation function $\notation{\Tsem{\cdot}{\theta}}{Denotational semantics of tree types}{n:tsem}$:
\begin{align*}
\Tsem{\cdot}{\cdot} &: \lcft \times (\dom{TVar} \rightarrow \powerset{\setofunrankedtrees}) \rightarrow \powerset{\setofunrankedtrees} \\
\Tsem{\emptyset}{\theta} & \eqdef \emptyset \\
\Tsem{()}{\theta} & \eqdef \{ () \}\\
\Tsem{X}{\theta} & \eqdef \theta(X) \\
\Tsem{l[T])}{\theta} & \eqdef \{l'(t) \mid l' \subtaggingrel l \et t \in \Tsem{T}{\theta} \} \\
\Tsem{T_1,T_2}{\theta} & \eqdef \{ t_1,t_2 \mid  t_1 \in \Tsem{T_1}{\theta} \et t_2 \in \Tsem{T_2}{\theta} \}\\
\Tsem{T_1 \tou T_2}{\theta}&\eqdef \Tsem{T_1}{\theta} \cup \Tsem{T_2}{\theta}\\
\Tsem{\typebind{X}{T} T}{\theta} & \eqdef\Tsem{T}{\leastfp{S}}\\
\end{align*}
where $\subtaggingrel$ is a global subtagging relation: a reflexive and transitive relation on labels\footnote{Subtagging goes beyond the expressive power of DTDs but a similar notion called ``substitution groups'' exists in XML Schemas (see \cite{hosoya-toplas05} for more details on subtagging).}, and $S(\theta')=\theta[X_i \mapsto \Tsem{T_i}{\theta'}]_{i\geq 1}$. Note that each function $S$ is monotone according to the ordering $\subseteq$ on  $\dom{TVar} \rightarrow \powerset{\setofunrankedtrees}$, and thus has a least fixpoint $\leastfp{S}$.

Types as defined above actually correspond to arbitrary context-free tree types, for which the decision problem for inclusion is known to be undecidable \cite{hopcroft00}. An additional restriction is imposed to reduce the expressive power of considered types so that they correspond to regular tree languages. The restriction (also used in \cite{hosoya-toplas05}) consists in a simple syntactic condition that allows unguarded (i.e. not enclosed by a label) recursive uses of variables, but restricts them to tail positions\footnote{For instance the type ``$\typebind{X,Y}{a[],Y}{b[],X \tou ()} X$'' is allowed.}. This condition ensures regularity, and the resulting class of regular tree languages is denoted $\notation{\lrtt}{Regular tree type expressions}{n:lrtt}$. 

\subsubsection{Document Type Definitions}


This subsection further details the connection between regular tree types and the widely used DTD standard. As they are defined in the W3C recommendation, DTDs \cite{xml} are local tree grammars\footnote{A local tree grammar is a regular tree grammar without \emph{competing} non-terminals. Two non-terminals $A$ and $B$ of a tree grammar are said to compete with each other if one production rule has $A$ in its left-hand side, one production rule has $B$ in its left-hand side, and these two rules share the same terminal symbol in the right-hand side.}, which are strictly less expressive than regular tree types. In the XML terminology, a type expression is called the \emph{content model}. DTD content models are described by the following syntax:
\smallsyntax{
\entry  T               [DTD tree type expression]
             l                     [label]
\oris        T_1 \tou T_2     [disjunction]
\oris        T_1,T_2               [sequence]
\oris        T?                    [optional occurrence]
\oris        T^*                   [zero, one or more occurrences]
\oris        T^+                   [one or more occurrences]
\oris        ()                    [empty sequence]
}
where $l \in \Sigma$. From the W3C specification, a DTD can be seen as a function that associates a content model to each label taken from a subset $\Sigma'$ of $\Sigma$, such that $\Sigma'$ gathers all labels used in content models. The set $\ldtd$ of tree types described by DTDs can thus be represented as follows:
\smallsyntax{
\entry    \notation{\ldtd}{DTD tree type expressions}{n:ldtd} \ni T               [DTD tree type expression]
             l                     [label]
\oris        T_1 \tou T_2     [disjunction]
\oris        T_1,T_2               [sequence]
\oris        T?                    [optional occurrence]
\oris        T^*                   [zero, one or more occurrences]
\oris        T^+                   [one or more occurrences]
\oris        ()                    [empty sequence]
\oris        \typebind{l}{T} T     [$n$-ary binder]}																
Note that $\ldtd \subseteq \lrtt$ is obvious, by associating a unique type variable to each label. 
In the following, DTDs are therefore not distinguished from general regular tree types anymore. 

\subsection{Binary Tree Types}
\label{foundations:binary-tree-types}
Section~\ref{foundations:unranked-and-binary-trees} presented a straightforward isomorphism between binary trees and sequences of unranked trees. There is also an isomorphism between unranked and binary tree types, which follows exactly the same intuition as for trees. 

Binary tree types are described by the following syntax:
\smallsyntax{
\entry    \notation{\lbtt}{Binary tree type expressions}{n:lbtt} \ni T               [binary tree type expression]
             \emptyset                [empty set of trees]
\oris       ()                        [empty sequence]
\oris       T_1 \tou T_2              [disjunction]
\oris       l(X_1, X_2)          [label]
\oris       \typebind{X}{T} T         [$n$-ary binder]
}
For any type, there is an equivalent binary type, and vice-versa. The translation function $\notation{\tsbin{\cdot}}{Mapping from unranked to binary tree types}{n:tsbin}$ shown on Figure~\ref{foundations:fig:Ttrans} (and adapted from the one found in \cite{hosoya-toplas05}) is used to convert a type into its corresponding binary representation. The function considers the environment $\theta : \dom{TVar} \rightarrow \lrtt$ for accessing the type bound to a variable $X_i$ by constructs of the form ``$\typebind{X}{T} T$''.

\begin{figure}[h]
\centering
\begin{align*}
\tsbin{\cdot} &: \lrtt \rightarrow \lbtt \\ 
\tsbin{\emptyset} & \eqdef \emptyset \\
\tsbin{()} & \eqdef  \epsilon \\
\tsbin{X} & \eqdef  \tsbin{\theta(X)} \\
\tsbin{l[T]} & \eqdef  \doubletypebind{X_1}{\tsbin{T}}{X_2}{\epsilon} l(X_1, X_2)\\
\tsbin{T_1 \tou T_2} & \eqdef  \tsbin{T_1} \tou \tsbin{T_2}\\
\tsbin{\typebind{X}{T} T} &\eqdef \typebindifree{X_i}{\tsbin{T_i}} \tsbin{T}\\
\tsbin{\emptyset, T} & \eqdef  \emptyset \\
\tsbin{(), T} & \eqdef  \tsbin{T} \\
\tsbin{X, T} & \eqdef  \tsbin{\theta(X),T}\\
\tsbin{l[T_1], T_2} & \eqdef  \doubletypebind{X_1}{\tsbin{T_1}}{X_2}{\tsbin{T_2}} l(X_1, X_2)\\
\tsbin{(T_1 \tou T_2), T_3} & \eqdef  \tsbin{T_1,T_3} \tou \tsbin{T_2,T_3} \\
\tsbin{(T_1, T_2), T_3} & \eqdef  \tsbin{T_1,(T_2,T_3)}\\
\tsbin{\typebind{X}{T} T, T'} & \eqdef  \typebindifree{X_i}{\tsbin{T_i}} \tsbin{T,T'}\\
\end{align*}
\caption{Binarization of Tree Types.} \label{foundations:fig:Ttrans}
\end{figure}

For example, Figure~\ref{foundations:DTDSample} gives a sample DTD that validates the well-formed XML document presented in Section~\ref{introduction:well-formed-sample-xml-doc} of Chapter~\ref{introduction}. The corresponding context-free tree type expression is presented on Figure~\ref{foundations:CFTSample}. It uses $14$ type variables (preceded by a dollar sign $\$$ by convention). Figure~\ref{foundations:BTTSample} shows its translation into binary tree type syntax.

\begin{figure}
\centering
\begin{verbatim}
<!ELEMENT plant (category?, tissue*, phylogeny?)>
<!ELEMENT category   (#PCDATA)>
<!ELEMENT tissue  (name+, def, note?)>
<!ELEMENT name    (#PCDATA)>
<!ELEMENT def   (#PCDATA)>
<!ELEMENT note (#PCDATA)>
<!ELEMENT phylogeny (plant+)>
\end{verbatim}
\caption{A Sample DTD.} \label{foundations:DTDSample}
\end{figure}

\begin{figure}
\centering
\begin{verbatim}
$Empty -> EMPTYSET
$Epsilon -> ()
$Any -> ()
$PCData -> ()
$name -> name($PCData)
$note -> note($PCData)
$1 -> $plant | $plant, $1
$phylogeny -> phylogeny($1)
$category -> category($PCData)
$def -> def($PCData)
$2 -> $name | $name, $2
$tissue -> tissue($2, $def, () | $note)
$3 -> () | $tissue, $3
$plant -> plant(() | $category, $3, () | $phylogeny)

Start symbol is $plant
14 type variables.
7 terminals.
\end{verbatim}
\caption{Sample Context-Free Tree Type Expression.} \label{foundations:CFTSample}
\end{figure}

\begin{figure}
\centering
\begin{verbatim}
$2  -> plant($1, $Epsilon) | plant($1, $2)
$7  -> EPSILON | note($Epsilon, $Epsilon)
$5  -> def($Epsilon, $7)
$3  -> name($Epsilon, $5) | name($Epsilon, $3)
$10 -> EPSILON | phylogeny($2, $Epsilon) | tissue($3, $10)
$1  -> EPSILON | phylogeny($2, $Epsilon) | 
       tissue($3, $10) | category($Epsilon, $10)
$plant -> plant($1, $Epsilon)

Start symbol is $plant
7 type variables.
7 terminals.
\end{verbatim}
\caption{Sample Binary Tree Type Expression.} \label{foundations:BTTSample}
\end{figure}

\subsection{Finite Tree Automata}
\label{foundations:fta}
Tree automata are a convenient operational formalism for expressing the notion of tree languages.  A language is recognizable if there exists an automaton which recognizes trees of the language. 
A detailed classification of tree automata and associated results on the recognizability of tree languages are presented in \cite{tata}.  This section presents the most basic results on finite tree automata needed for the remaining of this dissertation.

\paragraph{Bottom-Up Finite Tree Automata}
Formally, a bottom-up non-deter\-mi\-nis\-tic finite tree automaton (NFTA) over an alphabet $\Sigma$ of node labels is a tuple $(Q, Q_f, \Gamma)$ where $Q$ is the set of states, $Q_f \subseteq Q$ is a set of accepting states, and $\Gamma$ is a set of transitions. Transitions are either of the form $q \leftarrow \sigma$ or of the form $q'' \leftarrow \sigma(q,q')$, depending on the arity of the symbol $\sigma \in \Sigma$ (respectively a leaf or a binary constructor) and where $q, q', q''$ are automaton states belonging to $Q$. A bottom-up NFTA starts from the leaves and moves up the tree. At each step of the execution, a state is inductively associated with each subtree. The tree is accepted if the state labeled at the root is an accepting state.

\paragraph{Top-Down Finite Tree Automata}
There exists a symmetric counterpart of bottom-up NFTA called top-down NFTA, which correspond to the alternate direction used to recognize a tree. A top-down NFTA $(Q, Q_i, \Gamma)$ starts at the root and moves down to the leaves. Based on a state and a current node in the tree, a new state is inductively associated with each subtree. Transitions thus have the reverse form, and $Q_i$ is the set of initial states. The tree is accepted if every branch can be gone through this way. 

\paragraph{Determinism}
A deterministic finite tree automaton (DFTA) is one where no two transition rules have the same left-hand side. This definition matches the intuitive idea that for an automaton to be deterministic, one and only one transition must be possible for a given node. 

\paragraph{Expressive Power}

Top-down and bottom-up NFTA are equivalent (the transition rules are simply reversed, and the final states become the initial states). However, top-down DFTA are strictly less powerful than their deterministic bottom-up counterparts. This is because transition rules of tree automata can be seen as rewrite rules; and for top-down ones, the left-hand sides correspond to parent nodes. Consequently a deterministic top-down tree automaton will only be able to test for tree properties that are true in all branches, because the choice of the state to write into each child branch is determined at the parent node, without knowing the child branches contents.

Every bottom-up NFTA is equivalent to a bottom-up DFTA which can be obtained by the process of determinization. Determinization relies on the ``subset construction'' and the number of states of the equivalent DFTA can be exponential in the number of states of the given NFTA (see \cite{tata} for the detailed algorithm).
In the bottom-up paradigm, since NFTA and DFTA accept the same sets of tree languages, they are usually not distinguished and simply both referred as finite tree automata (FTA).

FTA are equivalent to regular tree types and therefore have the same expressiveness. 


\paragraph{FTA as XML Types}
Murata was the first to consider tree automata as a schema definition language \cite{murata-pddp98}. 
Since then, FTA were heavily used in many research works for modeling XML types \cite{neven-csl02}. In fact, the schema language Relax NG \cite{relax}, a competitor of XML Schema \cite{xml-schemas} (itself introduced as a replacement for DTDs \cite{xml}) is even directly inspired by FTA. A detailed comparison of these schema languages based on formal language theory is provided in \cite{murata-toit05}.

As a simple example, Figure~\ref{foundations:FTASample} illustrates a sample NFTA which accepts the set of trees defined by the DTD shown on Figure~\ref{foundations:DTDSample}. The NFTA accepts the set of all binary trees $\ntobin{t}$ such that the unranked tree $t$ is validated by the DTD of Figure~\ref{foundations:DTDSample}. Note that the NFTA can be seen as another notation for the binary tree type expression shown on Figure~\ref{foundations:BTTSample}. More interestingly, the DFTA obtained by determinization of this NFTA can be seen as the operational validator of the DTD.

\begin{figure}[h]
\centering
\begin{align*}
Q   & = \{q_1, q_2, q_3, q_5,  q_7, q_{10}, q_\epsilon, q_\text{plant}\}\\ 
Q_f & = \{q_\text{plant}\} \\
\Gamma & = \left\{
\begin{array}{lll}
q_2 &\leftarrow& \text{plant}(q_1, q_\epsilon) \\
q_2 &\leftarrow& \text{plant}(q_1, q_2) \\
q_7 &\leftarrow& \epsilon \\
q_7 &\leftarrow& \text{note}(q_\epsilon, q_\epsilon) \\
q_5 &\leftarrow& \text{def}(q_\epsilon, q_7) \\
q_3 &\leftarrow& \text{name}(q_\epsilon, q_5) \\
q_3 &\leftarrow& \text{name}(q_\epsilon, q_3) \\
q_{10} &\leftarrow& \epsilon \\
q_{10} &\leftarrow& \text{phylogeny}(q_2, q_\epsilon) \\
q_{10} &\leftarrow& \text{tissue}(q_3, q_{10}) \\
q_1 &\leftarrow& \epsilon \\
q_1 &\leftarrow& \text{phylogeny}(q_2, q_\epsilon) \\
q_1 &\leftarrow& \text{tissue}(q_3, q_{10}) \\
q_1 &\leftarrow& \text{category}(q_\epsilon, q_{10}) \\
q_\text{plant} &\leftarrow& \text{plant}(q_1, q_\epsilon) 
\end{array}\right\} 
\end{align*}
\caption{A Sample NFTA $(Q, Q_f, \Gamma)$.} \label{foundations:FTASample}
\end{figure}

\paragraph{Closure Properties}
\label{foundations:fta-closure}
One of the main advantages of FTA (compared to DTDs for instance) is their closure under set theoretic operations such as union, intersection, and complementation \cite{tata}.

The union of two tree automata is trivially built: let $A_1=(Q_1,Q_{f_1},\Gamma_1)$ and $A_2=(Q_2,Q_{f_2},\Gamma_2)$ be two FTA. Since states of a FTA may be renamed without loss of generality, it is assumed that $Q_1 \cap Q_2 = \emptyset$. It is then straightforward to verify that $A_1 \cup A_2=(Q,Q_{f},\Gamma)$ defined by: $Q=Q_1 \cup Q_2$, $Q_f=Q_{f_1} \cup Q_{f_2}$  and $\Gamma=\Gamma_1 \cup \Gamma_2$. 

Similarly, the intersection of two tree automata $A_1=(Q_1,Q_{f_1},\Gamma_1)$ and $A_2=(Q_2,Q_{f_2},\Gamma_2)$ is simply obtained by calculating a product automaton:
$$A_1 \cap A_2= (Q_1 \times Q_2, Q_{f_1} \times Q_{f_2}, \Gamma_1 \times \Gamma_2)$$

Complementation of a \emph{complete} DFTA simply consists in flipping accepting and rejecting states. Note that a DFTA $(Q, Q_f, \Gamma)$ is \emph{complete} if and only if there is a transition $q'' \leftarrow \sigma(q,q')$ for each $\sigma \in \Sigma$ and $(q,q',q'') \in Q^3$. Completing an automaton (e.g. adding new missing states and transitions, and then possibly updating the final set of states \cite{tata}) may be required before complementing it. The complement of a FTA $A$ is noted $\complement{(A)}$.


\paragraph{Containment for FTA}
By taking advantage of these closure properties, it is possible to check the containment of two FTA $A_1$ and $A_2$ (determining whether the set of trees accepted by $A_1$ is included into the set of trees accepted by $A_2$) as the emptiness check of the FTA $A_1 \cap \complement(A_2)$.

It can be decided in linear time whether the language accepted by a FTA is empty (see~\cite{tata} for details). However, complementation requires determinization of the tree automaton, which may cause an exponential increase of the number of states in the worst case \cite{tata}. Thus this technique has exponential time complexity. Essentially, there is no better way for checking containment between two FTA. As a result, the FTA containment problem is in EXPTIME\footnote{The complexity class EXPTIME is the set of all decision problems solvable by a deterministic Turing machine in $O(2^{p(n)})$ time, where $p(n)$ is a polynomial function of the input size $n$.} \cite{seidl-jc90}.

\section{Queries}
\label{foundations:queries}
Most queries used in the context of XML are either boolean or unary. Boolean queries give a yes/no answer on a tree (for instance the validation of an XML document w.r.t to a DTD is a boolean query). Unary queries select nodes from a document (for instance, finding the set of nodes selected by an XPath expression is a unary query).

Unary queries considered in this dissertation are among those defined by the powerful XPath standard introduced in Section~\ref{introduction:xpath-intro}. The static analysis of XPath queries is a hard problem that has recently attracted a lot of theoretical research attention. In particular, the computational complexity of the containment problem for XPath expressions has received much attention from the database community \cite{deutsch, wood2003, neven-icdt03, schwentick-sigmodrec04, suciu-miklau-jacm04}. The complexity of the emptiness problem for XPath expressions has also been studied in \cite{benedikt-pods05}. One source of difficulty for such decision problems is that they need to be checked on a possibly infinite quantification over a set of trees. A variety of factors also contribute to their complexity such as the operators allowed in XPath queries and the combination of them. For instance, one difficulty arises from the combination of upward and downward navigation on trees with recursion \cite{vardi-icalp98}.  Actually, when the whole XPath language is considered, decision problems such as containment and emptiness are undecidable. Therefore, in the literature, the focus was given to identifying major XPath features and studying their impact on the complexity of XPath decision problems. The distinctions between major features studied in the literature (extended from \cite{benedikt-pods05}) follow: 
\begin{itemize}
\item{positive vs. non-positive}: depending whether the negation operator is considered (positive) or not (non-positive) inside qualifiers.
\item{downward vs. upward}: depending whether queries specify downward or upward traversal of the tree, or both.
\item{recursive vs. non-recursive}: depending whether XPath transitive closure axes (for instance ``descendant'' or ``ancestor'') are considered or not.
\item{qualified vs. non-qualified}: depending whether queries allow filtering qualifiers or not.
\item{with vs. without data values}: depending whether comparisons of data values expressing joins are allowed or not.
\item{with vs. without counting}: depending whether counting of tree nodes is allowed or not.
\end{itemize}
Several XPath fragments combining only a few of these features have been studied: see \cite{schwentick-sigmodrec04} for an overview. From these results, it is known that containment and satisfiability for (reasonably) restricted XPath fragments, even without type constraints, ranges from EXPTIME to undecidable. However, techniques used for obtaining computational complexity bounds over specific subfragments do not scale when additional features are considered, and thus give no hints on how to address more realistic fragments. At the time of this dissertation, no relevant algorithm effectively able of answering realistic XPath decision problems in acceptable time and space bounds is known. XPath decision problems have been partially characterized from a strict computational complexity point of view, and remain unsolved in practice.





\subsection{Syntax of XPath Expressions}
\label{foundations:xpath-syntax}

In this dissertation, particular attention is paid at supporting a large XPath fragment, as realistic as possible, covering major features of the XPath 
standard \cite{xpath}. The syntax of considered XPath expressions is given on Figure~\ref{foundations:fig:xpath-abstract-syntax}.
The considered XPath fragment is non-positive, both downward and upward, recursive, qualified, and also includes union and intersection. It includes all axes. This is the largest fragment considered so far in the literature. It covers all major XPath features except counting and data values. The integration of counting is kept for future work, based on related work on logics for counting \cite{dal-zilio-popl04}. Data values are known to cause undecidability of XPath containment when combined with previous factors \cite{benedikt-pods05,schwentick-sigmodrec04}\footnote{Note however that the very recent work found in \cite{segoufin-pods06} obtained the theoretical decidability (between NEXPTIME and $3$-NEXPTIME) for a limited form of data value comparison. Integration of such restricted comparisons in the considered fragment and the effective algorithm presented in Chapter~\ref{xml-calculus:sec:algo} is one of the perspectives of this dissertation. At least an additional exponential time blow-up is however expected.}.  

\begin{figure}[h]
\smallsyntax{
\entry       \notation{\lxpath}{XPath expressions}{n:lxpath} \ni e     [XPath expression]
             /p              [absolute path]
\oris        p              [relative path]
\oris        e_1 \punion e_2 [union]
\oris        e_1 \cap e_2 [intersection] \\
\entry      \dom{Path} \quad p            [path]
             p_1/p_2      [path composition]
\oris        \qualif{p}{q} [qualified path]
\oris        \step{\axisvar}{\nodelabel} [step with node test]
\oris        \step{\axisvar}{*} [step] \\
\entry       \dom{Qualif} \quad  q            [qualifier]
    					q_1 \op{and} q_2 [conjunction]
\oris					q_1 \op{or} q_2  [disjunction]
\oris					\op{not}~q   [negation]
\oris					p            [path] \\
\entry       \dom{Axis} \quad  \axisvar [tree navigation axis (see Figure~\ref{introduction:fig:xpath-axes})]
             \axis{child}        []
\oris        \axis{self}         []
\oris        \axis{parent}       []
\oris        \axis{descendant}   []
\oris        \axis{descendant-or-self} []
\oris        \axis{ancestor}     []
\oris        \axis{ancestor-or-self}  []
\oris        \axis{following-sibling} []
\oris        \axis{preceding-sibling} []
\oris        \axis{following}    []
\oris        \axis{preceding}   []
}
\caption{XPath Abstract Syntax.}\label{foundations:fig:xpath-abstract-syntax}
\end{figure}

\subsection{XPath Denotational Semantics}
\label{foundations:xpath-denotational-semantics}
$\hiddennotation{\semanE{\cdot}{\cdot}}{Denotational semantics of XPath expressions}{n:semanE}$
In the classical denotational semantics of paths, first given in \cite{wadler}, the evaluation of an XPath expression over an XML document $t$ returns a set of nodes reachable from a context node $x$. The denotational semantics of the considered XPath fragment (adapted from \cite{wadler}) is given by the formal semantics function $\semanEFunc$ which defines the set of nodes returned by expressions, starting from a context node $x$ in the tree:
\begin{align*}
\semanE{\cdot}{\cdot} & :  \lxpath \rightarrow \dom{Node} \rightarrow \text{Set(\dom{Node)}} \\  
\semanE{/p}{x}{}                                & \eqdef \semanP{p}{\fun{root}{}}{} \\
\semanE{p}{x}{}                                 & \eqdef \semanP{p}{x}{} \\
\semanE{e_1 \shortmid e_2}{x} {}                & \eqdef \semanE{e_1}{x}{} \cup \semanE{e_2}{x}{}  \\
\semanE{e_1 \cap e_2}{x}{}                      & \eqdef \semanE{e_1}{x}{} \cap \semanE{e_2}{x}{}
\end{align*}
The formal semantics function $\semanPFunc$ defines the set of nodes returned by paths: 
\begin{align*}
\semanP{\cdot}{\cdot} &   :  \dom{Path} \rightarrow \dom{Node} \rightarrow \text{Set(\dom{Node)}} \\  
\semanP{p_1/p_2}{x}{}                           & \eqdef  \{ x_2 \; | \; x_1 \in \semanP{p_1}{x}{} \wedge x_2 \in \semanP{p_2}{x_1}{} \} \\
\semanP{p[q]}{x}{}                              & \eqdef \{ x_1 \; | \; x_1 \in \semanP{p}{x}{} \wedge \semanQ{q}{x_1}{} \} \\ 
\semanP{\step{a}{\sigma}}{x}{}                  & \eqdef  \{ x_1 \; | \; x_1 \in \semanA{a}{x}{} \wedge \fun{name}{x_1}=\sigma \} \\
\semanP{\step{a}{*}}{x}{}                       & \eqdef \{ x_1 \; | \; x_1 \in \semanA{a}{x}{} \}
\end{align*}

Note that the semantics of the $p_1/p_2$ construct corresponds to composition of unary queries. In this sense, XPath is fundamentally different from regular expressions patterns a la Hosoya \cite{hosoya-popl01} that rather use pattern-matching techniques. The function $\semanQFunc$ defines the semantics of qualifiers that basically state the existence or absence of one or more paths from a context node:
\begin{align*}
\semanQ{\cdot}{\cdot}  &  :  \dom{Qualifier} \rightarrow \dom{Node} \rightarrow \dom{Boolean} \\
\semanQ{q_1 \op{and} q_2}{x}{}     & \eqdef  \semanQ{q_1}{x}{} \wedge \semanQ{q_2}{x}{} \\
\semanQ{q_1 \op{or} q_2}{x}{}      & \eqdef \semanQ{q_1}{x}{} \vee \semanQ{q_2}{x}{} \\
\semanQ{\op{not}~q}{x}{}           & \eqdef \neg \; \semanQ{q}{x}{} \\
\semanQ{p}{x}{}                    & \eqdef  \semanP{p}{x}{} \neq \emptyset
\end{align*}
The semantics of paths relies on the navigational semantics of axes, given by the function $\semanAFunc$: 
\begin{align*}
\semanA{$\cdot$}{\cdot}  &  :  \dom{Axis} \rightarrow \dom{Node} \rightarrow \text{Set(\dom{Node)}} \\  
\semanA{child}{x}{}   & \eqdef  \fun{children}{x} \\
\semanA{parent}{x}{}  & \eqdef \fun{parent}{x} \\
\semanA{descendant}{x}{}   & \eqdef   \fun{children$^+$}{x} \\
\semanA{ancestor}{x}{}   & \eqdef   \fun{parent$^+$}{x} \\
\semanA{self}{x}{}   & \eqdef  \{x\} \\
\semanA{descendant-or-self}{x}{}   & \eqdef   \semanA{descendant}{x}{} \cup \semanA{self}{x}{} \\
\semanA{ancestor-or-self}{x}{}   & \eqdef   \semanA{ancestor}{x}{} \cup \semanA{self}{x}{} \\
\semanA{preceding}{x}{}   & \eqdef  \{y \; | \;  y \rbefore x\} \setminus \semanA{ancestor}{x}{} \\
\semanA{following}{x}{}   & \eqdef  \{y\; | \;  x \rbefore y\} \setminus \semanA{descendant}{x}{} \\
\semanA{following-sibling}{x}{}   & \eqdef  \{y\; | \;  y \in \fun{child}{\fun{parent}{x}} \et x \rbefore y\} \\
\semanA{preceding-sibling}{x}{}   & \eqdef  \{y\; | \;  y \in \fun{child}{\fun{parent}{x}} \et y \rbefore x\}
\end{align*}
Path and axis navigation (illustrated on a sample tree by Figure~\ref{introduction:fig:xpath-axes}) relies on a few assumed primitives over the XML tree data model: $\fun{root}{}$ returns the root of the tree; $\fun{children}{x}$ which returns the set of nodes which are children of the node x; $\fun{parent}{x}$ which returns the parent node of the node x; the relation $\rbefore$ which defines the ordering: $x \rbefore y $ holds if and only if the node $x$ is before the node $y$ in the depth-first traversal order of the $n$-ary XML tree; and finally $\fun{name}{}$ which returns the labeling of a node.

%
%
%
%
%
%
%
%
%



\section{Logical Formalisms: Two Yardsticks}
\label{foundations:2logics}

Unranked trees defined in Section~\ref{foundations:unranked-and-binary-trees} can be viewed as logical structures, in the sense of mathematical logic \cite{ebbinghaus-book05}. In this vision, the domain of a tree $t$, viewed as a structure, is the set of nodes of $t$, denoted by $\notation{\treedom{t}}{Domain of a tree (set of nodes)}{n:treedom}$. Formally, $\treedom{t}$ is the subset of $\N^*$ defined as follows: if $t=\sigma(t_1,...,t_n)$ with $\sigma \in \Sigma$, $n\ge 0$ and $t_1,...,t_n \in \setofunrankedtrees$, then $\treedom{t} = \{\epsilon\} \cup \setof{iu}{i \in \{1,...,n\}, u \in \treedom{t_i}}$. Thus, $\epsilon$ represents the root while $vj$ represents the $j^\text{th}$ successor of $v$. 

A relational vocabulary $(\notation{\infch}{Child relation between tree nodes}{n:infch}, \notation{\infsb}{Sibling relation between tree nodes}{n:infsb}, \setof{O_\sigma}{\sigma \in \Sigma})$  is often used \cite{neven-csl02,libkin-lics05,segoufin-pods06}. 
In this vocabulary, the $O_\sigma$ are unary relation predicates. For each $\sigma$ label in the alphabet $\Sigma$, $O_\sigma$ is the set of nodes that are labeled with $\sigma$. The symbols $\infch$ and $\infsb$ are binary predicates. The symbol $\infch$ is interpreted as the child relation: the set of pairs $(v, v\cdot i)$ where $v, v\cdot i \in \treedom{t}$. The symbol $\infsb$ is the sibling order: the set of pairs $(v \cdot i, v \cdot (i+1))$ where $v \cdot i, v\cdot (i+1) \in \treedom{t}$.

Classically, $\notation{\infchstar}{Transitive-reflexive closure of $\infch$}{n:infchstar}$ is defined as the transitive-reflexive closure of $\infch$ (the descendant/ancestor relationship between two nodes), and $\notation{\infsbstar}{Transitive-reflexive closure of $\infsb$}{n:infsbstar}$ as the transitive-reflexive closure of $\infsb$ (the linear ordering on siblings).


Most formalisms used in the context of XML are related to one of the two logics used over these relational structures: first-order logic, and monadic second order logic:
\begin{itemize}
\item
first-order logic and relatives are frequently used for query languages since they nicely capture their navigational features presented in the previous Section~\ref{foundations:xpath-denotational-semantics}. 
\item
monadic second order logic, which extends first-order logic by quantification over sets of nodes, is one of the most expressive (yet decidable) known logic. 
One of its main advantages in the context of XML is its ability to fully support XML types (regular tree languages).
\end{itemize}
The next sections are dedicated to these two logical formalisms, which are used as yardsticks logics in the XML setting. First-order logic is denoted by FO, and monadic second order logic by MSO. For XML applications, the relational vocabulary contains at least the labeling predicates $O_\sigma$ for $\sigma \in \Sigma$, which are thus omitted from notations in the remaining. The rest of the vocabulary is listed between brackets. For example, MSO$[\infch,\infsb]$ refers to the vocabulary $(\infch,\infsb, \setof{X_\sigma}{\sigma \in \Sigma})$. An important distinction between MSO and FO is that $\infchstar$ and $\infsbstar$ are definable from $\infch$ and $\infsb$ in MSO (using second-order quantification) but not in FO.



\section{First Order Logic}


Over a general relational structure, FO is undecidable, while its two-variable fragment is decidable \cite{mortimer-zlg75}. Therefore, restricting FO to its two-variable fragment, denoted FO$^2$, has become a classical idea when looking for decidability \cite{gradel-tcs99}. 
Furthermore, since $\infchstar$ and $\infsbstar$ are not definable from $\infch$ and $\infsb$ in FO, FO$^2[\infchstar, \infsbstar]$ is generally considered.

From the work found in \cite{geneves-tphols04} and \cite{marx-pods04}, it is known that XPath expressive power is close to FO$^2[\infchstar, \infsbstar]$ that captures its navigational behavior. Specifically, in \cite{geneves-tphols04}, a FO$^2[\infchstar, \infsbstar]$ interpretation of an XPath fragment is given and proven correct w.r.t. to XPath denotational semantics presented in Section~\ref{foundations:xpath-denotational-semantics}. The work found in \cite{marx-pods04} characterizes the navigational fragment of XPath (introduced as ``Core XPath'' in \cite{gottlob-tods05}) and shows how it can be extended in order to be complete with respect to FO$^2[\infchstar, \infsbstar]$.

The very recent work found in \cite{segoufin-pods06} proves the decidability of FO$^2[\infch, \infsb, \sim]$ where $\sim$ is a binary predicate such that $x\sim y$ holds for two nodes if they have the same data value. A consequence is the theoretical decidability of a limited form of comparison of data values in XPath. The corresponding decision procedure is observed to be between NEXPTIME and $3$-NEXPTIME, but unfortunately the approach gives no clue for a relevant effective algorithm \cite{segoufin-pods06}. 

FO nevertheless remains a convenient formalism for obtaining decidability results or theoretical characterizations of XPath queries. However, an argument in favor of MSO is that FO and its variants do not fully capture regular tree types \cite{segoufin-stacs05} which make them unsuited for dealing with XML types.

\section{Monadic Second-Order Logic}
\label{foundations:xml_mso_logic}  

MSO over trees is one of the most expressive -- yet decidable -- logic known. It is known since the 1960's that MSO exactly captures regular tree types. The appropriate MSO$[\infch,\infsb]$ variant over finite binary trees is named \emph{WS2S} which stands for \emph{weak monadic second-order logic of two successors}. WS2S was introduced in \cite{Thatcher68,Doner70}. In this calculus, first-order variables range over tree nodes. Second-order variables are interpreted as finite sets of tree nodes. \emph{Weak} means that the set variables are allowed to range only over finite sets. This is enough since XML documents have an unbounded depth but remain finite trees. \emph{Monadic} means that quantification is only allowed over unary relations (sets), not over polyadic relations. 
The \emph{two successors} refer to the left and right successors of a node in the binary tree. They are sufficient to consider general unranked XML trees without loss of generality, owing to the mapping $\ntobin{\cdot}$ presented in Section~\ref{foundations:logical_description_of_trees}.

This section progressively introduces WS2S in detail, and explains how it is decided through the automaton-logic connection \cite{Thatcher68,Doner70} using tree automata introduced in Section~\ref{foundations:fta}.

\subsection{Preliminary Definitions}
\label{foundations:logical_description_of_trees}

For notation consistency purposes, by convention, $0$ is used for denoting the left successor and $1$ for denoting the right successor of a node in a binary tree.  The definition of the domain of a finite binary tree is thus slightly updated as follows. For $t \in \setofbintrees, \treedom{t}$ is defined as the subset of $\{0,1\}$ such that if $t=\sigma(t_0,t_1)$ with $\sigma \in \Sigma$ and $t_0,t_1 \in \setofbintrees$, then $\treedom{t} = \{\epsilon\} \cup \setof{iu}{i \in \{0,1\}, u \in \treedom{t_i}}$. $\epsilon$ represents the root while $vj$ represents the $(j+1)^\text{th}$ successor of $v$, for $j \in \{0,1\}$. A node in the binary tree is thus a finite string over the alphabet $\{0,1\}$.

The notion of \emph{characteristic sets} is now defined, which further formalizes and generalizes the $O_\sigma$ unary predicates introduced in Section~\ref{foundations:2logics} for the labeling. A \emph{characteristic function} of a set $B$ is a function from $A$ to \{0,1\}, where $A$ is a superset of $B$. It returns 1 if and only if the element of $A$ is also an element of $B$:
$$\begin{array}{l}
B \subseteq A \\
f : A \rightarrow \{0,1\} \\
\forall a \in A, f(a) = \left \lbrace \begin{array}{l} 1, \text{if}~a \in B \\ 0, \text{if}~a \notin B \end{array}\right.
\end{array}$$
A \emph{characteristic set} is a subset of a set $A$ that contains all elements of $A$ for which the characteristic function returns 1:
$$\begin{array}{l}
X_f \subseteq A \\
X_f = \{ a \in A \; | \; f(a) = 1\}
\end{array}$$
In the following, characteristic sets of interest are subsets of $\treedom{t}$, which denote where a particular property holds in a tree. 
Particular attention is paid to the characteristic sets $X_{f_\sigma}$ which denote where a particular symbol $\sigma$ occurs. Consider for instance the binary tree $t=a(b(\epsilon, c(\epsilon,d)),\epsilon)$ over the alphabet $\Sigma = \{a, b, c,d\}$. It is identified by its tuple representation $\notation{\tuplerep{t}}{Tuple representation of the structure $t$}{n:tuplerep}=(X_{f_a},X_{f_b},X_{f_c},X_{f_d})$ where $X_{f_\sigma}$ is the characteristic set of the symbol $\sigma$:
\begin{align*}
X_{f_a} &= \{ \epsilon \} \\
X_{f_b} &= \{ 0 \} \\
X_{f_c} &= \{ 01 \} \\
X_{f_d} &=\{ 011 \}\\
\end{align*}
\label{foundations:shape}
The set $X_{f_a} \cup X_{f_b}\cup X_{f_c}\cup X_{f_d}$ of all positions contained in characteristic sets forms a \emph{shape}.

A node belongs to a characteristic set $X_{f_\sigma}$ (also noted $\notation{X_\sigma}{Characteristic set of the label $\sigma$}{n:characset}$) if and only if the node is labeled by $\sigma$. Note that in the example of Figure~\ref{foundations:fig:bijection}, one and only one symbol occurs at each position. In the general case however, there is no restriction on the content of characteristic sets. A given node may belong to several characteristic sets. In this case, a node may be labeled by several symbols. This can be used to encode other properties than XML labeling. On the opposite, a particular position may not be a member of any characteristic set. In this case, the overall structure contains a node which is not labeled by any symbol of the considered alphabet; therefore it is no longer a labeled tree on this alphabet. 
Chapter~\ref{containment} examines how XML trees can be encoded by constraining these structures using WS2S formulas introduced in the next section. 

\label{foundations:tree-restrictions}

\subsection{WS2S Formulas}
\label{foundations:ws2s-syntax}
From a syntactic point of view, WS2S formulas can be generated by a simple core language, whose abstract syntax follows:


\smallsyntax{
\entry  \notation{\lwss}{WS2S formulas}{n:lwss} \ni     \Phi, \Psi     [formula]
             X \subseteq Y               [inclusion]
\oris        X=Y-Z             [difference]
\oris        X=Y.0 [first successor]
\oris        X=Y.1 [second successor]
\oris        \neg \Phi [negation]
\oris        \Phi \et \Psi [conjunction]
\oris        \exists X.\Phi [existential quantification]}

where $X$, $Y$, and $Z$ denote arbitrary second-order variables.
Other usual logical connectives can be derived as syntactic sugars of the core:
\begin{align*}
\Phi \ou \Psi & \equalsdef  \neg ( \neg \Phi \et  \neg \Psi) \\
\Phi \ourimplies \Psi & \equalsdef \neg \Phi \ou \Psi \\
\Phi \Leftrightarrow \Psi & \equalsdef  \Phi \et \Psi \ou \neg \Phi \et \neg \Psi\\
\forall X.\Phi & \equalsdef  \neg \exists X. \neg \Phi
\end{align*}
Note that only second order variables appear in the core. This is because first order variables can be encoded as singleton second-order variables. A notation convention is adopted for simplifying the remaining part of the chapter: first-order variables are noted in lowercase and second-order variables in uppercase.


\subsection{WS2S Semantics}
\label{foundations:ws2s-sem}

This section gives an interpretation of WS2S formulas as finite subsets of $\{0,1\}^*$.
Given a fixed main formula $\phi$ with $k$ variables, its semantics is defined inductively. Let a tuple representation $\tuplerep{t}=(X_1,...,X_k) \in (\{0,1\}^*)^k$ be an interpretation of $\phi$. The notation $\tuplerep{t}(X)$ denotes the interpretation $X_i$ (such that $1 \leq i\leq k$) that $\tuplerep{t}$ associates to the variable $X$ occurring in $\phi$. The semantics of $\phi$ is inductively defined relative to $\tuplerep{t}$. The notation $\tuplerep{t} \vDash \phi$ (which is read: $\tuplerep{t}$ satisfies $\phi$) is used if the interpretation $\tuplerep{t}$ makes $\phi$ true:
\begin{align*}
\tuplerep{t} \vDash X \subseteq Y          & \text{ iff }               \tuplerep{t}(X) \subseteq \tuplerep{t}(Y) \\
\tuplerep{t} \vDash X=Y-Z                      & \text{ iff }            \tuplerep{t}(X) = \tuplerep{t}(Y) \setminus \tuplerep{t}(Z)\\
\tuplerep{t} \vDash X=Y.0                      & \text{ iff }             \tuplerep{t}(X) = \{p.0 \; | \; p \in \tuplerep{t}(Y) \} \\
\tuplerep{t} \vDash X=Y.1                      & \text{ iff }             \tuplerep{t}(X) = \{p.1 \; | \; p \in \tuplerep{t}(Y) \} \\
\tuplerep{t} \vDash \neg \phi                  & \text{ iff }             \tuplerep{t} \nvDash \phi \\
\tuplerep{t} \vDash \phi_1 \et \phi_2  & \text{ iff }                     \tuplerep{t} \vDash \phi_1 ~\text{and}~ \tuplerep{t} \vDash \phi_2 \\
\tuplerep{t} \vDash \exists X.\phi         & \text{ iff }                 \exists  I \subseteq \{0,1\}^*, \tuplerep{t}[X \mapsto I]\vDash \phi \\
\end{align*}
where the notation $\tuplerep{t}[X \mapsto I]$ denotes the tuple representation that interprets $X$ as $I$ and all other variables as $\tuplerep{t}$ does.
Note that the two successors of a particular position always exist in WS2S. 

A formula $\phi$ naturally defines a language $\notation{\lang{\phi}}{Language defined by the formula $\phi$}{n:langphi}=\{ \tuplerep{t} \; | \; \tuplerep{t} \vDash \phi\}$ over the alphabet $(\{0,1\}^*)^k$ , where $k$ is the number of variables of $\phi$.

\subsection{Equivalence of WS2S and FTA}
\label{foundations:deciding-ws2s}
It has been known since the 1960's that the class of regular tree languages is linked to decidability questions in formal logics. In particular, WS2S is decidable through the automaton-logic connection \cite{Thatcher68,Doner70}, using tree automata (introduced in Section~\ref{foundations:fta}). In 1968, Thatcher and Wright proved the following equivalence:
\begin{thm}[\cite{Thatcher68}]
WS2S is as expressive as finite tree automata.
\end{thm}
The proof works in two directions. First, it is shown that a WS2S formula can be created such that it simulates a successful run of a tree-automaton. Second, for any given WS2S formula a corresponding tree automaton can be built. 

Technically, the correspondence of WS2S formulas and tree automata relies on a convenient representation that links the truth status of a formula with the recognition operated by an automaton. This representation is a matricial vision of the tuple representation described in Section~\ref{foundations:shape}. Let $\tuplerep{t}$ be a tuple, its matricial representation $\notation{\matricialrep{t}}{Matricial representation of the structure $t$}{n:matricialrep}$ is indexed by variables indices and positions in the tree. Entries of $\matricialrep{t}$ correspond to values in $\{0,1\}$ of characteristic functions: an entry $(v,p)=1$ in $\matricialrep{t}$ means that the position $p$ belongs to the variable $X_v$.

Consider for instance the formula $\phi = (\exists X. \exists Y. \; Y=Z.0 \et X=Z.1)$ which has three variables $X$, $Y$, and $Z$. A typical matrix looks like:
$$\begin{array}{c|cccccc}
\;    & \epsilon & 0 & 00 & 01 & 010 & 1 \\ \hline
X     & 1  & 1 & 0   & 0  & 0 & 0 \\
Y     & 0  & 1 & 0   & 1  & 0 & 0\\
Z     & 0  & 0 & 1   & 0  & 0 & 1\\
\end{array}$$
Note that this matrix is finite since only finite trees are considered. It furthermore allows to capture finite trees of unbounded depth. As a counterpart, there is an infinite number of matrices that define the same interpretation: any number of columns of zeros may be appended at the right end of the matrix (for positions after the end of the tree). Let $\matricialrep{t}$ be the minimum matrix, without such empty suffix. Rows of the matrix are called tracks and give the interpretation of each variable, which is defined as the finite set $\{p \;|$ the bit for position $p$ in the $X_i$ track is $1\}$.

Each column of the matrix is a bit vector that indicates the membership status of a node to the variables of the formula. The automaton recognizes all the interpretations (matrices) that satisfy the formula.
A line by line reading of the matrix gives the interpretation of each variable (i.e. its associated set of positions), whereas an automaton processes the matrix column by column; it transits on each bit-vector.

\subsection{From Formulas to Automata}
\label{foundations:from-ws2s-to-automata}

Given a particular formula, a corresponding FTA can be built in order to decide the truth status of the formula. 

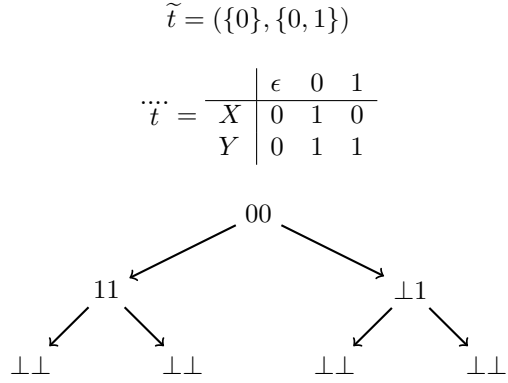
\begin{figure}[h]
\centering
\begin{tabular}{c}
$\tuplerep{t}=(\{0\}, \{0,1 \})$
\\ \\
$\matricialrep{t}=\begin{array}{l|lll}
  \;  & \epsilon & 0 & 1 \\ \hline
X & 0        & 1 & 0 \\
Y & 0        & 1 & 1 \\
\end{array}$
\\ 
\\
\begin{tikzpicture}[scale=1]
\draw (3,3) node(root) {$00$};
\draw (1,2) node(l) {$11$};
\draw (5,2) node(r) {$\bot1$};
\draw (0,1) node(ll) {$\bot\bot$};
\draw (2,1) node(lr) {$\bot\bot$};
\draw (4,1) node(rl) {$\bot\bot$};
\draw (6,1) node(rr) {$\bot\bot$};

\draw [thick, ->, black] (root) -- (l);
\draw [thick, ->, black] (root) -- (r);
\draw [thick, ->, black] (l) -- (ll);
\draw [thick, ->, black] (l) -- (lr);
\draw [thick, ->, black] (r) -- (rl);
\draw [thick, ->, black] (r) -- (rr);
\end{tikzpicture}

\end{tabular}
\caption{Representations of a Satisfying Interpretation of $X \subseteq Y$}
\label{foundations:term-symb}
\end{figure}

Let $\phi$ be a formula with $k$ second-order variables. As an interpretation of $\phi$, consider a tuple representation $\tuplerep{t}=(X_1,...,X_k) \in (\{0,1\}^*)^k$. The tree automaton that corresponds to $\phi$ is noted $\notation{\automaton{\phi}}{Tree automaton corresponding to $\phi$}{n:automaton}$. $\automaton{\phi}$ operates over the alphabet $\Sigma=\{0,1\}^k$, and can be seen as processing $\matricialrep{t}$ column by column.  Note however that there is an infinite number of matrices that defines the same interpretation. On one hand, any number columns of zeros can appear at the end of the matrix. On the other hand, a column of zeros can also appear for any position in the tree, before a non-empty column, denoting that this position is not a member of any interpretation.
The automaton therefore faces a problem: when recognizing a column of zeros, knowing if the recognition should stop (because the end of the tree has been reached) or continue. In other terms, the automaton needs to know the maximal depth of the tree as an additional information in order to know when to stop. To this end, a new termination symbol $\notation{\bot}{Termination symbol}{n:bot}$ is introduced. From a matricial point of view, this symbol appears as a component of a bit-vector whenever this component will not be $1$ anymore for the remaining bit-vectors to be processed.
Technically, $\automaton{\phi}$  recognizes the tree representation $\notation{\treerep{t}}{Tree representation of the structure $t$}{n:treerep}$ of $\tuplerep{t}$. $\treerep{t}$ is obtained from $\tuplerep{t}$ as follows:
\begin{enumerate}
\item the set of positions of $\treerep{t}$ is the prefix-closure of $X_1 \cup ... \cup X_k$
\item leaves of $\treerep{t}$ are labeled with $\bot^k$
\item binary constructors of the tree are labeled with an element of $\{\bot, 0,1 \}^k$ such that the $i^\text{th}$ component of a position $p$ in $\treerep{t}$ is marked: $1$ if and only if $p \in X_i$, $0$ if and only if $p \notin X_i$ and some extension of $p$ is in $X_i$, and $\bot$ otherwise
\end{enumerate}
Note that in this tree representation, $\bot$ appears as a component of a node label whenever no descendant node has a $1$ for the same component.
For example, Figure~\ref{foundations:term-symb} gives the tuple, the matrix, and the tree representation of a particular satisfying interpretation of the formula $X \subseteq Y$.

\begin{thm}[\cite{Thatcher68,Doner70}]
\label{foundations:logic-automaton}
For every formula $\phi$, there is an automaton $\automaton{\phi}$ such that:
$$\tuplerep{t} \vDash \phi \;\; \equiv \;\; \automaton{\phi} ~\text{accepts}~ \treerep{t}$$
\end{thm}

The automaton $\automaton{\phi}$ is calculated using an induction scheme. A basic bottom-up tree automaton corresponds to each atomic formula:
\begin{center}
$\begin{array}{lcl}
\vspace{1em}
\automaton{X \subseteq Y} & =  &\left(\left\{
\begin{array}{ll}
 q \leftarrow \bot\bot, & q \leftarrow \bot0(q, q) \\
  q \leftarrow \bot1(q, q), & q \leftarrow 00(q, q) \\
  q \leftarrow 01(q, q), & q \leftarrow 11(q, q) \\
\end{array}\right\}
, \{q \}\right) \\
\vspace{1em}
\end{array}$

$\begin{array}{lcl}
\automaton{X=Y-Z} & = &\left(\left\{
\begin{array}{ll}
   q \leftarrow \bot\bot\bot       , &
   q \leftarrow \bot\bot0(q, q), \\

   q \leftarrow \bot0\bot(q, q), &
   q \leftarrow \bot00(q, q), \\
   q \leftarrow \bot01(q, q), &

   q \leftarrow \bot11(q, q), \\

   q \leftarrow 0\bot\bot(q, q), &
   q \leftarrow 0\bot0(q, q), \\
   q \leftarrow 0\bot1(q, q), &

   q \leftarrow 00\bot(q, q), \\
   q \leftarrow 000(q, q), &
   q \leftarrow 001(q, q), \\

   q \leftarrow 011(q, q) , &

   q \leftarrow 11\bot(q, q), \\
   q \leftarrow 110(q, q),
 \end{array}\right\}
, \{ q\}\right)\end{array}$

$\begin{array}{lcl}
\automaton{X=Y.0} & = & \left(\left\{
\begin{array}{ll}
 q \leftarrow \bot\bot, &  q' \leftarrow 00(q, q') \\
  q' \leftarrow 00(q', q) & q' \leftarrow 01(q'', q) \\
  q'' \leftarrow 1\bot(q, q) & q'' \leftarrow 10(q,q) \\
\end{array}\right\}
, \{q' \}\right)
\end{array}$

$\begin{array}{lcl}
\automaton{X=Y.1} & = &\left(\left\{
\begin{array}{ll}
 q \leftarrow \bot\bot, &  q' \leftarrow 00(q, q') \\
  q' \leftarrow 00(q', q) & q' \leftarrow 01(q, q'') \\
  q'' \leftarrow 1\bot(q, q) & q'' \leftarrow 10(q,q) \\
\end{array}\right\}
, \{q' \}\right)%
\end{array}$
\end{center}
Logical connectives are then translated into automata-theoretic operations, taking advantage of the closure properties of tree automata (presented in Section~\ref{foundations:fta-closure}). Formula conjunction is translated into intersection of automata:
$$\automaton{\phi_1 \et \phi_2} = \automaton{\phi_1} \cap \automaton{\phi_2}$$ and negation is translated into automata complementation:
$$\automaton{\neg \phi} = \complement{(\automaton{\phi})}$$

Existential quantification relies on projection and determinization of tree automata. The automaton
$\automaton{\exists X.\phi}$ is derived from $\automaton{\phi}$ by projection. This means the alphabet of $\automaton{\exists X.\phi}$ has to be one element smaller than the alphabet of $\automaton{\phi}$. In every tuple of $\automaton{\phi}$ the X component is removed, so that its size is decreased by one. The rest of the automaton remains the same. Intuitively, $\automaton{\exists X.\phi}$ acts as $\automaton{\phi}$ except it is allowed to guess the bits for X. The automaton $\automaton{\exists X.\phi}$ may be non-deterministic even if $\automaton{\phi}$ was not \cite{tata}, that is why determinization is required.

As a result, for every formula $\phi$ it is possible to build an automaton $\automaton{\phi}$ in this manner, which defines the same language as $\phi$:
$$\lang{\automaton{\phi}}=\lang{\phi}$$

Analyzing the automaton $\automaton{\phi}$ allows to decide the truth status of the formula~$\phi$:
\begin{itemize}
\item
if $\lang{\automaton{\phi}}=\emptyset$ then $\phi$ is unsatisfiable;
\item
else $\phi$ is satisfiable. If $\lang{\complement(\automaton{\phi})}=\emptyset$ then $\phi$ is always satisfiable (valid). 
\end{itemize}

Possessing the full automaton corresponding to a formula is of great value, since it can be used for generating examples and counter-examples of the truth status of the formula. A relevant example (or counter-example) can be built by looking for an accepting run of the automaton (or its complement).

\subsection{WS2S Complexity}

\label{foundations:ws2s-complexity}
Two factors have a major impact on the cost of a WS2S decision procedure:
\begin{enumerate}
\item the number of second-order variables in the formula
\item the number of states of the corresponding automaton (automaton size)
\end{enumerate}

The number of second-order variables determines the alphabet size. More precisely, a formula with $k$ variables is decided by an automaton operating on the alphabet $\Sigma=\{0,1\}^k$. Representing the transition function $\delta$ of such an automaton can be prohibitive. Indeed, in the worst case, the representation of a complete FTA requires $2^k \cdot \left| Q\right|^3$ transitions where $Q$ is the set of states of the automaton.
A direct encoding with classical FTA such as the one described in Section~\ref{foundations:from-ws2s-to-automata} would lead to an impracticable algorithm. Modern logical solvers represent transition functions using BDDs \cite{bryant86} that can lead to exponential improvements \cite{mona-user-manual,tozawa-tableaux05}.

As seen in Section~\ref{foundations:from-ws2s-to-automata}, automaton construction is performed inductively by composing automata corresponding to each sub-formula. During this process, the number of states of intermediate automata may grow significantly. Automaton size depends on the nature of the 
automata-theoretic operation applied and the sizes of automata constructed so far. Each operation on tree automata particularly affects the size of the resulting automaton:
\begin{itemize}
\item
Automata intersection causes a quadratic increase in automaton size in the worst case, as well as all binary WS2S connectors ($\et$, $\ou$, $\Rightarrow$) that involve automata products \cite{mona-impl-secrets}.

\item when considering deterministic complete automata, automata complementation corresponding to WS2S negation is a linear-time algorithm that consists in flipping accepting and rejecting states.

\item
The major source of complexity originates from automata determinization which may cause an exponential increase of the number of states in the worst case \cite{tata}. 
Logical quantification involves automaton projection (c.f. Section~\ref{foundations:from-ws2s-to-automata}) which may result in a non-deterministic automaton,  thus involving determinization. Hopefully, a succession of quantifications of the same type can be combined as a single projection followed by a single determinization. However, any alternation of second-order quantifiers requires a determinization, thus possibly causing an exponential increase of the automaton size.
\end{itemize}

As a consequence, the number of states of the final automaton corresponding to a formula with $n$ quantifier alternations is in the worst case 
a tower of exponentials of height $c \cdot n$ where $c$ is some constant, and this is a lower bound \cite{sm73}. The translation from logical formulas to tree automata is thus \emph{non-elementary}\footnote{The term \emph{elementary} introduced by Grzegorczyk \cite{non-elementary} refers to functions obtained from some basic functions by operations of limited summation and limited multiplication. Consider the function $\fun{tower}{}$ defined by: $$\left\{ \begin{array}{l} \fun{tower}{n,0}=n \\ \fun{tower}{n,k+1}=2^{\fun{tower}{n,k}}
\end{array}\right.$$
Grzegorczyk has shown that every elementary function in one argument is bounded by $\lambda n. \fun{tower}{n,c}$ for some constant $c$. Hence, the term \emph{non-elementary} refers to a function that grows faster than any such function.}:
\begin{thm} \cite{meyer72,stockmeyer74} 
The satisfiability problem for WS2S formulas has an unbounded stack of exponentials as worst case lower bound.
\end{thm}
This high complexity, originating from the full construction and complementation of intermediate tree automata, is the counterpart of WS2S expressiveness and succinctness. Chapter~\ref{containment} of this dissertation investigates how it is possible to deal with this complexity in practice, proposes a decision procedure for XPath containment based on WS2S along with optimizations of the WS2S decision procedure in the XML setting.

\section{Temporal Logics}

Some temporal and fixpoint logics closely related to FO and MSO have been introduced and allow to avoid explicit automata construction.

\subsection{FO Relatives}
For query languages, Computational Tree Logic (CTL) has been proposed in \cite{clarke81}. CTL is equivalent to FO over tree structures \cite{libkin-lics05} and its satisfiability is in EXPTIME. The connection between XPath and FO relatives like CTL has been studied in \cite{Marx-edbt04,suciu-miklau-jacm04,libkin-lics05}. In particular, the work found in \cite{Marx-edbt04} characterizes a subset of XPath in terms of extensions of CTL, whose satisfiability is in EXPTIME. Authors of \cite{suciu-miklau-jacm04} also observed that a fragment of XPath can be embedded in CTL. 
However, regular tree languages are not fully captured by FO \cite{segoufin-stacs05}. These approaches are therefore not intended to support XML types. 

In a attempt to reach more expressive power, the work that is presented in \cite{marx-jacl05} proposes a variant of Propositional Dynamic Logic (PDL) \cite{fischer79} with an EXPTIME complexity, but whose exact expressive power (as a strict subset of MSO) is still under study. 

The goal of the XPath research presented so far is limited to establishing new theoretical properties and complexity bounds. 

The research presented in this dissertation differs in that it seeks, in addition to the previous goals, efficient implementation techniques and concrete design that may be directly applied to XML type-checking problems involving XPath queries and regular tree types.

\subsection{MSO Relatives}

The propositional modal $\mu$-calculus introduced in \cite{kozen83} has been shown to be as expressive as non-deterministic tree automata \cite{emerson-focs91}. From \cite{arnold-niwinski92,vardi-tacs99}, it is known that WS2S is exactly as expressive as the alternation-free fragment (AFMC) of the propositional modal $\mu$-calculus. The $\mu$-calculus subsumes all early logics such as CTL and PDL (see \cite{libkin-lics05} for a recent survey on tree logics). 
The $\mu$-calculus is trivially closed under negation, can be extended with converse programs, and still remains decidable in EXPTIME \cite{vardi-icalp98}.  The best known complexity for the resulting logic is $2^{O(n^4 \cdot\text{log}~n)}$ \cite{gradel-book02}. As a counterpart of its substantially inferior complexity, it looses the succintness of MSO. Fixpoint logics are indeed notorious for being difficult to understand, even for reasonably expert people, as pointed by \cite{bradfield-bookchapter01}. However, it is assumed in this dissertation that this is not a problem since the logic is only intended as a target for the compilation of XML concepts. 
As such, the $\mu$-calculus constitutes an interesting alternative for studying MSO-related problems. From a theoretical perspective, the AFMC with converse sounds as an appropriate logic for XML: it is expressive enough to capture a significant class of XPath decision problems, while offering an interesting balance between complexity and expressiveness.

The work found in \cite{tozawa-tableaux05} proposes a decision procedure for the AFMC, whose time complexity is $2^{O(n \cdot\text{log}~n)}$. However, models of the logic are Kripke structures (general infinite graphs), and the logic lacks the finite model property (i.e. there exist formulas which are satisfiable on Kripke structures and unsatisfiable on finite trees). In a preliminary work on XML type-checking, a logic for finite trees was presented \cite{tozawa-ppl04}, but the logic is not closed under negation. 

Chapter~\ref{analysis} of this dissertation studies how the recent AFMC decision procedure proposed in \cite{tozawa-tableaux05} can be used in the context of XML. Based on the outcome of these investigations, the final Chapters~\ref{xml-calculus:the-logic-for-xml} and \ref{xml-calculus:sec:algo} prove the decidability of a new logic for finite trees, derived from the $\mu$-calculus, in time $2^{O(n)}$ and propose an effective algorithm for checking its satisfiability in practice.

\section{Systems for XML Type-Checking}

This section presents other related work on XML type-checking frameworks, which do not definitely aim at supporting XPath. Actually, none of the approach presented in this section is able to effectively deal with the expressive power of the XPath fragment considered in this dissertation (and presented in Section~\ref{foundations:xpath-syntax}). Nevertheless, this section gathers the main approaches and ideas developed elsewhere for static type-checking in the XML setting. Although notably different, several approaches can be seen as complementary to the work proposed in this dissertation. Most techniques are based on regular tree languages and use tree automata introduced in Section~\ref{foundations:fta}.

\subsection{Formulations of the Static Validation Problem}

The paper \cite{audebaud-rr00} was influential in clearly defining the static validation problem. As an early attempt, it also proposes a set of typing rules to establish relationships between the input and output type of an XSLT transformation, but the method is only applicable to a tiny fragment of XSLT. The XML type-checking problem was later described in \cite{suciu-sigmodrec02}. A more recent survey work on the static type checkers for XML transformation languages can be found in \cite{moller-icdt05}. The remaining part of this section presents the major known frameworks and innovations around the type-checking of XML.

\subsection{Inverse Type Inference with Tree Transducers}

The paper \cite{suciu-sigmodrec02} describes how static type-checking can be performed using forward type inference. Forward type inference refers to the ability to automatically deduce the output type of the XML document derived from the evaluation of an XML transformation. This is usually done by inference rules, and corresponding type inference algorithms are generally polynomial in the XML setting \cite{tozawa-doceng01}. Type inference is used to do type-checking. For instance, if a program is assumed to return a type $T_\text{out}$; once the inferred output type $T_\text{out}^\text{inf}$ is known, type-checking can be performed by testing the inclusion $T_\text{out}^\text{inf} \subseteq T_\text{out}$. 
The work found in \cite{milo-jcss03,suciu-sigmodrec02} reveals an important limitation of forward type inference in the context of XML: unfortunately, forward type inference is not complete. This is because the output type of a program may actually be a non-regular tree language that cannot be infered. In that case, the infered regular type is typically a larger approximation of the actual type, and the type-checker rejects the correct program, because $T_\text{out}^\text{inf} \not\subseteq T_\text{out}$ (an example and details on this limitation can be found in \cite{suciu-sigmodrec02}).

The work found in \cite{milo-jcss03} introduces the technique of inverse type inference in an attempt to overcome this problem. 
Inverse type inference computes the allowed input language for a so-called $k$-pebble transducer given its output language. The resulting algorithm has non-elementary complexity. The paper \cite{martens-icdt03} investigates how the expressive power of tree transducers must be further restricted in order to allow a polynomial time decision algorithm. The practical relevance and usability of techniques based on tree transducers have not yet been demonstrated.

\paragraph{XSLT0} 
The paper \cite{tozawa-doceng01} examines a fragment of XSLT called XSLT0 which covers the structural recursion core of XSLT. It relies on inverse type inference  to perform exact static validation, in the manner of \cite{milo-jcss03} but with a more efficient (exponential time) algorithm. However, XSLT0 does not support XPath but only allows simple child steps in the recursion. Compiling XSLT into XSLT0 is thus possible for only the simplest transformations.

\subsection{XDuce, $\cduce$, Xtatic}
XDuce \cite{hosoya-toit03} was the first domain specific programming language with type-checking of XML operations. The most essential part of the type system is the subtyping relation, which is defined by inclusion of the values represented by the types (this is also called \emph{structural} subtyping\footnote{Structural subtyping is usually opposed to \emph{nominal} subtyping in which type compatibility and equivalence are not determined by the type's structure but through explicit declarations and names of the types. See \cite{su-popl02} and \cite{simeon-popl03} for more details on subtyping paradigms.}). The proposed algorithm for subtyping attempts to avoid the worst case exponential time complexity in practical cases. Instead of relying on tree automata determinization, it checks the inclusion relation by a top-down traversal of the original type expressions. XDuce's algorithm builds on the previous work found in \cite{aiken-fplca01}, and extends it with several implementation techniques. The resulting algorithm appears efficient in practice \cite{hosoya-toit03}. XDuce has provided the foundation for later languages, in particular the $\cduce$ \cite{benzaken-icfp03,frisch-thesis04} and XStatic \cite{GapeyevPierce03} languages. The $\cduce$ language attempts to extend XDuce towards being a general purpose functional language. To this end, $\cduce$ provides a more sophisticated type system featuring function types, intersection and negation types. It extends XDuce with higher-order functions, variations of pattern matching primitives, and parametric polymorphism \cite{hosoya-popl05}. 
Xtatic aims at integrating the main ideas from XDuce into C$^\#$. All these languages support pattern-matching through regular expression types but not XPath. As pointed in \cite{colazzo-icfp04}, a major difference is that pattern-matching implements a \emph{one-match} semantics, i.e. every pattern, instead of collecting every matched piece of data (as in standard query languages such as XPath), only binds the first match. Although some recent work shows how to translate parts of XPath into Xtatic \cite{GapeyevPierce2004}, the XPath fragment considered does not include reverse axes nor negation in qualifiers.

\subsection{Symbolic XML Schema Containment}
The work found in \cite{tozawa-ciaa03}  proposes a symbolic algorithm, based on binary decision diagrams \cite{bryant86}, in order to solve the containment between two XML schemas. The algorithm appears to be efficient in practice and favorably compares to the one used by XDuce. The idea of using symbolic techniques is similar to the one used in implementations presented in Chapters~\ref{analysis} and Chapter~\ref{xml-calculus:sec:algo}. The implicit encoding of FTA presented in \cite{tozawa-ciaa03} is however significantly simpler since it only considers XML types (XML types only use a simple form of tree navigation; they do not need upward nor multidirectional navigation in trees as XPath does). Nevertheless, this work was the first to reveal the interest of using implicit techniques in the context of XML. This work suggests and motivates further developments such as simplifications for particular cases of the more general symbolic techniques used in Chapters~\ref{analysis} and Chapter~\ref{xml-calculus:sec:algo}. 

\subsection{XJ}
The XJ \cite{harren-www05} language aims at integrating XML processing closely into Java. Types are regular expressions over XML Schema declarations. The type system has two levels: regular expression operators and XML Schema declarations. A peculiarity of XJ is that subtyping on the schema level is \emph{nominal}, i.e. type compatibility and containment is determined by explicit declarations and the name of the types (as in Java). This aspect contrasts with the structural subtyping systems used in XDuce (and in this dissertation). XJ subtyping on the regular expression level is defined as regular language inclusion on top of the schema subtyping. \cite{moller-icdt05} argues that an inherited drawback of the underlying nominal style of subtyping is that a given XML value may be tied too closely with its schema type, which thus makes certain transformations more complex than they could be. XJ nevertheless provides an interesting experiment of integration of type-safe processing in Java, and a detailed study of nominal subtyping in the context of XML can be found in \cite{simeon-popl03}.

\subsection{Approximated Approaches for XSLT}

Several approaches aim at proposing XSLT debugging features at compile-time by choosing to sacrifice exact decidability and to settle for pragmatic approximations instead. Among this line of work, the paper \cite{dong-adc04} aims at conservatively analyzing the flow of an XSLT transformation. It uses the control-flow information to detect unreachable templates and guarantee termination. The analysis is however less precise than the more recent one found in \cite{moller-rr05}. 
The work \cite{moller-rr05} presents a more complete approximated technique that is able to statically detect errors in XSLT stylesheets. Their approach could certainly benefit from using the exact algorithm proposed in Chapter~\ref{xml-calculus:sec:algo} instead of their conservative approximation.

\subsection{Path Correctness for $\mu$XQ Queries}

The work found in \cite{colazzo-jfp06} proposes a sound and complete type system for ensuring path correctness for XML queries. The notion of navigation correctness is similar to the emptiness problem formulated in chapter~\ref{analysis:xml-decision-pb} that can be used for detecting contradictions. The common idea is that if a subexpression of a query always yields an empty result then this should be considered as an error. The considered query language in \cite{colazzo-jfp06}, called $\mu$XQ, covers a minimal core of XQuery \cite{xquery} but ignores reverse navigation. In comparison, the XPath fragment considered in this dissertation includes all axes. The algorithm presented in Chapter~\ref{xml-calculus:sec:algo} may provide perspectives on how to extend the type system of \cite{colazzo-jfp06} to deal with reverse navigation.

\section{The Spatial Logic Perspective}

Spatial logics are formalisms traditionally used for describing the behavior and spatial structure of concurrent systems. The main ingredient of spatial logics is an operator called composition (or separation), which usually permits reasoning over concurrent and mobile processes \cite{boneva-lics05}. Spatial logics have recently been found useful in the study of semistructured data and related query languages as they allow to express properties about structures such as graphs \cite{cardelli-icalp02,dawar-rr04} and trees \cite{cardelli-mscs04}. 

The work found in \cite{cardelli-mscs04} proposes the TQL logic as the core of a query language for semistructured data represented as unranked trees and unordered trees. The TQL logic is based on the ambient logic \cite{cardelli-tcs00,cardelli-mscs06}. 
It is known that TQL is more expressive than MSO since it can express some counting properties about trees that can not be defined in MSO.
It has been shown that a fragment of the ambient logic contained in TQL is undecidable \cite{charatonik-tcs03}. Nevertheless, decidable fragments of TQL could be useful for building type systems for semistructured data such as the one proposed in \cite{calcagno-tldi03}, and also for testing emptiness and containment of queries, as suggested in \cite{cardelli-mscs04}. TQL thus provides an interesting foundation for further research.

The work found in \cite{boneva-lics05} considers a fragment of TQL called STL and characterize its expressiveness. STL satisfiability is shown undecidable but some syntactic restrictions over STL formulas allow to capture MSO.

The logic TL described in \cite{dal-zilio-popl04} is also based on the ambient logic. TL can be encoded into the so-called sheaves automata proposed in \cite{dal-zilio-rta03}, whose transitions are conditioned by Presburger formulas. 

The major difference between these spatial logics and the work presented in this dissertation is that spatial logics operates on \emph{unordered} trees, whereas this dissertation considers ordered trees (cf. Section~\ref{foundations:unranked-and-binary-trees}) such as structured documents. On one hand, the extension of TQL's data model with ordering is an interesting and important open issue \cite{conforti-webdb02}. On the other hand, extending the logic of ordered trees proposed in the Chapters~\ref{xml-calculus:the-logic-for-xml} and~\ref{xml-calculus:sec:algo} of this dissertation with counting constraints is also an interesting and promising perspective. These research directions can thus be seen as complementary and could certainly benefit from a reciprocal inspiration.

\subsection{The Sheaves Logic}

The work found in \cite{dal-zilio-aaecc06} introduces a modal logic for documents called GDL, inspired from TQL, and proves the decidability of a fragment of GDL called the Sheaves logic. The Sheaves logic (SL) operates on ordered trees, and combines regularity and counting constraints. SL provides an interleaving operator for dealing with mixed ordered and unordered content. One one hand SL lacks recursion, i.e. fixpoint operators which are needed for supporting query langages (cf. Chapter~\ref{analysis}); one the other hand SL allows to reason about numerical properties of the contents of elements, and may provide the inspiration for the integration of counting constraints in the logic presented in Chapter~\ref{xml-calculus}, kept for future work.

 \ifglobalcompil
  \else 
   \bibliographystyle{apalike}
   \bibliography{references}
   \end{document}
  \fi

\part*{Preliminary Investigations towards a Logic for XML}
\newif\ifglobalcompil\globalcompiltrue

  \ifglobalcompil
  \else 
    \input{Preamble} 
    \input{Markup}                       
    \newcommand{\chapterabstract}[1]{Chapter Abstract: #1}
    \begin{document} 
  \fi

\mychapter{Monadic Second-Order Logic for XML} 
\label{containment}


\section{Introduction}

This chapter first investigates how MSO can be used in the context of XML, despite its non-elementary complexity\footnote{It is well known that type inference for  higher-order typed lambda calculi can have non-elementary complexity, and is nevertheless effectively used by typed functional programming languages such as those of the ML family \cite{henglein-popl91}.}. 
A sound and complete decision procedure for containment of XPath queries is proposed based on MSO. Specifically, XPath queries are translated into equivalent formulas in WS2S introduced in Section~\ref{foundations:ws2s-syntax}. Using this translation, the logical formulation of the containment problem is constructed, and optimized, by taking into account XPath peculiarities. The containment formula is then decided using tree automata. When the containment relation does not hold between two XPath expressions, a counter-example XML tree is generated. A complexity analysis is provided, along with practical experiments. 


\paragraph{Chapter Outline}

Section~\ref{containment:xml-tree-ws2s} presents the encoding of XML trees into WS2S. Section~\ref{containment:xpath-ws2s} explains the translation of XPath queries to logical formulas. A complexity analysis and an optimization method are given in Section~\ref{containment:complexity-ws2s}. Experimental results and the outcome of this approach are respectively discussed in Sections~\ref{containment:implementation-and-experiments} and \ref{containment:outcome-ws2s}.

\section{Representation of XML Trees}
\label{containment:xml-tree-ws2s}
Section~\ref{foundations:logical_description_of_trees} presented how characteristic sets can be used for describing shapes. A shape is basically a second order variable, interpreted as a set of nodes, for which particular properties hold. Using WS2S, this section now expresses additional requirements that a shape should fulfill in order to be an XML tree.

The first requirements are structural. First, in order to be a tree, a shape  $X$  must be prefix-closed, that is, for any position in the tree, any prefix of this position is also in the tree:
$$\begin{array}{l}\text{PrefixClosed}(X) \; \equalsdef \; \forall x. \forall y. ((y=x.1 \; \ou \; y=x.0) \et y \in X) \ourimplies x \in X\end{array}$$
This ensures the shape is fully connected. Second, a predicate for the root of $X$ is defined:
$$\text{IsRoot}( X,  x) \equalsdef \; x \in X \et \neg(\exists z . z \in X \et (x=z.1 \ou x=z.0))$$
In order to be a tree and not a hedge, $X$ must have only one root with no sibling:
$$\begin{array}{l}\text{SingleRoot}(X) \; \equalsdef \; \forall x. \text{IsRoot}(X,x) \ourimplies x.1 \notin X\end{array}$$
Then, the labeling of the tree must be consistent with XML. The same symbol may appear at several locations in the tree with different arities: either as a binary constructor or as a leaf. However, one and only one symbol is associated with a position in the shape. Assume that the set of characteristic sets forms a partition:
$$\begin{array}{lll}\text{Partition}(X,X_1,...,X_n) & \; \equalsdef \; & X=\bigcup_{i=1}^n X_i \et \text{Disjoint}(X_1,...,X_n) \\
\text{Disjoint}(X_1,...,X_n) &\; \equalsdef \; & \bigwedge_{i \neq j} X_i \cap X_j = \emptyset\end{array}$$
this prevents a node to have multiple labels, but it also prevents a tree to be labeled using an infinite alphabet. 
The problem comes from declaring $X=\bigcup_{i=1}^n X_i$ that prevents any other symbol to occur in the tree.
Consider instead that the characteristic sets must be disjoint, then a position in the tree may not be a member of any of the considered characteristic sets. That is how labeling from an infinite alphabet is emulated.
As a result, an XML tree is encoded in the following way:
$$\begin{array}{rl}
\text{XMLTree}(X,X_1,...,X_n)  \; \equalsdef \; &  \text{PrefixClosed}(X) \\
 \et & \text{SingleRoot}(X) \\
 \et & \text{Disjoint}(X_1,...,X_n) \\
 \et & X \neq \emptyset 
\end{array}$$
where $X$ is the tree (non-empty in order not to get degenerated results) and the $X_i$s are the characteristic sets.
Figure~\ref{containment:fig:sample-tree-mona} introduces how this is formulated in MONA Syntax~\cite{mona-user-manual}, for the case of two characteristic sets of interest named \texttt{Xbook} and \texttt{Xcitation}.
The only difference is that the shape $X$ is declared as a global free variable named \texttt{\$} together with associated restrictions, instead of being passed as a parameter to predicates. In MONA syntax, ``\texttt{var2}'' is the keyword for declaring a free second-order variable; ``\texttt{all1}'' is the universal quantifier for first-order variables; and ``\&'' and ``$|$'' respectively stand for the ``$\et$'' and ``$\ou$'' connectives.

\begin{figure}
\centering
\begin{boxedverbatim}
ws2s;
# Data Model
var2 $ where ~empty($) 
 & (all1 x : all1 y : ((y=x.1 | y=x.0) 
   & (y in $)) => x in $) 
 & all1 r : (r in $ & ~(ex1 z : z in $ 
   & (r=z.1 | r=z.0))) 
              => r.1 notin $;
             
# Characteristic sets
var2 Xbook, Xcitation;

# Partition
((all1 x : x in Xbook =>x notin Xcitation)
&(all1 x : x in Xcitation =>x notin Xbook));
\end{boxedverbatim}
\caption{Sample XML Tree in MONA WS2S Syntax.}\label{containment:fig:sample-tree-mona}
\end{figure}

\section{Interpretation of XPath Queries}
\label{containment:xpath-ws2s}



This section explains how an XPath expression can be translated into an equivalent WS2S formula. This logical interpretation basically consists in considering a query as a relation that connects two tree nodes: a context node from which the query is applied, and a result node (selected by the query).

\subsection{Navigation and Recursion}
\label{containment:modeling-recursion-ws2s}
As a first step toward a WS2S encoding of XPath expressions, the navigational primitives over binary trees must be expressed.
Considering binary trees involves recursion for modeling the usual child relation on unranked trees (c.f. Figure~\ref{foundations:fig:bijection} and the isomorphism between binary and unranked trees detailed in Section~\ref{foundations:unranked-and-binary-trees}). Recursion is not available as a basic construct of WS2S. Recursion can be defined via a transitive closure formulated using second-order quantification.

%
%
%
%
%
%

The following-sibling relation is first expressed in WS2S. Consider a second-order variable $F$ as the set of nodes of interest. The following-sibling relation is defined as an induction scheme. The base case just captures that the immediate right successor of $x$ is effectively its first following sibling:
$$(x.1 \in F)$$
Then the induction step states that the immediate right successor of every position in $F$ is also among the following siblings, and formulates this as a transitive closure:
$$\forall z. (z \in F \ourimplies z.1 \in F)$$
The global requirement for a node $y$ to be one of the following siblings of $x$ is now formulated. The node $y$ must belong to the set $F$ which is closed under the following-sibling relation starting from $x.1$:
$$(x.1 \in F \et \forall z. z \in F \ourimplies z.1 \in F) \ourimplies y \in F$$
Note that this formula is satisfied for multiple sets $F$. For instance, the set of all tree nodes satisfies this implication. Actually, only the smallest set $F$ for which the formula holds is of interest: the set which contains all and only all following siblings. A way to express this is to introduce a universal quantification over $F$. Indeed, ranging over all such set of nodes notably takes into account the particular case where $F$ is minimal, i.e. the set of interest. If the global formula holds for every $F$, $y$ is also in the minimal set that contains only the following siblings of $x$. Therefore, the XPath ``following-sibling'' axis is defined as the WS2S predicate:
\begin{multline*}
\text{followingsibling}(X,x,y) \equalsdef \forall F . F \subseteq X \ourimplies \\ ((x.1 \in F \et \forall z. z \in F \ourimplies z.1 \in F) \ourimplies y \in F)
\end{multline*}
that expresses the requirements for a node $y$ to be a following sibling of a node $x$ in the tree $X$.
XPath ``descendant'' axis can be modeled in the same manner. The set $D$ of interest is initialized with the left child of the context node, and is closed under both successor relations:
\begin{multline*}
\text{descendant}(X, x, y) \equalsdef \forall D . D \subseteq X \ourimplies \\ (x.0 \in D \et \forall z . (z \in D \ourimplies z.1 \in D \et z.0 \in D) \ourimplies y \in D)
\end{multline*}
Considering these two relations as navigational primitives, more complex ones can be built out of them:
\begin{align*} 
\axis{child}( X,  x,  y) & \equalsdef  y=x.0 \; \ou \; \text{followingsibling}(X,x.0,y) \\
\axis{following}( X,  x,  y) & \equalsdef  \exists z. z \in X \et z.1 \in X \et \axis{ancestor}(X,x,z) \\
                             &  \quad \et  \axis{descendant}(X,z.1,y) \\
\axis{self}( X,  x,  y) & \equalsdef  x=y \\
\axis{descendantorself}(X, x, y) & \equalsdef \axis{self}(X,x,y)  \; \ou \; \axis{descendant}(X,x,y)
\end{align*}
Eventually, the other XPath axes are defined as syntactic sugars by taking advantage of XPath symmetry:
\begin{align*}
\axis{ancestor}( X,  x,  y) & \equalsdef \axis{descendant}(X,y,x) \\
\axis{parent}( X,  x,  y) & \equalsdef  \axis{child}(X,y,x) \\
\axis{precedingsibling}( X,  x,  y) & \equalsdef \axis{followingsibling}(X, y,x) \\
\axis{ancestororself}( X,  x,  y) & \equalsdef  \axis{descendantorself}(X,y,x) \\
\axis{preceding}( X,  x,  y) & \equalsdef  \axis{following}(X,y,x)
\end{align*}

\subsection{Logical Composition of Steps} \label{containment:logical-path-composition-ws2s}
$\hiddennotation{\logicE{\cdot}{\cdot}{\cdot}}{WS2S Interpretation of XPath}{n:logicE}$
This section describes how path composition operators are translated into logical connectives. The translation is formally specified as a ``derivor'' shown on Figure~\ref{containment:fig:logical-translation} and written $\logicE{e}{x}{y}$ where:
\begin{itemize}
\item the parameter $e$ (surrounded by special ``syntax'' braces $\llbracket \rrbracket$) is the source language parameter that is rewritten; 
\item the additional parameters $x$ and $y$ are respectively the context and the result node of the query.
\end{itemize}

\begin{figure}
\centering
\begin{align*}
\logicE{\cdot}{\cdot}{\cdot} &: \dom{Expression} \rightarrow \dom{Node} \rightarrow \dom{Node} \rightarrow \lwss \\
\logicE{/p}{x}{y}                                &   \eqdef   \exists z . \fun{isroot}{z} \wedge \logicP{p}{z}{y} \\ 
\logicE{p}{x}{y}                                &    \eqdef  \logicP{p}{x}{y} \\ 
\logicE{e_1 \shortmid e_2}{x}{y}                                &  \eqdef \logicE{e_1}{x}{y} \vee \logicE{e_2}{x}{y}  \\
\logicE{e_1 \cap e_2}{x}{y}                                &  \eqdef \logicE{e_1}{x}{y} \wedge \logicE{e_2}{x}{y} \\ \\
\logicPFunc &: \dom{Path} \rightarrow \dom{Node} \rightarrow \dom{Node} \rightarrow \lwss \\
\logicP{p_1/p_2}{x}{y}                                &  \eqdef \exists z . \logicP{p_1}{x}{z} \wedge \logicP{p_2}{z}{y} \\
\logicP{\qualif{p}{q}}{x}{y}                                   &  \eqdef \logicP{p}{x}{y} \wedge \logicQ{q}{y}  \\
\logicP{\step{$a$}{\sigma}}{x}{y}                         &  \eqdef a(x,y) \wedge y \in X_\sigma \\
\logicP{\step{$a$}{*}}{x}{y}                         &  \eqdef  a(x,y) \\ \\
\logicQFunc &: \dom{Qualifier} \rightarrow \dom{Node} \rightarrow \lwss \\
\logicQ{q_1 \op{and} q_2}{x}    &  \eqdef  \logicQ{q_1}{x} \wedge \logicQ{q_2}{x} \\
\logicQ{q_1 \op{or} q_2}{x}     &  \eqdef \logicQ{q_1}{x} \vee \logicQ{q_2}{x} \\
\logicQ{\op{not}~q}{x}     &  \eqdef \neg \; \logicQ{q}{x} \\
\logicQ{p}{x}                   &  \eqdef \exists y . \logicP{p}{x}{y} \\
\end{align*}
\caption{Translating XPath into WS2S.}
\label{containment:fig:logical-translation}
\end{figure}

The compilation of an XPath expression to WS2S relies on $\logicPFunc$ in charge of translating paths into formulas, and the dual derivor $\logicQFunc$ for translating qualifiers into formulas. The basic principle is that $\logicP{p}{x}{y}$ holds for all pairs $x,y$ of nodes such that $y$ is accessed from $x$ through the path $p$. Similarly, $\logicQ{q}{x}$ holds for all nodes $x$ such that the qualifier $q$ is satisfied from the context node $x$.

The interpretation of path composition $\logicP{p_1/p_2}{x}{y}$ consists in checking the existence of an intermediate node that connects the two paths, and therefore requires a new fresh variable to be inserted. The same holds for $\logicE{/p}{x}{y}$ that restarts from the root to interpret $p$, whatever the current context node $x$ is. 

Paths can occur inside qualifiers therefore $\logicEFunc$, $\logicPFunc$ and $\logicQFunc$ are mutually recursive. Since the interpretations of paths and qualifiers are respectively dyadic and monadic formulas, the translation of a path inside a qualifier $\logicQ{p}{x}$ requires the insertion of a new fresh variable whose only purpose consists in testing the existence of the path.

Eventually, the translation of steps relies on the logical definition of axes: $a(x,y)$ denotes the WS2S predicate defining the XPath axis $a$, as described in Section~\ref{containment:modeling-recursion-ws2s}. For instance, Figure~\ref{containment:fig:query_in_mona} presents the WS2S translation of the XPath expression: 
\xpath{\step{child}{\name{book}}/\qualif{\step{descendant}{\name{citation}}}{\step{parent}{\name{section}}}}

\begin{figure}
\centering
\begin{boxedverbatim}
# Translated XPath expression: 
# child::book/descendant::citation[parent::section]
ws2s;
# Data Model
var2 $ where ~empty($) 
 & (all1 x : all1 y : ((y=x.1 | y=x.0) 
    & (y in $)) => x in $) 
 &  all1 r : (r in $ & ~(ex1 z : z in $ 
    & (r=z.1 | r=z.0))) 
               => r.1 notin $;

# Characteristic sets
var2 Xbook, Xcitation, Xsection;

# Partition
((all1 x: x in Xbook => x notin Xcitation 
                      & x notin Xsection)&
(all1 x: x in Xcitation => x notin Xbook 
                         & x notin Xsection)&
(all1 x: x in Xsection => x notin Xbook 
                          & x notin Xcitation));

# Query (parameters are context and result nodes)
pred xpath1 (var1 x, var1 y)= 
  ex1 x1 : child(x,x1) & x1 in Xbook 
  & descendant(x1,y) & y in Xcitation 
  & ex1 x2 : parent(y,x2) & x2 in Xsection;
\end{boxedverbatim}
\caption{WS2S Translation of a Sample XPath in MONA Syntax.} \label{containment:fig:query_in_mona}
\end{figure}









\subsection{Formulating XPath Containment}
\label{containment:xpath-containment-ws2s}


The XPath containment problem can now be expressed in terms of a logical formula. Given two XPath expressions $e_1$ and $e_2$, the WS2S formula corresponding to checking their containment is built in two steps.
First, each XPath expression is translated into a WS2S logical relation that connects two nodes in the tree, as presented in Section~\ref{containment:logical-path-composition-ws2s}. 
Then the data model is unified. Each translation yields a set of characteristic sets. The union of them is built, so that characteristic sets that correspond to symbols used in both expressions are identified. 

From a logical point of view, $e_1 \subseteq e_2$ means that each pair of nodes $(x,y)$ such that $x$ and $y$ are connected by the logical relation corresponding to $e_1$ is similarly connected by the logical relation obtained from $e_2$:
\begin{equation}
 \forall x.  \; \forall y.  \; \logicE{e_1}{x}{y}  \ourimplies  \logicE{e_2}{x}{y}
\label{containment:containment-pb-logic-ws2s}
\end{equation}
The containment relation holds between expressions $e_1$ and $e_2$ if and only if the WS2S formula (\ref{containment:containment-pb-logic-ws2s}) is satisfied for all trees.
With respect to the notations of Section~\ref{containment:xml-tree-ws2s}, the containment between expressions $e_1$ and $e_2$ is thus formulated as:
\begin{multline*}
\forall X .\; \text{XMLTree}(X, X_1, ..., X_n) \ourimplies (\forall x \in X. \; \forall y \in X. \; \logicE{e_1}{x}{y}  \ourimplies  \logicE{e_2}{x}{y})
\end{multline*}
where the $X_i$ are members of the union of all characteristic sets detected for each expression. 
Consider for instance the two XPath expressions:
\begin{align*}
e_1 & \eqdef \step{child}{\name{book}}/\qualif{\step{descendant}{\name{citation}}}{\step{parent}{\name{section}}} \\
e_2 & \eqdef \qualif{\step{descendant}{\name{citation}}}{\step{ancestor}{\name{book}} \op{and} \step{ancestor}{\name{section}}}
\end{align*}
Figure~\ref{containment:fig:containment-in-mona} presents the generated WS2S formula for checking containment between $e_1$ and $e_2$, in MONA syntax.
The formula is determined valid (which means $e_1 \subseteq e_2$) in less than 0.2 seconds, the time spent to build the corresponding automaton and analyze it.
The formula for the reciprocal containment check between $e_2$ and $e_1$ is satisfiable, which means $e_2 \not \subseteq e_1$. 
The total running time of the decision procedure is less than 0.9 seconds, including the generation of the counter-example, shown below:
\begin{verbatim}
<book>
  <section>
    <other>
      <citation/>
  	</other>
  </section>
</book>
\end{verbatim}

\begin{figure}
\centering
\begin{boxedverbatim}
ws2s;
# Checking XPath Containment between 
#'child::book/descendant::citation[parent::section]'
# and 'descendant::citation[ancestor::book 
#                            and ancestor::section]'

# Data Model
var2 $ where ~empty($) 
 & (all1 x : all1 y : ((y=x.1 | y=x.0) 
    & (y in $)) => x in $) 
 & all1 r : (r in $ & ~(ex1 z : z in $ 
    & (r=z.1 | r=z.0))) 
              => r.1 notin $;

# Characteristic sets
var2 Xbook, Xcitation, Xsection;

# Queries (parameters are context and result nodes)
pred xpath1 (var1 x, var1 y)= 
   ex1 x1 : child(x,x1) & x1 in Xbook 
   & descendant(x1,y) & y in Xcitation 
   & ex1 x2 : parent(y,x2) & x2 in Xsection;
pred xpath2 (var1 x, var1 y)= 
   descendant(x,y) & y in Xcitation 
   & ex1 x1 : (ancestor(y,x1) & x1 in Xbook) 
   & ex1 x2 :  (ancestor(y,x2) & x2 in Xsection);

# Problem formulation 
((all1 x: x in Xbook => x notin Xcitation
                      & x notin Xsection)&
(all1 x: x in Xcitation => x notin Xbook 
                         & x notin Xsection)&
(all1 x: x in Xsection => x notin Xbook 
                          & x notin Xcitation))
=>
(all1 x: all1 y: (xpath1(x,y)=> xpath2(x,y)));
\end{boxedverbatim}
\caption{Sample WS2S Formula for XPath Containment in MONA Syntax.} \label{containment:fig:containment-in-mona}
\end{figure}







\subsection{Soundness and Completeness}



Soundness and completeness of the decision procedure for XPath Containment are ensured by construction. Indeed, consider the initial definition of the containment problem: provided a XML tree, checking containment between two XPath $e_1$ and $e_2$ consists in determining if the following proposition holds:
\begin{equation}
 \forall x,  \semanE{e_1}{x}{}  \subseteq  \semanE{e_2}{x}{}
\label{containment:containment-pb-ws2s}
\end{equation}
By definition, (\ref{containment:containment-pb-ws2s}) is logically equivalent to: 
\begin{equation}
 \forall x,  \forall y,  y \in \semanE{e_1}{x}{}  \ourimplies y \in \semanE{e_2}{x}{}
\label{containment:containment-pb2-ws2s}
\end{equation}
Then the last step remaining to prove is the equivalence between~(\ref{containment:containment-pb2-ws2s}) and (\ref{containment:containment-pb-logic-ws2s}).
To this end, the compilation of XPath expressions into WS2S formulas must preserve XPath denotational semantics, which means:
\begin{thm}
\label{containment:sem-equiv-ws2s}
The logical translation of XPath expressions is equivalent to XPath denotational semantics:
\begin{equation}
\logicP{e}{x}{y} \equiv  y \in \semanP{e}{x}{}   
\label{containment:sem-eq-p-ws2s}
\end{equation}
\end{thm}



\paragraph{Proof (Sketch)}
The proof uses an induction over the structure of paths. Since the definition of paths and qualifiers is cross-recursive, a mutual induction scheme is used. The scheme relies on the dual property for qualifiers that also needs to be proved:
\begin{equation}
\forall p, \forall x, (\semanQ{q}{x} \equiv \logicQ{q}{x})
\label{containment:sem-eq-q-ws2s}
\end{equation}
Specifically (\ref{containment:sem-eq-p-ws2s}) is proved by taking (\ref{containment:sem-eq-q-ws2s}) as assumption, and reciprocally (\ref{containment:sem-eq-q-ws2s}) is proved under (\ref{containment:sem-eq-p-ws2s}) as assumption.  
Both equivalences (\ref{containment:sem-eq-p-ws2s}) and (\ref{containment:sem-eq-q-ws2s}) are proved inductively for each compositional layer. The idea basically consists in associating corresponding logical connectives to each set-theoretic composition operator used in the denotational semantics. 
XPath qualifier constructs trivially correspond to logical WS2S connectives. Path constructs involves set-theoretic union and intersection operations which are respectively mapped to logical disjunction and conjunction. Two path constructs: $p_1/p_2$ and $p[q]$ require specific attention in the sense their denotational semantics introduce particular compositions over sets of nodes. They are recalled below:
\begin{align*}
\semanP{p_1/p_2}{x}{} &  \eqdef \{ x_2 \; | \; x_1 \in \semanP{p_1}{x}{} \wedge x_2 \in \semanP{p_2}{x_1}{} \} \\
\semanP{p[q]}{x}{} &  \eqdef \{ x_1 \; | \; x_1 \in \semanP{p}{x}{} \wedge \semanQ{q}{x_1}{} \}
\end{align*}
Auxiliary lemmas are introduced in order to clarify how these constructs are mapped to WS2S.
The XPath construct $p_1/p_2$ is generalized as a function $\fun{product}{}$, whereas the XPath construct $p[q]$ is generalized as $\fun{filter}{}$:
\begin{align*}
\fun{product}{} & : \text{Set(\dom{Node)}} \rightarrow (\dom{Node} \rightarrow \text{Set(\dom{Node})}) \rightarrow \text{Set(\dom{Node})} \\
\fun{filter}{}  & : \text{Set(\dom{Node)}} \rightarrow (\dom{Node} \rightarrow \dom{Boolean}) \rightarrow \text{Set(\dom{Node})}
\end{align*}
$\fun{product}{}$ is characterized by the lemmas (\ref{containment:product1-ws2s}) and (\ref{containment:product2-ws2s}), in which $y$ and $z$ are nodes, and $S$ is a set of nodes. These lemmas abstract over XPath navigational functionalities performed by axes by letting $f$ denoting a function that returns a set of nodes provided a current node:
\begin{equation}
\forall y, \forall z, \forall S, \forall f: \dom{Node} \rightarrow \text{Set(\dom{Node})},
z \in S \ourimplies  y \in (f z) \ourimplies y \in \fun{product}{S, f} \label{containment:product1-ws2s}
\end{equation}
\begin{equation}
\forall y, \forall S, \forall f: \dom{Node} \rightarrow \text{Set(\dom{Node})},
y \in \fun{product}{S,f} \ourimplies \exists z, z \in S  \et y \in (f z).
\label{containment:product2-ws2s}
\end{equation}
The function $\fun{filter}{}$ is in turn characterized by the following lemma:  
\begin{equation}
\forall y, \forall g: \dom{Node} \rightarrow \dom{Boolean},
 y \in \fun{filter}{S, g} \ourimplies y \in S
 \label{containment:filter-ws2s}
 \end{equation}
The auxiliary lemmas (\ref{containment:product1-ws2s}), (\ref{containment:product2-ws2s}), and (\ref{containment:filter-ws2s}) are also proved by induction.
Developing the proof in constructive logic involves the (trivial) decidability of set-theoretic inclusion and of the denotational semantics of qualifiers.
The full formal proof is detailed in \cite{geneves-tphols04}. It has been mechanically checked by the machine using the Coq formal proof management system \cite{CoqTutorialV8}.

%
%
 

\section{Complexity Analysis and Optimization}
\label{containment:complexity-ws2s}

The translation of an XPath query to its logical representation is linear in the size of the
input query. Indeed, each expression is decomposed then translated inductively in one pass without
any duplication, as shown by the formal definition of $\logicEFunc$ in Section~\ref{containment:logical-path-composition-ws2s}. 

The second step is the decision procedure, which, compared to the translation, represents the major part of the cost. The truth status of a WS2S formula is decided throughout the logic-automaton connection as described in Sections~\ref{foundations:deciding-ws2s} and~\ref{foundations:from-ws2s-to-automata} of previous Chapter~\ref{foundations}. This translation from logical formulas to tree automata, while effective, is unfortunately non-elementary. This bound may sound discouraging. Fortunately, the worst-case scenario which corresponds to complex formulas, is not likely to occur for small instances of the containment in practice. 
Furthermore, recent works on MSO solvers - especially those using BDDs techniques \cite{bryant86} such as MONA \cite{mona-user-manual} - suggest that in particular practical cases the explosiveness of this technique can be effectively controlled. 

In practice, the implementation relies on MONA \cite{mona-user-manual} that implements the WS2S decision procedure along with various optimizations. Additionally, a significant optimization that takes advantage of XPath peculiarities for combating automaton size explosion is described in the following subsection. 

\subsection{Optimization Based on Guided Tree Automata}

A major source of complexity arises from the translation of composed paths. Each translation of the form $\logicP{p_1/p_2}{x}{y}$ introduces an existentially quantified first-order variable which ranges over all possible tree positions (c.f. Figure~\ref{containment:fig:e1_translation}). 

The idea in this section is to take advantage of XPath navigational peculiarities for attempting to reduce the scope associated to such variables. 
XPath navigates the tree step by step: each step selects a set of nodes which is in turn used to select a new one by the next step. The interpretation of a variable inserted during the translation of $p_1/p_2$ corresponds to the intermediate node which is a result of $p_1$ and the context node of $p_2$. 
The truth status of the formula is determined by the existence of such an intermediate node at a particular position in the tree. 
If one can determine regions in the tree in which such a node may appear from those where it cannot appear, valuable positional knowledge is gained that can be used to reduce the variable scope. It is interesting to try to identify the region in the tree (or even some larger approximation) in which the node must be located in order for the formula to be satisfied.
XPath sequential structure of steps makes it possible to exploit such positional knowledge. Indeed, consider for instance
the expression: \xpath{e_3\eqdef/\step{child}{\name{book}}/\qualif{\step{descendant}{\name{*}}}{\step{child}{\name{citation}}}} 
$e_3$ navigates from the document root through its ``book'' children elements and then selects all descendant nodes provided they have at least one child named ``citation''.
Several conditions must be satisfied by a tree $t_1$ in order to yield a result for $e_3$: 
\begin{itemize}
\item $t_1$ must have at least one ``book'' element as a child of the root; 
\item $t_1$ must have at least one element that must be a descendant of the ``book'' element; 
\item for this node to be selected it must have at least one child named ``citation''. 
\end{itemize}
\begin{figure}
\centering
\begin{boxedverbatim}
e1(x,y) = ex1 x1 : isroot(x1) & x1 in $ 
 & ex1 x2 : child(x1,x2) &  x2 in Xbook 
 & descendant(x2,y) & y in $ 
 & ex1 x3 : child(y,x3) & x3 in Xcitation;
\end{boxedverbatim} 
\caption{WS2S Translation of $e_3$ in MONA Syntax.}\label{containment:fig:e1_translation}
\end{figure}
This is made explicit by the logical translation $\logicE{e_3}{x}{y}$ in MONA syntax shown on Figure~\ref{containment:fig:e1_translation}.
In this translation, \texttt{x1}, \texttt{x2}  and \texttt{x3} denote the respective positions of the root node, a ``book'' child, and a ``citation'' child of the selected position \texttt{y}. These variables actually only range over a particular set of positions in the tree. By definition, the root can only appear at depth level $0$, the ``book'' element can only occur at level $1$ and its descendants occur at any depth level $l$ greater or equals to $2$. Eventually, the  ``citation'' element should occur at level $l+1$. This is because each step introduces its particular positional constraint which can be propagated to the next steps.

\begin{figure}
\centering
\includegraphics[width=10cm, keepaspectratio=true]{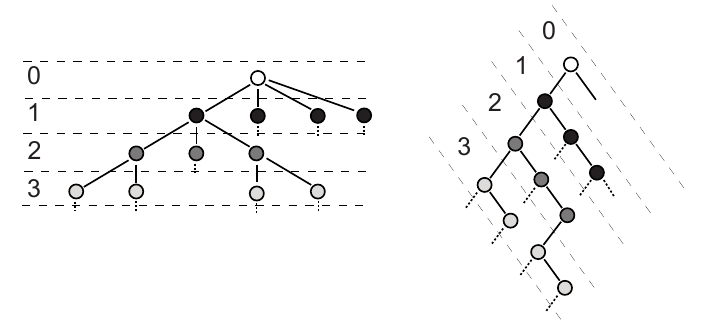}
\caption{Depth Levels in the Unranked and Binary Cases.}\label{containment:fig:depth}
\end{figure}

The idea of taking advantage of positional knowledge is even more general.
Theoretically, normal bottom-up FTA are sufficient for deciding validity of a WS2S formula (as presented in Section~\ref{foundations:deciding-ws2s} of Chapter~\ref{foundations}). 
However composition of such automata is particularly sensitive to state space explosion, as presented in Section~\ref{foundations:ws2s-complexity}.
Guided tree automata (GTA) \cite{mona-gta-algos} have been introduced in order to combat such state space explosion by following the divide and conquer approach. A GTA is just an ordinary FTA equipped with an additional deterministic top-down tree automaton called the guide. The latter is introduced to take advantage of positional knowledge, and used for partitioning the FTA state space into independent subspaces. Top-down deterministic automata are strictly less powerful than ordinary (bottom-up or non-deterministic top-down) FTA \cite{tata}. However, this is not a problem since the guide is only intended to provide additional auxiliary information used for optimization purposes. As a consequence, the more precise is the guide, the more efficient is the decision procedure, but an approximation is sufficient.  
The guide basically splits the state space of the FTA in independent subsets. Therefore the transition relation of the bottom-up automaton is split into a family of transition functions, one for each state space name. A state space name corresponds to a particular depth level or a set of depth levels.
GTA can be composed in the same way than ordinary FTA as explained in Section~\ref{foundations:deciding-ws2s} of Chapter~\ref{foundations}. A GTA can be seen as an ordinary tree automaton, where the state space has been factorized according to the guide. A GTA with only one state space is just an ordinary tree automaton. A detailed description of GTA can be found in \cite{mona-gta-algos}. GTA-based optimization may lead to exponential improvements of the decision procedure \cite{BRICS-EP-00-SME-CTDC}. 


A tree partitioning based on the depth levels is now introduced. It is depicted by Figure~\ref{containment:fig:depth} for a $n$-ary sample tree and its binary counterpart. Based on this partitioning, a positional constraint (a restricted set of depth levels) is associated to each node variable. Indeed, a node referred by an XPath can occur at several depth levels since some axes involve transitive closure (c.f. Section~\ref{foundations:xpath-denotational-semantics} of Chapter~\ref{foundations}). Moreover, the set of depth levels can even be infinite since XPath offers recursion in unbounded trees.

\begin{figure*}
\centering
\begin{align*}
\levelEFunc &: \lxpath \rightarrow \text{Set(\dom{Int)}} \rightarrow \text{Set(\dom{Int)}} \\
\levelE{/p}{N}                                & \eqdef  \levelP{p}{\{0\}} \\
\levelE{p}{N}                                &  \eqdef  \levelP{p}{\N} \\
\levelE{e_1 \shortmid e_2}{N}                                &  \eqdef \levelE{e_1}{N} \cup \levelE{e_2}{N}  \\
\levelE{e_1 \cap e_2}{N}                                &  \eqdef \levelE{e_1}{N} \cap \levelE{e_2}{N} \\ \\
\levelPFunc &: \dom{Path} \rightarrow \text{Set(\dom{Int)}} \rightarrow \text{Set(\dom{Int)}} \\
\levelP{p_1/p_2}{N}                                &  \eqdef \levelP{p_2}{\levelP{p_1}{N}} \\
\levelP{\qualif{p}{q}}{N}                                   &  \eqdef \levelP{p}{N}  \\ 
\levelP{\step{self}{n}}{N}   & \eqdef N \\
\levelP{\step{child}{n}}{N}   &  \eqdef \{n+1 \; | \; n \in N \} \\
\levelP{\step{parent}{n}}{N}  &  \eqdef  \{n-1 \; | \; n \in N \} \\
\levelP{\step{descendant}{n}}{N}&  \eqdef  \{n' \; | \; n \in N \et n'>n \}   \\
\levelP{\step{descendant-or-self}{n}}{N}&  \eqdef  \{n' \; | \; n \in N \et n'>=n \}   \\
\levelP{\step{ancestor}{n}}{N}  & \eqdef  \{n' \; | \; n \in N \et n'>=0 \et n'<n \}\\
\levelP{\step{ancestor-or-self}{n}}{N}  &  \eqdef  \{n' \; | \; n \in N \et n'>=0 \et n'<=n \}\\
\levelP{\step{following}{n}}{N}  &  \eqdef \N - \{0\}\\
\levelP{\step{preceding}{n}}{N} 	&  \eqdef \N - \{0\}\\
\levelP{\step{following-sibling}{n}}{N}  &  \eqdef N\\
\levelP{\step{preceding-sibling}{n}}{N} 	&  \eqdef N
\end{align*}
\caption{Computation of the Depth Levels of Nodes Selected by a Path.}\label{containment:fig:level-computation}
\end{figure*}
$\hiddennotation{\levelE{\cdot}{\cdot}}{Calculation of a set of depth levels}{n:levelEfunc}$
The computation of sets of depth levels is calculated by the function shown on Figure~\ref{containment:fig:level-computation}, and written $\levelE{e}{N}$ where $e$ is the XPath expression to be analyzed and  $N$ is the set of positional constraints corresponding to the context node from which $e$ is applied.
Again, the algorithm proceeds inductively on the structure of XPath expressions. XPath steps are base cases for which the set of levels is effectively calculated from the previous one. Transitive closure axes such as ``$\axis{descendant}$'' turn the set of depth levels into an infinite one, even if the previous was finite. Path composition basically propagates the level calculations by combining with the base cases. 
Note that an important precision can be gained with absolute XPath expressions. In this case, the initial set of depth levels is the singleton $\{0\}$ as opposed to relative XPath expressions for which the context node is not known and the initial set of depth levels is subsequently $\N$.

\begin{figure*}
\centering
\begin{align*}
\optlogicEFunc &: \lxpath \rightarrow \dom{Node} \rightarrow \dom{Node} \rightarrow \text{Set(\dom{Int)}} \rightarrow \lwss \\
\optlogicE{/p}{x}{y}{N}                                & \eqdef  \exists \restrict{z}{\{0\}}.\fun{isroot}{z} \wedge \optlogicP{p}{z}{y}{\{0\}} \\ 
\optlogicE{p}{x}{y}{N}                                & \eqdef  \optlogicP{p}{x}{y}{\N} \\ 
\optlogicE{e_1 \shortmid e_2}{x}{y}{N}                                & \eqdef  \optlogicE{e_1}{x}{y}{N} \vee \optlogicE{e_2}{x}{y}{N}  \\
\optlogicE{e_1 \cap e_2}{x}{y}{N}                                & \eqdef  \optlogicE{e_1}{x}{y}{N} \wedge \optlogicE{e_2}{x}{y}{N} \\ \\ 
\optlogicPFunc &: \dom{Path} \rightarrow \dom{Node} \rightarrow \dom{Node} \rightarrow \text{Set(\dom{Int)}} \rightarrow \lwss \\
\optlogicP{p_1/p_2}{x}{y}{N}                                & \eqdef  \exists \restrict{z}{\levelP{p_1}{N}}. \optlogicP{p_1}{x}{z}{N} \wedge \optlogicP{p_2}{z}{y}{N} \\
\optlogicP{\qualif{p}{q}}{x}{y}{N}                                   & \eqdef  \optlogicP{p}{x}{y}{N} \wedge \optlogicQ{q}{y}{N}  \\
\optlogicP{\step{$a$}{\sigma}}{x}{y}{N}                         & \eqdef   a(x,y) \wedge y \in X_\sigma \\
\optlogicP{\step{$a$}{*}}{x}{y}{N}                        & \eqdef  a(x,y) \\  \\
\optlogicQFunc &: \dom{Qualifier} \rightarrow \dom{Node} \rightarrow \text{Set(\dom{Int)}} \rightarrow \lwss \\
\optlogicQ{q_1 \op{and} q_2}{x}{N}    & \eqdef \optlogicQ{q_1}{x}{N} \wedge \optlogicQ{q_2}{x}{N} \\
\optlogicQ{q_1 \op{or} q_2}{x}{N}     & \eqdef \optlogicQ{q_1}{x}{N} \vee \optlogicQ{q_2}{x}{N} \\
\optlogicQ{ \op{not}~q}{x}{N}     & \eqdef  \neg \; \optlogicQ{q}{x}{N} \\
\optlogicQ{p}{x}{N}                   & \eqdef \exists \restrict{y}{\levelP{p}{N}}. \optlogicP{p}{x}{y}{N}
\end{align*}
\caption{Translating XPath into WS2S with Restricted Variable Scopes.}\label{containment:fig:opt_translation}
\end{figure*}
$\hiddennotation{\optlogicEFunc\llbracket\cdot\rrbracket}{Optimized WS2S Interpretation of XPath}{n:optLogicE}$
The optimized compilation of XPath expressions to WS2S formulas is given on Figure~\ref{containment:fig:opt_translation}. $\optlogicEFunc$, $\optlogicPFunc$ and $\optlogicQFunc$ are respective optimized versions of $\logicEFunc$, $\logicPFunc$ and $\logicQFunc$, which convey a set of depth levels as an additional parameter passed to $\levelEFunc$ and $\levelPFunc$. These functions compute the restrictions on variable scope that are inserted by $\optlogicPFunc$ and $\optlogicQFunc$. ``$\exists \restrict{z}{D}$'' denotes the fact that the existentially quantified first-order variable $z$ is restricted to appear at a depth level among the set of depth levels $D$.
In practice, $\levelEFunc$ and $\levelPFunc$ can be merged into $\optlogicEFunc$ and can be implemented in a single pass over the XPath expression. Thus the translation and the depth level computation remain linear in the size of the query.

MONA provides an implementation of GTA. The application of the previous algorithm to $e_3$ leads to the logical formulation shown on Figure~\ref{containment:fig:optimized_e1_translation} in MONA syntax.

\begin{figure}
\centering
\begin{boxedverbatim}
guide l0 -> (l1, epsilon),
      l1 -> (l2, l1),
      l2 -> (l3, l2),
      l3 -> (lothers, l3),
      lothers -> (lothers, lothers),
      epsilon -> (epsilon, epsilon);
		  
e1(x,y)= ex1 [l0] x1 : (isroot(x) & x=x1 & x in $) 
& ex1 [l1] x2 : child(x1,x2) & x2 in Xbook
& descendant(x2,y) & y in $ 
& ex1 [l3, lothers] x3 : child(y,x3) 
& x3 in Xcitation;
\end{boxedverbatim}
\caption{Optimized WS2S Translation of $e_3$ in MONA Syntax.} \label{containment:fig:optimized_e1_translation}
\end{figure}

The guide obtained in this translation means that the root is labeled with ``l0''; its left and right successor nodes are labeled with ``l1'' and ``epsilon'' respectively. The ``epsilon'' is a dummy state space reflecting the fact that the underlying shape is a tree and not a hedge. No variable is associated with this state space. The ``lothers'' state space represents any tree node occurring at a depth level greater than $3$. Such a state space is associated with variables whose scope is of unbounded depth. The size of the guide 
depends on the maximum depth level found among the computed restrictions. Formally, a guide for a maximum depth level $n$ is a top-down deterministic tree automaton with $  \{q_0, ..., q_{n+1} \} \cup \{ q_\epsilon \} $ as set of states, $q_0$ as the single initial state, and the following set of transitions:
$$ \begin{array}{ll} 
& \{ q_0 \rightarrow (q_1, q_\epsilon) \} \\
\cup  &\{ q_i \rightarrow (q_{i+1}, q_i)  \; | \;  i \in [1... n]  \}  \\
\cup & \{  q_{n+1} \rightarrow (q_{n+1}, q_{n+1})\} \\
\cup & \{ q_\epsilon \rightarrow (q_\epsilon, q_\epsilon) \} 
\end{array}$$ 
where $q_i$ ($i \in [0... n]$) denotes the state space name corresponding to the depth level $i$, and $q_{n+1}$ represents all depth levels greater or equal to $n+1$ . For formulating the XPath containment, the guide is computed from the two XPath expressions. Specifically, the deepest (and thus the most precise) guide is chosen as the guide for both expressions. 

Eventually, each variable is restricted with a list of state spaces that represents the regions in the tree where its valuation must be searched.
For instance, ``\texttt{ex1 [l1] x2}'' means the scope of the variable \texttt{x2} is limited to tree nodes occurring at depth level $1$.

This optimization is useful for both kinds of XPath expressions: absolute and relative. More precise restrictions can be computed for absolute XPath expressions (for which the initial set of depth levels is the singleton $\{0\}$).





%

\section{Implementation and Experiments}
\label{containment:implementation-and-experiments}

The approach has been implemented. A compiler (written in Java) takes XPath expressions and translates them into WS2S formulas. A Java interface controls the C++ implementation of the MONA WS2S solver, and in addition provides precise runtime statistics on the decision procedure. 

The evolution of the intermediate automata (in terms of states, number of BDD nodes involved, the minimizations, products, projections...) are reported in realtime during a run of the decision procedure. For example, Figure~\ref{containment:fig:stats} shows detailed statistics on the intermediate automata built during the comparison of the following two XPath expressions $e_4$ and $e_5$:
\begin{align*}
e_4 &\eqdef \name{a}/\name{b}[\step{descendant}{\name{c}}]/\step{following-sibling}{\name{d}}/\name{e} \\
e_5 &\eqdef \name{a}/\name{d}[\step{preceding-sibling}{\name{b}}]/\name{e}
\end{align*}
The horizontal axes of charts of Figure~\ref{containment:fig:stats} correspond to the number of automata operations. In that case, $380$ operations were needed to complete the XPath containment test. Once the decision procedure terminates, the result of the comparison is displayed in the console:
\begin{verbatim}
"a/b[descendant::c]/following-sibling::d/e" is contained in 
"a/d[preceding-sibling::b]/e" [Total Time: 00:00:00.18]
\end{verbatim}

\begin{figure}
\centering
\includegraphics[width=120mm]{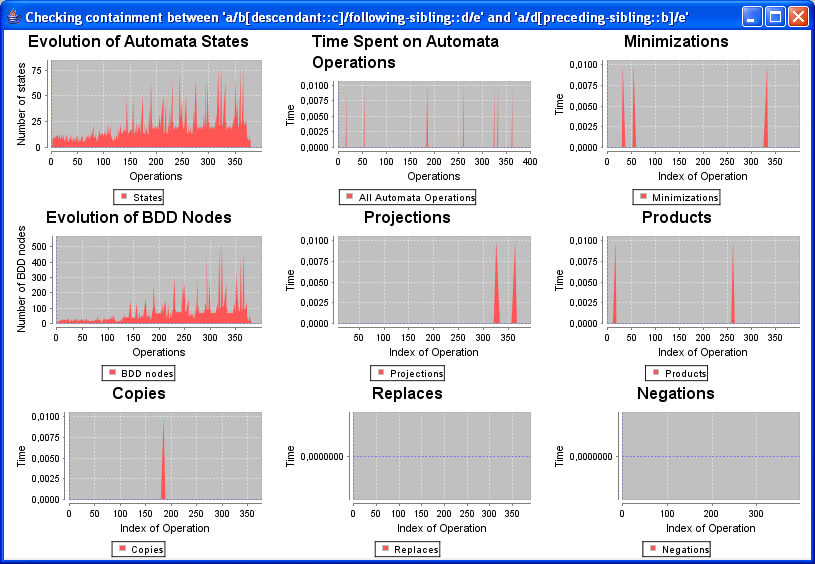}
\caption{Statistics on Intermediate Automata for a Containment Check.}\label{containment:fig:stats}
\end{figure}

Extensive tests have been carried out with the implementation. Tests have been reported in \cite{geneves-RR01-2006}. They are not detailed here, since it is difficult to come up with a clear conclusion based on the observed practical behavior of this decision procedure on a few instances. Instead, only the major lessons learned from the practical experiments are summarized:

\begin{itemize}
\item The GTA-based optimization has been observed to be particularly useful as guides cause a small overhead compared to the significant performance gains they provide on many instances. Some containment instances cannot be solved without this optimization.

\item For small expressions (that are most likely to occur in practice in XSLT transformations, as suggested by \cite{moller-rr05}), it has been observed over many instances that the implementation can run in acceptable time and space bounds. Since this approach is sound and complete over a large XPath fragment, it provides an interesting alternative to the less complex but incomplete decision procedure over a very restricted XPath fragment previously studied in the literature \cite{suciu-miklau-jacm04}.

\item For larger XPath expressions however, intermediate tree automata constructed can be so large that blow-ups are observed, even using GTA. Practical experiments notably suggest that the WS2S decision procedure implemented in MONA is particularly sensitive to the alphabet size, which clearly makes the approach inappropriate for XPath expressions that use a large number of tag names.

\item The explosiveness of the approach is very difficult to control in practice. It is possible to find relatively small expressions for which blow-ups cannot be controlled, even by the GTA-based optimization. Subsequently, there exist relatively small XPath containment instances for which containment cannot be decided in acceptable time and space bounds.

\item As a result, no clear conclusion can be drawn from the experiments, concerning the maximum size and complexity of XPath expressions for which this procedure could offer practical guarantees. Such a characterization is made very difficult by the huge number of parameters that must be taken into account, due to all the optimizations implemented in MONA \cite{mona-impl-secrets}. It is thus very difficult to estimate up to which XPath expression size and complexity this decision procedure can be used in practice. Observed results on tested instances suggest that this approach may be efficient for XPath expressions using less than $10$ tag names, and indicate that it cannot be reasonably used with larger alphabets.
\end{itemize}


\section{Outcome}
\label{containment:outcome-ws2s}

An approach based on MSO has been proposed for the XPath containment problem: query containment is formulated in terms of a WS2S formula, which is then decided using tree automata. An optimization method based on guided tree automata is proposed in an attempt to take advantage of XPath peculiarities in order to improve time and space requirements of the complex decision procedure.

An advantage of the approach is that it provides a sound and complete decision procedure for a large XPath fragment. Another advantage of this technique is to allow generation of tree examples and counter-examples of the truth status of the formula. 

The major drawback of this approach, however, is that the decision procedure is based on the full construction and complementation of the intermediate automata. This makes the explosiveness of the approach very hard to control in practice and unfortunately restricts its use to only small XPath expressions.

Surprisingly enough, the full construction and determinization of intermediate FTA often seems unnecessary. Indeed huge intermediate automata are almost always reduced by following projection operations. This can been observed on most practical scenarios owing to the detailed statistics reported by the implementation (see for instance the peaks in the evolution of intermediate automata states on Figure~\ref{containment:fig:stats}). The determinization of huge intermediate automata is the source of uncontrollable blow-ups in practice. On many instances, it has been observed that the memory representation of intermediate automata may require several hundreds of megabytes (or even several gigabytes which is not affordable on most current machines), even if this appears to be unnecessary since the final resulting automaton is only of several kilobytes in size.

One direction of future work is to search for tree automata guides that produce a finer-grained partition of the automaton state space, in order to enhance the scalability of the decision procedure. Another perspective is to search for approaches that do not construct unnecessary parts of intermediate automata, or even do not construct automata at all. This is the motivation that underlies investigations presented in the next chapter.

 \ifglobalcompil
  \else 
   \bibliographystyle{apalike}
   \bibliography{references}
   \end{document}
  \fi

\newif\ifglobalcompil\globalcompiltrue

  \ifglobalcompil
  \else 
    \input{Preamble} 
    \input{Markup}                       
    \begin{document} 
  \fi
  
\mychapter{XML and the Modal $\mu$-Calculus}
\label{analysis}

\section{Introduction}
Investigations presented in this chapter are motivated by a search for automata theoretic approaches that avoid explicit construction of tree automata. In this direction, this chapter attempts to build efficient decision procedures for XML problems by using the alternation-free modal $\mu$-calculus. This logic is as expressive as WS2S, less succinct, but has a lower complexity (exponential time).

This chapter shows how XPath can be linearly translated into the $\mu$-calculus. In addition, regular tree types (including DTDs) are also linearly embedded in the $\mu$-calculus. XPath decision problems (containment, emptiness, equivalence, overlap, coverage) in the presence or absence of XML types are expressed as formulas in this logic. A state of the art decision procedure for $\mu$-calculus satisfiability is used to solve the generated formula and to construct relevant example and/or counter-example XML trees. The system has been fully implemented. 

\paragraph{Chapter Outline}
The chapter is organized as follows: in Section~\ref{analysis:afmc-intro} the $\mu$-calculus is introduced; Section~\ref{analysis:kripke-xml-trees} explains how general graph models of this logic can be restricted so that they represent XML trees. The translation of XPath queries into this logic is described in Section~\ref{analysis:xpath-embedding}. Section~\ref{analysis:xml-types-embedding} embeds regular XML types into the logic. Based on these translations, Section~\ref{analysis:xml-decision-pb} explains how to formulate and solve the considered decision problems. A complexity analysis is presented in Section~\ref{analysis:complexity-analysis}, along with implementation principles of the system. Finally, the outcome of this approach is discussed in Section~\ref{analysis:outcome}.

\section{The $\mu$-Calculus}
\label{analysis:afmc-intro}

The \emph{propositional $\mu$-calculus} is a propositional modal logic extended with least and greatest fixpoint operators \cite{kozen83}.
A \emph{signature} $\notation{\signature}{Signature for the $\mu$-calculus}{n:sign}$ for the $\mu$-calculus consists of a set $\notation{\dom{Prop}}{Set of atomic propositions}{n:atomicprops}$ of atomic propositions, a set $\notation{\dom{Var}}{Set of propositional variables}{n:propvars}$ of propositional variables, and a set $\notation{\dom{FProg}}{Set of atomic programs}{n:atomicprogs}$ of atomic programs. In the XML context, atomic propositions represent the symbols of the alphabet $\Sigma$ used to label XML trees. Atomic programs allow navigation in trees.

The $\mu$-calculus with converse\footnote{The $\mu$-calculus with converse is also known as the \emph{full} $\mu$-calculus, or alternatively as the \emph{two-way} $\mu$-calculus in the literature.} \cite{vardi-icalp98} augments the propositional $\mu$-calculus by associating with each atomic program $a$ its converse $\notation{\overline{a}}{Converse of an atomic program}{n:conv}$ (such that $\overline{\overline{a}}=a$). A \emph{program} $\alpha$ is either an atomic program or its converse. $\notation{\dom{Prog}}{Set of programs}{n:programs}$ denotes the set $\dom{FProg} \cup \{\overline{a} \mid a \in \dom{FProg} \}$. This is the only difference with the propositional $\mu$-calculus that lacks converse programs. Equipping the logic with converse programs is useful for supporting query langages that allow both forward and backward navigation in trees (see Section~\ref{analysis:converse-useful}). Converse programs generally provide a mean to reason about the past, which also proved to be  useful in the context of program verification \cite{vardi-icalp98}. The interaction of converse programs with other constructs of the logic is known to be quite subtle. In particular, in $\mu$-calculus it is known that converse programs interact with recursion in such a way that the finite model property is lost \cite{vardi-icalp98}. The decidability of the $\mu$-calculus extended with converse was proved to be in EXPTIME in \cite{vardi-icalp98}, by introducing a new class of alternating two-way automata on infinite trees.

The set $\fullmucalculus$ of formulas of the $\mu$-calculus with converse over the signature $\signature$ is defined as follows: \label{analysis:mu-syntax}

\smallsyntax{
\entry       \notation{\fullmucalculus}{Set of $\mu$-calculus formulas}{n:fullmu} \ni \phi, \psi     [formula]
             \true              [true]
\oris        p            [atomic proposition]
\oris        \neg \phi [negation]
\oris        \phi \et \psi [conjunction]
\oris        \umod{\alpha} \phi [universal modality]
\oris        X [variable]
\oris        \mu X . \phi [least fixpoint]
}
\noindent
where $p \in \dom{Prop}$, $X \in \dom{Var}$ and $\alpha$ is a program. Note that $X$ should not occur negatively in $\mu X.\phi$. The following abbreviations are defined:
\begin{align*}
\false & \eqdef \neg \true \\
\phi_1~\ou~\phi_2 & \eqdef \neg (\neg \phi_1~\et~ \neg \phi_2) \\
\emod{\alpha} \phi & \eqdef \neg \umod{\alpha} \neg \phi \\
\nu X . \phi & \eqdef \neg \mu X. \neg \phi\{\subst{\neg X}{X}\}
\end{align*}

$\emod{\alpha} \phi$ is called the existential modality and $\nu X . \phi$ the greatest fixpoint.
The semantics of the full $\mu$-calculus is given with respect to a \emph{Kripke structure} $K = \kripke{W, R, L}$ where $W$ is a set of nodes, $R : \dom{Prog} \rightarrow \powerset{W \times W}$ assigns to each atomic program a transition relation over $W$, and $L$ is an interpretation function that assigns to each atomic proposition a set of nodes.
$\hiddennotation{\musem{\phi}{\cdot}{\cdot}}{Interpretation of $\mu$-calculus formula $\phi$}{n:musem}$
The formal semantics function $\musem{\phi}{K}{V}$ shown on Figure~\ref{analysis:fig:mu-sem} defines the semantics of a $\mu$-calculus formula $\phi$ in terms of a Kripke structure $K$ and a valuation $V$. A valuation $V : \dom{Var} \rightarrow \powerset{W} $ maps each variable to a subset of $W$. For a valuation $V$, a variable $X$, and a set of nodes $W' \subseteq W$, $V[X/W']$ denotes the valuation that is obtained from $V$ by assigning $W'$ to $X$.

\begin{figure}[h]
\centering
\begin{align*}
\musem{\any}{K}{V} &: \fullmucalculus \longrightarrow \powerset{W} \\
\musem{\true}{K}{V} & \eqdef W \\
\musem{\false}{K}{V} & \eqdef \emptyset \\
\musem{p}{K}{V} & \eqdef L(p) \\
\musem{\neg \phi}{K}{V} & \eqdef W \setminus \musem{\phi}{K}{V}  \\
\musem{\phi_1 \ou \phi_2}{K}{V} & \eqdef \musem{\phi_1}{K}{V} \cup \musem{\phi_2}{K}{V} \\
\musem{\phi_1 \et \phi_2}{K}{V} & \eqdef \musem{\phi_1}{K}{V} \cap \musem{\phi_2}{K}{V}\\
\musem{\umod{\alpha} \phi}{K}{V} & \eqdef \{ w  : \forall w'  (w,w') \in R(\alpha) \ourimplies w' \in \musem{\phi}{K}{V} \}\\
\musem{\emod{\alpha} \phi}{K}{V} & \eqdef  \{ w  : \exists w'  (w,w') \in R(\alpha) \et w' \in \musem{\phi}{K}{V} \} \\
\musem{\mu X . \phi}{K}{V} & \eqdef \bigcap \{ W' \subseteq W : \musem{\phi}{K}{V[X/W']} \subseteq W' \} \\
\musem{\nu X . \phi}{K}{V} & \eqdef \bigcup \{ W' \subseteq W : \musem{\phi}{K}{V[X/W']} \supseteq W' \} \\
\musem{X}{K}{V} & \eqdef V(X)
\end{align*}
\caption{Semantics of the $\mu$-Calculus.} \label{analysis:fig:mu-sem}
\end{figure}


Note that if $\phi$ is a sentence (i.e. all propositional variables occurring in $\phi$ are bound), then no valuation is required. For a node $w \in W$ and a sentence $\phi$, $K,w \models \phi $ iff $w \in \musem{\phi}{K}{}$ denotes that $\phi$ holds at $w$ in $K$.

The two modalities $\left\langle a \right\rangle \phi$ (possibility) and $\left[ a \right] \phi$ (necessity) are operators for navigating the structure.


\label{analysis:nnf}
In order to avoid redundancy, only a subset of $\fullmucalculus$ composed of formulas in negation normal form is of interest. A formula is in \emph{negation normal form} if and only if all negations in the formula appear only before atomic propositions. Every formula is equivalent to a formula in negation normal form \cite{kozen83}, which can be obtained by expanding negations using De Morgan's rules together with standard dualities for modalities and fixpoints (cf. Figure~\ref{analysis:fig:positive-normal-form-rules}). For readability purposes, however, translations of XPath expressions given in Section~\ref{analysis:xpath-embedding} are not given in negation normal form.

 \begin{figure}[h]
 \centering
 \begin{align*}
 \neg \umod{\alpha} \phi &= \emod{\alpha} \neg \phi \\
 \neg \emod{\alpha} \phi &=  \umod{\alpha} \neg \phi \\
 \neg \mu X.\phi &=  \nu X. \neg \phi \{\subst{X}{\neg X}\} \\
 \neg \nu X.\phi &= \mu X. \neg \phi \{\subst{X}{\neg X}\} \\
 \neg (\phi_1 \et \phi_2) &=  \neg \phi_1 \ou \neg \phi_2\\
 \neg (\phi_1 \ou \phi_2) &= \neg \phi_1 \et \neg \phi_2\\
 \neg \neg \phi &=  \phi
 \end{align*}
 \caption{Dualities for Negation Normal Form.} \label{analysis:fig:positive-normal-form-rules}
 \end{figure}

For reasoning on XML trees, only a specific subset of $\fullmucalculus$, namely the alternation-free modal $\mu$-calculus with converse over finite binary trees is of interest. 

A $\fullmucalculus$ formula $\phi$ in negation normal form is \emph{alternation-free} whenever the following condition holds\footnote{For instance, $\nu X. (\mu Y. \someFCverifies Y \et p) \ou \someNSverifies X$ is alternation-free but $\nu X. (\mu Y. \someFCverifies Y \et X) \ou p$ is not since $X$ bound by $\nu$ appears freely in the scope of $\mu Y$.}: if $\mu X . \phi_1$ (respectively $\nu X.\phi_1$) is a subformula of $\phi$ and $\nu Y . \phi_2$ (respectively $\mu Y.\phi_2$) is a subformula of $\phi_1$ then $X$ does not occur freely in $\phi_2$.

The following section now introduces the additional restrictions of $\fullmucalculus$ related to finite binary trees.

\section{Kripke Structures and XML Trees}
\label{analysis:kripke-xml-trees}
In this section, the satisfiability problem of $\fullmucalculus$ over Kripke structures is restricted to the satisfiability problem over finite binary trees.

The propositional $\mu$-calculus has the \emph{finite tree model property}: a formula that is satisfiable, is also satisfiable on a finite tree \cite{kozen88}.
Unfortunately, the introduction of converse programs causes the loss of the finite model property \cite{vardi-icalp98}. Therefore, the finite model property must be reinforced along with some other properties to ensure finite binary models that encode XML structures.  

First, each XML node has at most one $\Sigma$-label, i.e. $p \et p'$ never holds for distinct atomic propositions $p$ and $p'$. This can be easily incorporated in a $\mu$-calculus satisfiability solver.

Second, for navigating binary trees, only two atomic programs $\fc$ and $\ns$ are used, together their associated relations $R(\fc) = \fcrel$ and $R(\ns) = \nsrel$ whose meaning is to respectively connect a node to its left child and to its right child. 
For any $(x,y) \in W \times W$,  $x \fcrel y$ holds iff $y$ is the left child of $x$ (i.e. the first child in the unranked tree representation) and $x \nsrel y$ holds iff $y$ is the right child of $x$ in the binary tree representation (i.e. the next sibling in the unranked tree representation).


For each atomic program $a \in \{ \fc, \ns\}$,  $R(\overline{a})$ is defined to be the relational inverse of $R(a)$, i.e., $R(\overline{a}) = \{ (v,u) : (u,v) \in R(a)\}$. 
Thus programs $\alpha \in \{ \fc, \ns, \invfc, \invns \}$ are considered inside modalities for navigating downward and upward in trees.

Restrictions for a Kripke structure to form a finite binary tree are now defined.
A Kripke structure $T=\kripke{W, R, L}$ is a finite binary tree if it satisfies the following conditions:
\begin{enumerate}
\item[(1)] $W$ is finite
\item[(2)] the set of nodes $W$ together with the accessibility relation $\fcrel \cup \nsrel$ define a tree
\item[(3)] $\fcrel$ and $\nsrel$ are partial functions, i.e. for all $m \in W$ and $j \in \{\fc,\ns \}$ there is at most one $m_j \in W$ such that $(m, m_j) \in R(j)$.
\end{enumerate}

A finite binary tree $T=\kripke{W, R, L}$ satisfies $\phi$ if $T, r \models \phi$ where $r \in W$ is the root of the tree $T$.

The previous restrictions are now expressed in $\fullmucalculus$. For accessing the root, the $\fullmucalculus$ formula $$\notation{\phiroot}{$\mu$-formula satisfied at a root}{n:phiroot} = \allinvFCverify \false \et \allinvNSverify \false \et \neg \someNSverifies \true$$ is used. Its meaning is to select a node provided it has no parent and no sibling.

The property for ensuring finiteness relies on K\"onig's lemma which states that \emph{a finitely branching infinite tree has some infinite path} or, in other words, a finitely branching tree in which every branch is finite is finite.
The expression $\nu X . \someFCverifies X \ou \someNSverifies X$ is only satisfied by structures containing infinite or cyclic paths.
To prevent the existence of such paths, the previous formula is negated and, propagating negation using the rules presented on Figure~\ref{analysis:fig:positive-normal-form-rules}, yields the following formula: $$\phift = \mu X. \allFCverify X \et \allNSverify X$$
$\notation{\phift}{$\mu$-formula ensuring structure finiteness}{n:phift}$ states that all descending branches are finite from the current context node ($\phift$ is vacuously satisfied at the leaves). $\phift$ must hold at the root (i.e. $\phiroot \et \phift$ must hold), in order to ensure structure finiteness. This is for condition (1) to be satisfied. 

Properties (2) and (3) still need to be enforced. This is done by rewriting existential modalities in such a way that if a successor is supposed to exist, then there exists at least one, and if there are many all verify the same property. This is a way to overcome the difficulty that in $\mu$-calculus, one cannot naturally express a property like ``a node has exactly $n$ successors''. 
Technically, $\phi^{\text{FBT}}$ denotes the formula $\phi$ where all occurrences of $\emod{\alpha} \psi$ are replaced by $\emod{\alpha} \true \et \umod{\alpha} \psi^{\text{FBT}}$. 
Furthermore, a node cannot be both a left child and a right child: the formula $(\neg \someinvFCverifies \true \ou \neg \someinvNSverifies \true)$ must be satisfied at each node.

%
\begin{thm}[\cite{tozawa-tableaux05}]
\label{analysis:fbtm-prop}A $\fullmucalculus$ formula $\phi$ is satisfied by a finite binary tree model if and only if the formula $\phiroot \et \mu X. (\neg \someinvFCverifies \true \ou \neg \someinvNSverifies \true) \et \allFCverify X \et \allNSverify X \et \phi^{\text{FBT}}$ is satisfied by a Kripke structure.\end{thm}
The proof of the ``if'' part iteratively constructs a tree model and proceeds by induction on the structure on $\phi$. The ``only if'' part is almost immediate. Theorem~\ref{analysis:fbtm-prop} gives the adequate framework for formulating decision problems on XML structures in terms of a $\mu$-calculus formula.

\section{XPath Embedding}
\label{analysis:xpath-embedding}
This section explains how an XPath expression can be translated into an equivalent formula in $\fullmucalculus$. Navigation as performed by XPath in unranked trees is translated in terms of navigation in the binary tree representation (using the isomorphism  presented in Section~\ref{foundations:unranked-and-binary-trees}). The translation adheres to XPath formal semantics in the sense that the translated formula holds for nodes which are selected by the XPath query. 

\subsection{Logical Interpretation of Axes}

The formal translations of navigational primitives (namely XPath axes) are 
formally specified on Figure~\ref{analysis:fig:axes}. The translation function noted 
``$\mucalcA{\axisvar}{\chi}$'' takes an XPath axis \(\axisvar\) as input, and 
returns its $\fullmucalculus$ translation, in terms of a $\fullmucalculus$ formula 
$\chi$ given as a parameter. This parameter represents a context and allows to 
compose formulas, which is needed for translating path composition.  
$\mucalcA{\axisvar}{\chi}$ holds for all nodes that can be accessed through 
the axis $\axisvar$ from some node verifying $\chi$. 

For instance, the translated formula $\mucalcA{\axis{child}}{\chi}$ is satisfied by children of the context $\chi$. These nodes are composed of the first child and the remaining children. From the first child, the context must be reached immediately by going once upward via $\overline{1}$. From the remaining children, the context is reached by going upward (any number of times) via $\overline{2}$ and then finally once via $\overline{1}$.


\begin{figure}
\centering
\begin{align*}
\mucalcA{\cdot}{\cdot} &: \dom{Axis} \rightarrow \fullmucalculus \rightarrow \fullmucalculus \\
\mucalcA{\axis{self}}{\chi} & \eqdef       \chi            \\
\mucalcA{\axis{child}}{\chi} & \eqdef  \mu Z. \someinvFCverifies \chi \ou 
\someinvNSverifies Z  \\
\mucalcA{\axis{following-sibling}}{\chi} & \eqdef  \mu Z. \someinvNSverifies \chi \ou \someinvNSverifies Z \\ 
\mucalcA{\axis{preceding-sibling}}{\chi} & \eqdef   \mu Z. \someNSverifies \chi \ou 
\someNSverifies Z \\
\mucalcA{\axis{parent}}{\chi} & \eqdef  \someFCverifies \mu Z. \chi \ou 
\someNSverifies Z  \\
\mucalcA{\axis{descendant}}{\chi} & \eqdef  \mu Z. \someinvFCverifies (\chi  
\ou Z) \ou \someinvNSverifies Z   \\ 
\mucalcA{\axis{descendant-or-self}}{\chi} & \eqdef  \mu Z. \chi \ou \mu Y. \someinvFCverifies (Y \ou Z) \ou \someinvNSverifies Y\\
\mucalcA{\axis{ancestor}}{\chi} & \eqdef  \someFCverifies \mu Z. \chi \ou 
\someFCverifies Z \ou \someNSverifies Z  \\
\mucalcA{\axis{ancestor-or-self}}{\chi} & \eqdef  \mu Z. \chi \ou \someFCverifies \mu Y. Z \ou \someNSverifies Y\\ 
\mucalcA{\axis{following}}{\chi} & \eqdef  
\mucalcA{\axis{descendant-or-self}}{\eta_1(\chi)}\\
\mucalcA{\axis{preceding}}{\chi} & \eqdef \mucalcA{\axis{descendant-or-self}}{\eta_2(\chi)} \\
 \eta_1(\chi) & \eqdef \mucalcA{\axis{following-sibling}}{\mucalcA{\axis{ancestor-or-self}}{\chi}}\\
 \eta_2(\chi) & \eqdef \mucalcA{\axis{preceding-sibling}}{\mucalcA{\axis{ancestor-or-self}}{\chi}}
\end{align*}
\caption{Translation of XPath Axes.}\label{analysis:fig:axes}
\end{figure}

\subsection{Logical Interpretation of Expressions}

\begin{figure}
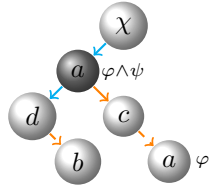

\centering
\begin{align*}
\mucalcE{\cdot}{\cdot} &: \lxpath \rightarrow \fullmucalculus \rightarrow \fullmucalculus \\
\mucalcE{/p}{\chi}                              & \eqdef \mucalcP{p}{\left(\phirootexpr  \et \mu Y. \chi \ou \someFCverifies Y \ou \someNSverifies Y\right)}  \\ 
\mucalcE{p}{\chi}                               & \eqdef  \mucalcP{p}{\left(\chi \right)} \\
\mucalcE{e_1 \shortmid e_2}{\chi}              & \eqdef    \mucalcE{e_1}{\chi}  \ou \mucalcE{e_2}{\chi} \\
\mucalcE{e_1 \cap e_2}{\chi}                   & \eqdef  \mucalcE{e_1}{\chi}  \et \mucalcE{e_2}{\chi}  \\ \\
\mucalcP{\cdot}{\cdot}   &: \dom{Path} \rightarrow \fullmucalculus \rightarrow \fullmucalculus \\
\mucalcP{p_1/p_2}{\chi}                                & \eqdef  \mucalcP{p_2}{\left(\mucalcP{p_1}{\chi}\right)} \\
\mucalcP{\qualif{p}{q}}{\chi}                          & \eqdef \mucalcP{p}{\chi} \et \mucalcQex{q}{\true} \\ 
\mucalcP{\step{\axisvar}{\nodelabel}}{\chi}                     & \eqdef  \nodelabel \et \mucalcA{\axisvar}{\chi} \\
\mucalcP{\step{\axisvar}{*}}{\chi}                         & \eqdef  \mucalcA{\axisvar}{\chi}
\end{align*}
\caption{Translation of Expressions and Paths.}\label{analysis:fig:exprs}
\end{figure}

Figure~\ref{analysis:fig:exprs} gives the translation of XPath expressions into 
$\fullmucalculus$. The translation function ``$\mucalcE{e}{\chi}$'' takes an XPath 
expression \(e\) and a $\fullmucalculus$ formula \(\chi\) (denoting a particular context) as input, and returns 
the corresponding $\fullmucalculus$ translation. The translation of relative XPath expressions use the current context $\chi$. The translation of absolute 
expressions navigates from $\chi$ to the root which is taken as initial context for the expression.

For example, Figure~\ref{analysis:fig:bin} illustrates the translation of the XPath 
expression ``$\step{child}{a}[\step{child}{b}]$''. This expression selects all 
``$a$'' child nodes of a given context which have at least one ``$b$'' child.  
The translated $\fullmucalculus$ formula holds for ``$a$'' nodes which are 
selected by the expression. The first part of the translated formula, 
\(\phi\), corresponds to the step ``$\step{\axis{child}}{a}$'' which selects 
candidates ``$a$'' nodes.  The second part, \(\psi\), navigates downward in 
the subtrees of these candidate nodes to verify that they have at least one 
``$b$'' child.

\begin{figure}[h]
\centering
\begin{tikzpicture}[scale=0.6]

\draw (0,-0.5) node(example) {Translated Query:};
\draw (3.5,-0.44) node(xpathpart1) [text=purple]{$\step{child}{a}$};
\draw (5.6,-0.5) node(xpathpart2) [text=violet]{$[\step{child}{b}]$};

\draw (1.2,-2) node(xpathtrans1) [text=purple]{$\underbrace{a \et (\mu Z . \emod{\overline{1}}\chi \ou \emod{\overline{2}}Z)}_{\phi}$};
\draw (4.6,-1.6) node(xpathtrans2) {$\et$};
\draw (6.9,-2) node(xpathtrans3) [text=violet]{$\underbrace{\emod{1}\mu Y .b \ou \emod{2}Y}_{\psi}$};

\draw (2,4) node(x) [ball color=white,circle,text=black] {$\chi$};

\draw (3,1) node(c3) [ball color=white,circle,text=black]{$a$};
\draw (3.75,1) node(c3label) {$_{\phi}$};

\draw (2,2) node(c2) [ball color=white,circle,text=black]{$c$};
\draw [thick, dashed, ->, orange] (c2) -- (c3);

\draw (1,3) node(c1) [ball color=gray,circle,text=black]{$a$};
\draw [thick, ->, orange] (c1) -- (c2);

\draw [thick, ->, cyan] (x) -- (c1);

\draw (0,2) node(b1) [ball color=white,circle,text=black]{$d$};
\draw [thick, ->, cyan] (c1) -- (b1);

\draw (1,1) node(b2) [ball color=white,circle,text=black]{$b$};
\draw [thick, dashed, ->, orange] (b1) -- (b2);

\draw (2,3) node(c1selectedlabel) {$_{\phi \et \psi}$};
\end{tikzpicture}
 \caption{XPath Translation Example.}\label{analysis:fig:bin}
\end{figure}

\label{analysis:converse-useful}
Note that without converse programs it would have been impossible to differentiate 
selected nodes from nodes whose existence is tested, since properties must be stated 
on both the ancestors and the descendants of the selected node. Equipping the 
$\fullmucalculus$ logic with both forward and converse programs is therefore 
crucial for supporting XPath\footnote{One may ask whether it is possible to 
eliminate upward navigation at the XPath level but it is well known that such 
XPath rewriting techniques cause exponential blow-ups of expression sizes 
\cite{symmetry}.}. Logics without converse programs may only be used for 
solving XPath emptiness but cannot be used for solving other decision problems 
such as containment efficiently.

XPath most essential construct $p_1/p_2$ translates into formula composition 
in $\fullmucalculus$, such that the resulting formula holds for all nodes accessed 
through $p_2$ from those nodes accessed from $\chi$ by $p_1$. The translation 
of the branching construct $p[q]$ significantly differs. The resulting formula 
must hold for all nodes that can be accessed through $p$ and from which $q$ 
holds. To preserve semantics, the translation of $p[q]$ stops the ``selecting 
navigation'' to those nodes reached by $p$, then filters them depending on
whether $q$ holds or not. This is expressed by introducing a dual formal 
translation function for XPath qualifiers, noted $\mucalcQex{q}{\chi}$ and 
defined in Figure~\ref{analysis:fig:qualifs}, that performs ``filtering'' instead of 
navigation.  Specifically, $\mucalcP{\cdot}{\cdot}$ can be seen as the 
``navigational'' translating function: the translated formula holds for target 
nodes of the given path.
On the opposite, $\mucalcQex{\cdot}{\cdot}$ can be seen as the ``filtering'' 
translating function: it states the existence of a path \emph{without moving 
to its end}. The translated formula $\mucalcQex{q}{\chi}$ (respectively 
$\mucalcPex{p}{\chi}$) holds for nodes from which there exists a qualifier $q$ 
(respectively a path $p$) leading to a node verifying $\chi$.

\begin{figure}
\centering
\begin{align*}
\mucalcQex{\cdot}{\cdot} &: \dom{Qualif} \rightarrow \fullmucalculus \rightarrow \fullmucalculus \\
\mucalcQex{q_1 \op{and} q_2}{\chi}    & \eqdef   \mucalcQex{q_1}{\chi} \et \mucalcQex{q_2}{\chi}  \\
\mucalcQex{q_1 \op{or} q_2}{\chi}     & \eqdef   \mucalcQex{q_1}{\chi} \ou \mucalcQex{q_2}{\chi} \\
\mucalcQex{\op{not}~q}{\chi}          & \eqdef  \neg~\mucalcQex{q}{\chi} \\
\mucalcQex{p}{\chi}                   & \eqdef   \mucalcPex{p}{\chi} \\ \\
\mucalcPex{\cdot}{\cdot} &: \dom{Path} \rightarrow \fullmucalculus \rightarrow \fullmucalculus \\
\mucalcPex{p_1/p_2}{\chi}                               & \eqdef  \mucalcPex{p_1}{\left(\mucalcPex{p_2}{\chi}\right)}  \\
\mucalcPex{\qualif{p}{q}}{\chi}                         & \eqdef  \mucalcPex{p}{\left(\chi \et \mucalcQex{q}{\true}\right)} \\
\mucalcPex{\step{\axisvar}{\nodelabel}}{\chi}                    & \eqdef  \mucalcAex{\axisvar}{\left(\chi \et \nodelabel\right)} \\
\mucalcPex{\step{\axisvar}{*}}{\chi}                         & \eqdef  \mucalcAex{\axisvar}{\chi} \\ \\
\mucalcAex{\cdot}{\cdot} &: \dom{Axis} \rightarrow \fullmucalculus \rightarrow \fullmucalculus \\
\mucalcAex{\axisvar}{\chi} & \eqdef  \mucalcA{\fun{symmetric}{\axisvar}}{\chi}
\end{align*}
\caption{Translation of Qualifiers.}\label{analysis:fig:qualifs}
\end{figure}

XPath translation is based on these two translating ``modes'', the first one 
being used for paths and the second one for qualifiers. Whenever the 
``filtering'' mode is entered, it will never be left. 

Translations of paths inside qualifiers are also given on 
Figure~\ref{analysis:fig:qualifs}. They use the specific translations for axes inside 
qualifiers, based on XPath symmetry: $\fun{symmetric}{\axisvar}$ denotes the 
symmetric XPath axis corresponding to the axis $\axisvar$ (for instance 
$\fun{symmetric}{\axis{child}}=\axis{parent}$).

\subsection{Correctness and Complexity}



The translation of XPath in $\fullmucalculus$ can be proven correct with respect to XPath denotational semantics.
First, a Wadler-like semantics of XPath expressions is defined with respect to Kripke structures that are XML trees. Let $\kripkexmltrees$ be the set of Kripke structures that are finite binary trees (as defined in Section~\ref{analysis:kripke-xml-trees}) and $\kripkexmltreenodes=\setof{w \in W}{\kripke{W, R, L} \in \kripkexmltrees}$ the set of nodes of such structures. Given a finite binary tree $T=\kripke{W, R, L} \in \kripkexmltrees$ and some node $x \in W$ of $T$, the functions $\teinterp{\cdot}{T}{x}$, $\tpinterp{\cdot}{T}{x}$, $\tqinterp{\cdot}{T}{x}$ and $\tainterp{$\cdot$}{T}{x}$ respectively define the semantics of XPath expressions, paths, qualifiers, and axes:

\begin{align*}
\teinterp{\cdot}{\cdot}{\cdot} & : \lxpath \rightarrow \kripkexmltrees \rightarrow \kripkexmltreenodes \rightarrow \powerset{\kripkexmltreenodes}\\
 \teinterp{/p}{T}{x} & \eqdef  \tpinterp{p}{T}{\kripkexmltreeroot{T}} \\
 \teinterp{p}{T}{x} & \eqdef  \tpinterp{p}{T}{x}\\
 \teinterp{e_1 \shortmid e_2}{T}{x} & \eqdef \teinterp{e_1}{T}{x} \cup \teinterp{e_2}{T}{x}\\
 \teinterp{e_1 \cap e_2}{T}{x} &  \eqdef \teinterp{e_1}{T}{x} \cap \teinterp{e_2}{T}{x}
\end{align*}

\begin{align*}\tpinterp{\cdot}{\cdot}{\cdot} & : \dom{Path} \rightarrow \kripkexmltrees \rightarrow \kripkexmltreenodes \rightarrow \powerset{\kripkexmltreenodes}\\
 \tpinterp{p_1/p_2}{T}{x} &  \eqdef \setof{z \in \tpinterp{p_2}{T}{y}}{y \in \tpinterp{p_1}{T}{x}} \\
 \tpinterp{\qualif{p}{q}}{T}{x} &  \eqdef \setof{y \in \tpinterp{p}{T}{x}}{\tqinterp{q}{T}{y}}\\ 
 \tpinterp{\step{\axisvar}{\nodelabel}}{\kripke{W,R,L}}{x} &  \eqdef  \setof{y \in \tainterp{\axisvar}{\kripke{W,R,L}}{x}}{y \in L(\nodelabel)}\\ 
 \tpinterp{\step{\axisvar}{*}}{T}{x} &  \eqdef \left\{y \in \tainterp{\axisvar}{T}{x}\right\}
\end{align*}

\begin{align*}\tqinterp{\cdot}{\cdot}{\cdot} & : \dom{Qualif} \rightarrow \kripkexmltrees \rightarrow \kripkexmltreenodes \rightarrow \{\text{true}, \text{false} \}\\
 \tqinterp{q_1 \op{and} q_2}{T}{x} &  \eqdef \tqinterp{q_1}{T}{x} \ou \tqinterp{q_2}{T}{x}\\
 \tqinterp{q_1 \op{or} q_2}{T}{x} &  \eqdef \tqinterp{q_1}{T}{x} \et \tqinterp{q_2}{T}{x}\\
 \tqinterp{\op{not}~q}{T}{x} &  \eqdef \neg \tqinterp{q}{T}{x} \\
 \tqinterp{p}{T}{x} &  \eqdef \tpinterp{p}{T}{x} \neq \emptyset     
\end{align*}

\begin{align*}  \tainterp{$\cdot$}{\cdot}{\cdot} & : \dom{Axis} \rightarrow \kripkexmltrees \rightarrow \kripkexmltreenodes \rightarrow \powerset{\kripkexmltreenodes}\\
 \tainterp{\axis{self}}{T}{x} &  \eqdef \{x\} \\
 \tainterp{\axis{child}}{\kripke{W,R,L}}{x} &  \eqdef \setof{y \in W}{x \fcrel y} \cup \setof{z \in W}{x \fcrel y \et y \nsrelplus z}\\
 \tainterp{\axis{following-sibling}}{\kripke{W,R,L}}{x} & \eqdef \setof{z \in W}{x \nsrelplus z}\\
 \tainterp{\axis{preceding-sibling}}{\kripke{W,R,L}}{x} &  \eqdef \setof{z \in W}{z \nsrelplus x}\\ 
 \tainterp{\axis{parent}}{\kripke{W,R,L}}{x} &  \eqdef \setof{p \in W}{p \fcrel x} \cup \setof{p \in W}{p \fcrel y \et y \nsrelplus x} \\
 \tainterp{\axis{descendant}}{T}{x} &  \eqdef \tainterp{\axis{child}}{T}{x} \\& \quad ~\cup \setof{z \in \tainterp{\axis{descendant}}{T}{y}}{y \in \tainterp{\axis{child}}{T}{x}}\\
 \tainterp{\axis{descendant-or-self}}{T}{x} &  \eqdef \tainterp{\axis{descendant}}{T}{x} \cup \tainterp{\axis{self}}{T}{x}\\
 \tainterp{\axis{ancestor}}{T}{x} &  \eqdef  \tainterp{\axis{parent}}{T}{x} \\ & \quad ~\cup \setof{z \in \tainterp{\axis{ancestor}}{T}{y}}{y \in \tainterp{\axis{parent}}{T}{x}} \\
 \tainterp{\axis{ancestor-or-self}}{T}{x} &   \eqdef \tainterp{\axis{ancestor}}{T}{x} \cup \tainterp{\axis{self}}{T}{x}\\
  \end{align*}

\begin{align*}
 \tainterp{\axis{following}}{T}{x} &  \eqdef \setof{z \in \tainterp{\axis{descendant-or-self}}{T}{y}}{y \in \xaf{T}{x}}\\
 \tainterp{\axis{preceding}}{T}{x} &  \eqdef \setof{z \in \tainterp{\axis{descendant-or-self}}{T}{y}}{y \in \xap{T}{x}} \\ 
 \xaf{T}{x} & \eqdef \setof{ y \in \tainterp{\axis{following-sibling}}{T}{w}}{w \in \xaa{T}{x}}\\
 \xap{T}{x} & \eqdef \setof{ y \in \tainterp{\axis{preceding-sibling}}{T}{w}}{w \in \xaa{T}{x}} \\
 \xaa{T}{x} & \eqdef \{ w \in \tainterp{\axis{ancestor-or-self}}{T}{x}\}
\end{align*}

The auxiliary function $\kripkexmltreeroot{T}$ returns the root of $T$, and the relation symbol $\nsrelplus$ used in the semantics of axes denotes the transitive closure of the relation $\nsrel$ defined in Section~\ref{analysis:kripke-xml-trees}.

The correctness of the translation of XPath into $\fullmucalculus$ can now be stated:

\begin{thm}[Translation Correctness]
\label{analysis:thm-correctness}

For any finite binary tree $T \in \kripkexmltrees$, nodes $x$ and $y$ of $T$, property $\chi \in \fullmucalculus$, expression $e \in \lxpath$, and path $p \in \dom{\emph{Path}}$, the following equivalences hold:

\begin{align}
(\forall \chi \in \fullmucalculus ~~ T,x \models \chi ~~\ourimplies~~ T,y \models \mucalcE{e}{\chi}) \quad & \Longleftrightarrow \quad y \in \teinterp{e}{T}{x} \\
T,y \models \mucalcE{e}{\chi} \quad  &\Longleftrightarrow \quad y \in \bigcup_{\setof{x}{T, x \models \chi }} \teinterp{e}{T}{x}  \\
(\forall \chi \in \fullmucalculus ~~ T,x \models \chi ~~\ourimplies~~ T,y \models \mucalcP{p}{\chi}) \quad & \Longleftrightarrow \quad y \in \tpinterp{p}{T}{x}   \\
(\forall \chi \in \fullmucalculus ~~ T,y \models \chi ~~\ourimplies~~ T,x \models \mucalcPex{p}{\chi}) \quad & \Longleftrightarrow \quad y \in \tpinterp{p}{T}{x}
\end{align}

\end{thm} 

\begin{mypfo}
Each equivalence is proved by a straightforward structural induction that ``peels off'' the compositional layers of each set of rules.
\end{mypfo}
This result links XPath decision problems to satisfiability in $\fullmucalculus$.  
Note that the size of a translated formula $\mucalcE{e}{\chi}$ is linear in the length of the XPath expression $e$ since there is no duplication of subformulas of arbitrary length in the formal translations\footnote{Formulas in which the formal parameter $\chi$ appears twice (see Figure~\ref{analysis:fig:exprs} and Figure~\ref{analysis:fig:qualifs}) do not cause such duplication since at this stage $\chi$ carries a constant. Section~\ref{analysis:gamma-dot} explains how $\chi$ is initialized with a constant at the expression level.}.

%

\section{Translation of Regular Tree Languages}
\label{analysis:xml-types-embedding}
The translation of regular tree types into $\mu$-calculus is now introduced. It is based on the binary representation of types introduced in Chapter~\ref{foundations}.
In order to simplify translations, a notation for a $n$-ary least fixpoint binder is introduced: 
$$\afletmu{X}{\phi} \psi$$
This notation is actually a syntactic sugar for $\psi$ where all free occurrences of $X_i$ have been replaced by $\mu X_i. \phi_i$ until $\psi$ becomes closed (that is all $X_i$ in $\psi$ are in scope of their corresponding unary $\mu$-binder).
This provides a shorthand for denoting a $\fullmucalculus$ formula which would be of exponential size if expressed using only the unary least fixpoint construct. Such a naive expansion contains unnecessary duplicate formulas whereas the satisfiability solver operates only on a single copy of them (see Section~\ref{analysis:complexity-analysis}). 
Therefore, the $n$-ary binder is a useful compact notation for representing $\fullmucalculus$ translations of recursive types, without introducing useless blow-ups between representation of formulas and their satisfiability test. 

The translation from binary regular tree types into $\fullmucalculus$ formulas is given by the following function $\notation{\Ttomu{\cdot}}{Translation of tree types into $\mu$-calculus}{n:rtttomu}$ :

\begin{align*}
\Ttomu{\cdot} &: \lbtt \rightarrow \fullmucalculus \\
\Ttomu{\emptyset} & \eqdef \false \\
\Ttomu{\epsilon} & \eqdef \false \\ 
\Ttomu{T_1 \tou T_2} & \eqdef \Ttomu{T_1} \ou \Ttomu{T_2} \\
\Ttomu{l(X_1, X_2)} & \eqdef \et \Tsucc{X_1}{1} \et \Tsucc{X_2}{2}\\
\Ttomu{\typebind{X}{T} T} & \eqdef \afletmuifree{X_i}{\Ttomu{T_i}} \Ttomu{T} \\
\end{align*}
where there is an implicit bijective correspondence between $\lbtt$ variables from $\dom{TVar}$ and $\fullmucalculus$ variables from $\dom{Var}$. Note that the translations of the empty tree type and the empty tree are the same since empty trees should not be explicitly mentioned in satisfiability results.  The function $\Tsucc{\cdot}{\cdot}$ sets the tree frontier accordingly:
\begin{align*}
\Tsucc{\cdot}{\cdot} &: \dom{Prog}\times  \dom{TVar} \rightarrow \fullmucalculus \\
\Tsucc{X}{\alpha} & \eqdef \left\{ \begin{array}{ll}\umod{\alpha}X & \text{\emph{ if }} \nullable{X} \\ \emod{\alpha}X & \text{\emph{ if not }} \nullable{X}\end{array}\right.\\
\end{align*}
The predicate $\nullable{\cdot}$ indicates if a type contains the empty tree:
\begin{align*}
\nullable{\cdot} & : \dom{TVar} \cup \lbtt \rightarrow \{\text{true},\text{false}\} \\
\nullable{X} & \eqdef \nullable{\theta(X)} \\
\nullable{\emptyset} & \eqdef \text{false} \\
\nullable{\epsilon} & \eqdef \text{true} \\
\nullable{l} & \eqdef \text{false} \\
\nullable{T_1 \tou T_2} & \eqdef \nullable{T_1} \ou \nullable{T_2} \\
\nullable{l(X_1, X_2)} & \eqdef \text{false} \\
\nullable{\typebind{X}{T} T} & \eqdef \nullable{T} \\
\end{align*}

%
%
%


\section{Solving XML Decision Problems}
\label{analysis:xml-decision-pb}
Both XPath over unranked trees, and regular unranked tree types have been translated in the unifying $\fullmucalculus$ logic over binary trees.
Owing to these translations, XML decision problems (such as XPath containment, equivalence, emptiness, overlap and coverage) in the presence or absence of XML types are now reduced to satisfiability in $\fullmucalculus$.

\paragraph{Correlating Context Nodes for Path Comparison}
\label{analysis:gamma-dot}
In order to correlate two different paths when performing any kind of mutual-relationship checking, a special atomic proposition $\startatom$ is introduced.
This atomic proposition marks the initial context node(s) from which an XPath expression is applied. $\startatom$ is used as initial value of the $\chi$ parameter of the translating function $\mucalcE{\cdot}{\chi}$. For an XPath expression $e \in \lxpath$, $\mucalcE{e}{\startatom}$ is thus a sentence, that is denoted by $\phi_e$ in the remaining. Owing to the introduction of $\startatom$, formulas may refer to the same context multiple times. This allows to compare different XPath expressions applied to the \emph{same} initial context that can be any node in any tree.



\paragraph{Formulating of XML Problems}
Some simplified notations are first introduced: $\setofdocs$ denotes the set of trees: by default, $\setofdocs=\setofunrankedtrees$, and whenever an optional DTD $d \in \ldtd$ is specified $\setofdocs$ = $\Tsem{d}{\emptyset}$.  Additionally, $\phitype$ denotes the $\fullmucalculus$ embedding of the tree language $\setofdocs$. In the absence of DTDs $\phitype=\true$, and $\phitype=\Ttomu{\tsbin{d}}$ in the presence of $d \in \ldtd$.

Several decision problems needed in applications can be expressed in terms of $\fullmucalculus$ formulas:

\begin{itemize}

\item XPath containment
\begin{itemize}
\item Input: $e_1, e_2 \in \lxpath$  and optional $d \in \ldtd$
\item Problem: Does $e_2$ always select all nodes selected by $e_1$? 
\item Definition: $\forall t \in \setofdocs, \forall x \in t, \containmentpb$
\item Tested $\fullmucalculus$ formula: $\phi_{e_1} \et \neg \phi_{e_2}$
\end{itemize}

\item XPath equivalence
\begin{itemize}
\item Input: $e_1, e_2 \in \lxpath$ and optional $d \in \ldtd$
\item Problem: Does $e_2$ always select exactly the same nodes as $e_1$? 
\item Definition: $\forall t \in \setofdocs, \forall x \in t, \equivalencepb$
\item Equivalence can be tested by two successive and separate containment checks
\end{itemize}

\item XPath emptiness
\begin{itemize}
\item Input: $e \in \lxpath$ and optional $d \in \ldtd$
\item Problem: Will $e$ ever return a non-empty set of nodes? 
\item Definition: $\forall t \in \setofdocs, \forall x \in t,  \satisfiabilitypb$
\item Tested $\fullmucalculus$ formula: $\phi_e$
\end{itemize}

\item XPath overlap
\begin{itemize}
\item Input: $e_1, e_2 \in \lxpath$ and optional $d \in \ldtd$
\item Problem: May $e_1$ and $e_2$ select common nodes? 
\item Definition: $\forall t \in \setofdocs, \forall x \in t,   \overlappb$
\item Tested $\fullmucalculus$ formula: $\phi_{e_1} \et \phi_{e_2} $
\end{itemize}

\item XPath coverage
\begin{itemize}
\item Input: $e_1, e_2,..., e_n \in \lxpath$  and optional $d \in \ldtd$
\item Problem: Are nodes selected by $e_1$ always selected by one of the $e_2, ..., e_n$? 
\item Definition: $\forall t \in \setofdocs, \forall x \in t, \coveragepb$
\item Tested $\fullmucalculus$ formula: $\phi_{e_1} \et \bigwedge_{2\leq i \leq n} \neg  \phi_{e_i}$
\end{itemize}

\end{itemize}

Note that for the containment problem, the unsatisfiability of $\phi_{e_1} \et \neg \phi_{e_2}$ is tested. Indeed, checking that an XPath expression $e_1$ is contained into another expression $e_2$ consists in checking that the implication $\phi_{e_1} \Rightarrow \phi_{e_2}$ holds for all trees. In other terms, there exists no tree for which the results of $e_1$ are not included in those of $e_2$, i.e. the negated implication $\phi_{e_1} \et \neg \phi_{e_2}$ is unsatisfiable. 

Since the finite binary tree model property must be enforced (as seen in Section~\ref{analysis:fbtm-prop}), decision problems are formulated from the root, and the actually checked formula becomes:
\begin{equation} 
\phiroot \et \phift \et (\phitype \et \mu X. \phitested \ou \someFCverifies X \ou \someNSverifies X)^\text{FBT} 
\label{analysis:xpath-dpb-formula}
\end{equation}
where $\phi_{\text{tested}}$ corresponds to a particular XPath decision problem from those given above.
Intuitively, the fixpoint is introduced for ``plunging'' XPath navigation performed by $\phitested$ at any location in the tree. It is for example necessary for relative XPath expressions that involve upward navigation in the tree.

It is important to note that formula~(\ref{analysis:xpath-dpb-formula}) is always alternation-free since both embeddings of XPath and tree types produce alternation-free formulas, and the negation of an alternation free sentence remains alternation-free. In practice, negated sentences introduced by XPath embeddings are turned into negation normal form, by applying the rules given on Figure~\ref{analysis:fig:positive-normal-form-rules}.

\section{Complexity Analysis and Implementation Principles}
\label{analysis:complexity-analysis}

The proposed approach has been implemented. A compiler takes XPath expressions as input, and translates them into $\fullmucalculus$ formulas. Another compiler takes regular tree types as input (DTDs) and outputs their $\fullmucalculus$ translation. The formula of a particular decision problem is then composed, normalized and solved. 

The $\mu$-calculus satisfiability solver is specialized for the alternation-free $\mu$-calculus with converse. It is closely inspired from the tableau methods described in \cite{tozawa-tableaux05} and \cite{vardi-jancl06}. A detailed description of the AFMC solver is beyond the scope of this chapter (see \cite{tozawa-tableaux05} for more details on an AFMC solver; and Chapter~\ref{xml-calculus:sec:algo} for a detailed description of a logical solver specialized for XML). The focus here is rather given to the AFMC solver aspects which allow to establish precise complexity results for the considered XML decision problems with the $\mu$-calculus approach. The algorithm relies on a top-down tableau method which attempts to construct satisfying Kripke structures by a fixpoint computation. Nodes of the tableau are specific subsets of a set called the Lean \cite{vardi-jancl06}. Given a formula $\psi \in \fullmucalculus$, the Lean is the subset of the Fischer-Ladner closure \cite{fischer79} of $\psi$ composed of atomic and modal subformulas of $\psi$ \cite{vardi-jancl06}. The algorithm starts from the set of all possible nodes, and repeatedly removes inconsistent nodes until a fixpoint is reached. At the end of the computation, if $\psi$ is present in a node of the fixpoint, then $\psi$ is satisfiable.
In this case, the fixpoint contains a satisfying model that can be easily extracted and used as a satisfying example XML tree. 

The complexity of the addressed XML decision problems can now be stated:

\begin{prop}
XPath containment, equivalence, emptiness, overlap and coverage decision problems, in the presence or absence of regular tree constraints, can be solved in time complexity $2^{O(n\cdot \text{log}~n)}$, where $n$ is the Lean size of the corresponding $\fullmucalculus$ formula. 
\end{prop}

This upper-bound is derived from:
\begin{enumerate}
\item the linear translations of XPath and regular tree types into the $\mu$-calculus;
\item the $2^{O(n\cdot \text{log}~n)}$ time complexity of the solver, which corresponds to the best known complexity for deciding alternation-free $\mu$-calculus with converse over Kripke structures \cite{tozawa-tableaux05}. Note that this complexity is smaller than the best known complexity for the whole $\mu$-calculus with converse \cite{vardi-icalp98} which is $2^{O(n^4\cdot \text{log}~n)}$ \cite{gradel-book02}. 
\end{enumerate}
The key observation for the linear translation of regular tree types is that only distinct atomic and modal subformulas of the translated formula are present in the Lean, even for a $n$-ary binder $\phi=\afletmu{X}{\phi} X_k$. More precisely, the Lean corresponding to the translation of $\phi$ contains at most:
\begin{itemize}
\item
 the two eventualities $\emod{a}\true$ for $a=1,2$
 \item  $2\cdot m$ universalities $\umod{a}\phi$ where $m$ is the number of binary tree type variables in the binder and the constant factor corresponds to the downward programs $a=1,2$
 \item the atomic propositions representing the alphabet symbols used in $\phi$
\end{itemize}

Deriving complexity from properties of the closure of a formula was first used by Fischer and Ladner for establishing decidability of PDL in single exponential time \cite{fischer79}. Analog observations have also been made for the modal logic K \cite{vardi-jancl06}, and the $\mu$-calculus over general Kripke structures \cite{tozawa-tableaux05}. 
These results can be seen as an application of this technique to the case where regular tree types are combined with XPath bidirectional queries over finite trees. 

Keys for the efficiency of the method on large practical instances are as follows:
\begin{enumerate}
\item Nodes of the tableau contain only modal formulas and exactly one atomic proposition (for XML), which greatly reduces the number of enumerated nodes for large alphabets.
\item Negation in the $\mu$-calculus is rather straightforward compared to automata techniques. Indeed, handling $\fullmucalculus$ formulas in negation normal form simply reduces to checking membership of atomic propositions in tableau nodes. This contrasts with tree automata techniques which require for every negation the full construction and complementation of automata with an exponential blow-up. As pointed out in \cite{baader-ijcar01} and \cite{vardi-jancl06}, tableau methods for logics with the tree model property can be viewed as implementations of the automata-theoretic approach which avoids an explicit automata construction.
\item The implementation relies on representing sets of nodes and operating on them symbolically using Binary Decision Diagrams (BDDs) \cite{bryant86}. BDDs provide a canonical representation of boolean functions. Their effectiveness is well known in the domain of formal verification of systems \cite{clarke-book99}. BDD variables encode truth status of Lean formulas. The cost of BDD operations is very sensitive to variable ordering.  Finding the optimal variable ordering is known to be NP-complete \cite{hojati-iccd96}. However, several heuristics are known to perform well in practice \cite{clarke-book99}. Choosing a good initial variable order does significantly improve performance. Preserving locality of the initial problem happens to be essential. It can be easily observed that the variable order determined by the breadth-first traversal of the initial formula (thus keeping sister subformulas in close proximity while ordering Lean formulas) yields better results in practice.
\end{enumerate}

There are still areas for improvements though. In particular, a large amount of time is spent in the $\mu$-loop detection performed by the solver for avoiding cycles and infinite paths in the case of finite recursion \cite{tozawa-tableaux05}. From this perspective, transforming the $\mu$-calculus formula at the syntactic level (as presented in Section~\ref{analysis:kripke-xml-trees}) and then relying on loop detection to enforce the finite model property is overkill. The approach may be improved by considering XML finite tree structures as models of the logic, and building an appropriate satisfiability solver for such structures.


\section{Outcome}
\label{analysis:outcome}
An approach for solving XPath decision problems by reduction to satisfiability of alternation-free modal $\mu$-calculus with converse over general Kripke structures has been proposed. XPath queries and regular tree types are linearly translated into the AFMC. XML decision problems are expressed as formulas in this logic, then decided using a solver for AFMC satisfiability. With respect to MSO, this yields much more efficient (exponential time) decision procedures for XML decision problems.
Nevertheless, this approach may still be greatly improved, since models of the logic are too general for the XML setting, and one has to pay extra costs for restricting them appropriately. One direction of future work consists in designing a more appropriate calculus where models are finite trees instead of general Kripke structures. This is what is achieved in the remaining of this dissertation.








 \ifglobalcompil
  \else 
   \bibliographystyle{apalike}
   \bibliography{references}
   \end{document}
  \fi

\part*{A Fixpoint Modal Logic with Converse for XML}
\newif\ifglobalcompil\globalcompiltrue
\newif\iffull\fulltrue 

  \ifglobalcompil
  \else 
    \input{Preamble} 
    \input{Markup}                       
    \newcommand{\chapterabstract}[1]{Chapter Abstract: #1}
    \begin{document} 
  \fi

\mychapter{A Fixpoint Modal Logic with Converse for XML}
\label{xml-calculus}
\label{xml-calculus:the-logic-for-xml}

\section{Introduction}
This chapter and the following introduce the final results of this thesis, based on the lessons learned from the investigations reported in previous chapters.

The decidability of a new logic with converse for finite and ordered trees is proved. The logic is sufficiently expressive to support XPath bidirectional navigation in finite trees along with regular tree languages. The logic is derived from the $\mu$-calculus and inherits some of its desirable properties, while improving the best known complexity for finite trees. These discoveries are naturally applied to the static analysis of XML specifications, for which they yield sound, complete and efficient decision procedures. The proof method is based on two auxiliary results. First, XML regular tree types and XPath expressions have a linear translation to \emph{cycle-free} formulas.  Second, the least and greatest fixpoints are equivalent for finite trees, hence the logic is closed under negation.

\paragraph{Chapter Outline}
This chapter presents focused trees in Section~\ref{xml-calculus:sec:focus} as a convenient data model for XML. The logic is then introduced in Section~\ref{xml-calculus:sec:logic}, and translations of XML concepts into the logic are presented in Section~\ref{xml-calculus:xml-concepts}.  

\section{Focused Trees}
\label{xml-calculus:sec:focus}

In this chapter, a less conventional approach is used to represent XML trees, called 
\emph{focused trees}. Focused trees are directly inspired by Huet's Zipper data structure \cite{huet-jfunc97}, and are closely related to pointed trees introduced in \cite{podelski-tal92,nivat-dm93}, which were extended to pointed hedges and applied to the XML setting in \cite{murata-pods01}. Focused trees not only describe a tree but also its context: its previous siblings and its parent, recursively. Exploring such a structure has the advantage to preserve all information, which is quite useful 
when considering languages such as XPath that allow forward and backward axes of navigation.

Formally, an alphabet $\alphabet$ of labels, ranged over by 
\(\nodelabel\) is assumed.
\smallsyntax{
\singleentry \treevar        [] \tree{\nodelabel}{\treelistvar}              [tree]\\
\entry       \treelistvar    [list of trees]
             \treelistnil              [empty list]
\oris        \treelistcons{\treevar}{\treelistvar} [cons cell]\\
\entry       \contextvar     [context]
             \contexttop {\treelistvar}{\treelistvar}    [root of the tree]
            \oris \contextnode{\treelistvar}{\contextvar}{\nodelabel}{\treelistvar} 
            [context node]\\
\singleentry \focusedtreevar [] \focusedtree{\treevar}{\contextvar} [focused tree]\\
}

In order to deal with XPath containment, it is needed to represent in a focused 
tree the place where the evaluation was started using a \emph{context mark}.  
To do so, we consider focused trees where a single tree or a single context 
node is marked, as in \(\tree{\starttrue{\nodelabel}}{\treelistvar}\) or 
\(\contextnode{\treelistvar}{\contextvar}{\starttrue{\nodelabel}}{\treelistvar}\).  
When the presence of the mark is unknown, it is written as 
\(\tree{\notation{\startunk{\nodelabel}}{Node with unknown context mark}{n:unkcontext}}{\treelistvar}\).

\(\notation{\focusedtrees}{Set of finite focused trees}{n:setoffctrees}\) denotes the set of finite focused trees with a single 
mark.  The \emph{name} of a focused tree is defined as 
\(\fname{\focusedtree{\tree{\startunk{\nodelabel}}{\treelistvar}}{\contextvar}} 
= \nodelabel\).
Navigation in focused trees is now described, in binary style. Four 
directions can be followed: for a focused tree \(\focusedtreevar\), 
\(\godown{\focusedtreevar}\) changes the focus to the children of the current 
tree, \(\goright{\focusedtreevar}\) changes the focus to the next sibling of 
the current tree, \(\goup{\focusedtreevar}\) changes the focus to the parent 
of the tree \emph{if the current tree is a leftmost sibling}, and 
\(\goleft{\focusedtreevar}\) changes the focus to the previous sibling.

Formally:
\begin{align*}
  \godown{\focusedtree{\tree{\startunk{\nodelabel}}{\treelistcons{\treevar}{\treelistvar}}}{\contextvar}} 
  &\eqdef 
  \focusedtree{\treevar}{\contextnode{\treelistnil}{\contextvar}{\startunk{\nodelabel}}{\treelistvar}}\\
  \goright{\focusedtree{\treevar}{\contextnode{\treelistvar_l}{\contextvar}{\startunk{\nodelabel}}{\treelistcons{\treevar'}{\treelistvar_r}}}}
  &\eqdef
  \focusedtree{\treevar'}{\contextnode{\treelistcons{\treevar}{\treelistvar_l}}{\contextvar}{\startunk{\nodelabel}}{\treelistvar_r}}\\
  \goup{\focusedtree{\treevar}{\contextnode{\treelistnil}{\contextvar}{\startunk{\nodelabel}}{\treelistvar}}} 
  &\eqdef
  \focusedtree{\tree{\startunk{\nodelabel}}{\treelistcons{\treevar}{\treelistvar}}}{\contextvar}\\
  \goleft{\focusedtree{\treevar'}{\contextnode{\treelistcons{\treevar}{\treelistvar_l}}{\contextvar}{\startunk{\nodelabel}}{\treelistvar_r}}}
  &\eqdef
  \focusedtree{\treevar}{\contextnode{\treelistvar_l}{\contextvar}{\startunk{\nodelabel}}{\treelistcons{\treevar'}{\treelistvar_r}}}
\end{align*}

When the focused tree does not have the required shape, these operations are 
not defined.

\section{Formulas of the Logic}
\label{xml-calculus:sec:logic}


The logic to which XPath expressions and XML regular tree types 
are going to be translated is introduced. It is a sub-logic of the alternation free modal 
\(\mu\)-calculus with converse. Next, a restriction on the  considered
formulas is introduced, and an interpretation of formulas as sets of finite 
focused trees is given. Then, it is shown that the logic has a single fixpoint for these 
models and that it is closed under negation.

In the following definitions, \(a \in \domProg\) are \emph{programs} and atomic 
propositions  \(\atomprop\) correspond to labels from \(\alphabet\). It is also assumed that \(\overline{\overline{a}} = a\).

\begin{figure}
\smallsyntax{
\mulogic \ni \entry       \phi,\psi     [formula]
             \true              [true]
\oris        \atomprop \quad | \quad \neg \atomprop             [atomic prop (negated)]
\oris        \startatom \quad | \quad \neg \startatom [context (negated)]
\oris        X [variable]
\oris        \phi \ou \psi [disjunction]
\oris        \phi \et \psi [conjunction]
\oris        \emod{a}\phi \quad | \quad \neg \emod{a}\true [existential 
(negated)]
\oris        \murec{X_i}{\phi_i}{\psi} [least $n$-ary fixpoint]
\oris        \nurec{X_i}{\phi_i}{\psi} [greatest $n$-ary fixpoint]
}
\caption{Logic formulas}
\label{fig:logic}
\end{figure}

Formulas, defined in Fig. \ref{fig:logic} include the truth predicate, atomic propositions (denoting the name of the tree in focus), start propositions (denoting the presence of the start mark), disjunction and conjunction of formulas, formulas under an existential (denoting the existence a subtree satisfying the sub-formula), and least and greatest nary fixpoints. We chose to include a nary version of the latter because regular types are often defined as a set of mutually recursive definitions, making their translation in our logic more succinct. In the following we write ``\(\mu X.\phi\)'' for ``\(\mu X.\phi \text{ in } \phi\)''.

\begin{figure}
\begin{align*}
\finterp{\true}{V} &\eqdef \focusedtrees &
\finterp{\atomprop}{V} &\eqdef 
\setof{\focusedtreevar}{\fname{(\focusedtreevar)} = \atomprop} \\
\finterp{X}{V} &\eqdef V(X) &
\finterp{\neg \atomprop}{V} &\eqdef 
\setof{\focusedtreevar}{\fname{(\focusedtreevar)} \neq \atomprop}\\
\finterp{\phi \ou \psi}{V} &\eqdef \finterp{\phi}{V} \cup \finterp{\psi}{V} &
\finterp{\startatom}{V} &\eqdef \setof{\focusedtreevar}{\focusedtreevar = 
\focusedtree{\tree{\starttrue{\nodelabel}}{\treelistvar}}{\contextvar}} \\
\finterp{\phi \et \psi}{V} &\eqdef \finterp{\phi}{V} \cap \finterp{\psi}{V} &
\finterp{\neg \startatom}{V} &\eqdef \setof{\focusedtreevar}{\focusedtreevar 
= \focusedtree{\tree{\nodelabel}{\treelistvar}}{\contextvar}}
\end{align*}
\begin{align*}
\finterp{\emod{a} \phi}{V} &\eqdef \setof{\focusedtreevar\emod{\overline 
a}}{\focusedtreevar \in \finterp{\phi}{V} \land 
\isdefined{\focusedtreevar\emod{\overline a}}}\\
\finterp{\neg \emod{a} \true}{V} &\eqdef 
\setof{\focusedtreevar}{\isundefined{\focusedtreevar\emod{a}}}\\
\finterp{\murec{X_i}{\phi_i}{\psi}}{V} &\eqdef \text{let \(T_i = \left(\bigcap 
\setof{\overline{T_i} \subseteq 
\overline{\focusedtrees}}{\finterp{\overline{\phi_i}}{V[\overline{T_i/X_i}]} 
\subseteq \overline{T_i}}\right)_i\)}
\\
&\qquad \text{ in } \finterp{\psi}{V[\overline{T_i/X_i}]}\\
\finterp{\nurec{X_i}{\phi_i}{\psi}}{V} &\eqdef \text{let \(T_i = \left(\bigcup 
\setof{\overline{T_i} \subseteq \overline{\focusedtrees}}{ \overline{T_i} 
\subseteq \finterp{\overline{\phi_i}}{V[\overline{T_i/X_i}]}}\right)_i\)}
\\
&\qquad \text{ in } \finterp{\psi}{V[\overline{T_i/X_i}]}
\end{align*}
\caption{Interpretation of formulas}
\label{xml-calculus:fig:interp}
\end{figure}

An interpretation of formulas as sets of finite focused trees 
with a single start mark is now given on Figure~\ref{xml-calculus:fig:interp}. $\hiddennotation{\finterp{\phi}{\cdot}}{Interpretation of the formula $\phi$}{n:finterp}$
The interpretation of the nary fixpoints first compute the smallest or largest 
interpretation for each \(\phi_i\) then returns the interpretation of \(\psi\) 
using these bindings.

The set of valid formulas is now restricted to \emph{cycle-free formulas}, i.e. formulas that have a bound on the number of \emph{modality cycles} independently of the number of unfolding of their fixpoints. A modality cycle is a subformula of the form \(\emod{a}\phi\) where \(\phi\) contains a \emph{top-level} existential of the form \(\emod{\overline{a}}\psi\). ``Top-level'' means under an arbitrary number of conjunctions or disjunctions, but not under any other construct. For instance, the formula ``\(\mu X.\emod{1} (\phi \ou \emod{\overline{1}} X) \text{ in } X\)'' is not cycle free: for any integer \(n\), there is an unfolding of the formula with \(n\) modality cycles. On the other hand, the formula ``\(\mu X.\emod{1}(X \ou Y), \; Y.\emod{\overline{1}}(Y \ou \true) \text{ in } X\)'' is cycle free: there is at most one modality cycle.

Cycle-free formulas have a very interesting property, which can now be described. To test whether a tree satisfies a formula, one may define a straightforward inductive relation between trees and formulas that only holds when the root of the tree satisfies the formula, unfolding fixpoints if necessary. Given a tree, if a formula \(\phi\) is cycle free, then every node of the tree will be tested a finite number of time against any given subformula of \(\phi\). The intuition behind this property, which holds a central role in the proof of lemma \ref{xml-calculus:lem:finite_unfold}, is the following. If a tree node is tested an infinite number of times against a subformula, then there must be a cycle in the navigation in the tree, corresponding to some modalities occurring in the subformula, between one occurrence of the test and the next one. As trees are considered, the cycle implies there is a modality cycle in the formula (as cycles of the form \(\emod{1}\emod{2}\emod{\overline{1}}\emod{\overline{2}}\) cannot occur). Hence the number of modality cycles in any expansion of \(\phi\) is unbounded, thus the formula is not cycle free.

Figure~\ref{xml-calculus:fig:cyclefree} gives an inductive relation that decides whether a formula is cycle free.

\begin{figure}
  \begin{mathpar}
    \inferrule{ \phi = \true, \atomprop, \neg \atomprop, \startatom, \text{ or } \neg 
    \startatom}{\cyclefree{\varenv}{\varcf}{\varignore}{\varrecurse}{\phi}} \and
    \inferrule{ \cyclefree{\varenv}{\varcf}{\varignore}{\varrecurse}{\phi} \\ \cyclefree{\varenv}{\varcf}{\varignore}{\varrecurse}{\psi} }{ 
    \cyclefree{\varenv}{\varcf}{\varignore}{\varrecurse}{\phi \ou \psi} } \and
    \inferrule{ \cyclefree{\varenv}{\varcf}{\varignore}{\varrecurse}{\phi} \\ \cyclefree{\varenv}{\varcf}{\varignore}{\varrecurse}{\psi} }{ 
    \cyclefree{\varenv}{\varcf}{\varignore}{\varrecurse}{\phi \et \psi} } \and
    \inferrule{ }{ \cyclefree{\varenv}{\varcf}{\varignore}{\varrecurse}{\neg \emod{a}\true} } \and
    \inferrule{ \cyclefree{\varenv}{(\cfinject{\varcf}{\emod{a}}) }{\varignore}{\varrecurse}{ \phi } }{
    \cyclefree{\varenv}{\varcf}{\varignore}{\varrecurse}{\emod{a} \phi} } \and
    \inferrule{ \forall X_j \in \overline{X_i}. \left(\cyclefree{(\varenv + \overline{\cfbind{X_i}{\phi_i}})}{(\cfextend{\varcf}{\overline{\cfbind{X_i}{\cfnil}}})}{\varignore \setminus \overline{X_i}}{\varrecurse \setminus \overline{X_i}}{\phi_j}\right) \\ 
 \cyclefree{\varenv}{\varcf}{\varignore \cup \overline{X_i}}{\varrecurse \setminus \overline{X_i}}{\psi} }{ 
    \cyclefree{\varenv}{\varcf}{\varignore}{\varrecurse}{\murec{X_i}{\phi_i}{\psi}} }
\and
	    \inferrule{ \forall X_j \in \overline{X_i}. \left(\cyclefree{(\varenv + \overline{\cfbind{X_i}{\phi_i}})}{(\cfextend{\varcf}{\overline{\cfbind{X_i}{\cfnil}}})}{\varignore \setminus \overline{X_i}}{\varrecurse \setminus \overline{X_i}}{\phi_j}\right) \\ 
	 \cyclefree{\varenv}{\varcf}{\varignore \cup \overline{X_i}}{\varrecurse \setminus \overline{X_i}}{\psi} }{ 
	    \cyclefree{\varenv}{\varcf}{\varignore}{\varrecurse}{\nurec{X_i}{\phi_i}{\psi}} }
\and
    \inferrule[NoRec]{X \in \varrecurse \\ \varcf(X) = \emod{a} }{ \cyclefree{\varenv}{\varcf}{\varignore}{\varrecurse}{X} }
		\and
		    \inferrule[Rec]{X \not\in \varrecurse \\ \cyclefree{\varenv}{\varcf}{\varignore}{\varrecurse \cup \{X\}}{\varenv(X)} }{ \cyclefree{\varenv}{\varcf}{\varignore}{\varrecurse}{X} }
				\and
				    \inferrule[Ign]{X \in \varignore}{ \cyclefree{\varenv}{\varcf}{\varignore}{\varrecurse}{X} }
  \end{mathpar}
  \caption{Cycle-free formulas}
  \label{xml-calculus:fig:cyclefree}
\end{figure}

In the judgement  \(\cyclefree{\varenv}{\varcf}{\varignore}{\varrecurse}{\phi}\) of Fig.~\ref{xml-calculus:fig:cyclefree}, \(\varenv\) is an environment binding some recursion variables to their formulas, \(\varcf\) binds variables to modalities, \(\varrecurse\) is a set of variables that have already been expanded (see below), and \(\varignore\) is a set of variables already checked.

The environment \(\varcf\) used to derive the judgement consists of bindings 
from variables (from enclosing fixpoint operators) to modalities. A modality 
may be \(\cfnil\), no information is known about the variable, \(\emod{a}\), 
the last modality taken \(\emod{a}\) was consistent, or \(\cfbot\), a cycle 
has been detected.  A formula is not cycle free if an occurrence of a variable 
under a fixpoint operator is either not under a modality (in this case 
\(\varcf(X) = \cfnil\)), or is under a cycle (\(\varcf(X) = \cfbot\)).  Cycle 
detection uses an auxiliary operator to detect modality cycles:

\begin{equation*}
  \cfinject{\varcf}{\emod{a}} \eqdef \{ 
  \cfbind{X}{(\cfinject{\varcf(X)}{\emod{a}})} \}
\end{equation*}
where
\[
\begin{array}{c|c c c c|}
  \cfinject{\cdot}{\cdot} & \emod{\hasleftsucc} & \emod{\hasrightsucc} & 
  \emod{\isleftsucc} & \emod{\isrightsucc} \\ \hline
  \cfnil & \emod{\hasleftsucc} & \emod{\hasrightsucc} & \emod{\isleftsucc} & 
  \emod{\isrightsucc} \\
  \emod{\hasleftsucc} & \emod{\hasleftsucc} & \emod{\hasrightsucc} & \cfbot & 
  \emod{\isrightsucc} \\
  \emod{\hasrightsucc} & \emod{\hasleftsucc} & \emod{\hasrightsucc} & 
  \emod{\isleftsucc} & \cfbot \\
  \emod{\isleftsucc} & \cfbot & \emod{\hasrightsucc} & \emod{\isleftsucc} & 
  \emod{\isrightsucc} \\
  \emod{\isrightsucc} & \emod{\hasleftsucc} & \cfbot & \emod{\isleftsucc} & 
  \emod{\isrightsucc} \\
  \cfbot &
  \cfbot &
  \cfbot &
  \cfbot &
  \cfbot
\end{array}
\]

To check that mutually recursive formulas are cycle-free, one proceeds the following way. When a mutually recursive formula is encountered, for instance  
\(\murec{X_i}{\phi_i}{\psi}\), every recursive binding is checked. Because of mutual recursion, formulas cannot be checked independently and a variable must be expanded the first time it is encountered (rule \hname{Rec}). However there is no need to expand it a second time (rule \hname{NoRec}). When checking \(\psi\), as the formulas bound to the enclosing recursion have been checked to be cycle free, there is no need to further check these variables (rule \hname{Ign}). To account for shadowing of variables, newly bound recursion variables are removed from \(\varignore\) and \(\varrecurse\) when checking a recursion. One may easily prove that if \(\cyclefree{\varenv}{\varcf}{\varignore}{\varrecurse}{\phi}\) holds, then \(\varignore \cap \varrecurse = \emptyset\).

This relation decides whether a formula is cycle free because, if it is not, there must be a recursive binding of \(X_i\) to \(\phi_i\) such that \(\phi_i\{\subst{\phi_i}{X_i}\}\{\subst{\overline{\phi_j}}{\overline{X_j}}\}\) exhibits a modality cycle above \(X_i\), where the \(X_j\) are recursion variables being defined (either in the recursion defining \(X_i\) or in an enclosing recursion definition).

With these definitions, a first result can now be shown: in the finite focused-tree 
interpretation, the least and greatest fixpoints coincide for cycle-free 
formulas. To this end, a stronger result is proved, which states that a given 
focused tree is in the interpretation of a formula if it is in a finite 
unfolding of the formula.  In the base case, the formula \(\atomprop 
\et \neg \atomprop\) is used as ``false''.
\hiddennotation{\finiteunfold{\phi}}{Finite unfolding of a formula $\phi$}{n:finiteunfold}
\begin{defn}[Finite unfolding]
  A \emph{finite unfolding} of a formula \(\phi\) belongs to the set 
  \(\finiteunfold{\phi}\) inductively defined as
  \begin{align*}
    \finiteunfold{\phi} &\eqdef \{ \phi \} \quad \text{for \(\phi = \true, 
    \atomprop, \neg \atomprop, \startatom, \neg \startatom, X, \neg 
    \emod{a}\true\)} \\
    \finiteunfold{\phi \ou \psi} &\eqdef \setof{\phi' \ou \psi'}{\phi' \in 
    \finiteunfold{\phi}, \psi' \in \finiteunfold{\psi}}\\
    \finiteunfold{\phi \et \psi} &\eqdef \setof{\phi' \et \psi'}{\phi' \in 
    \finiteunfold{\phi}, \psi' \in \finiteunfold{\psi}}\\
    \finiteunfold{\emod{a}{\phi}} &\eqdef \setof{\emod{a}{\phi'}}{\phi' \in 
    \finiteunfold{\phi}}\\
    \finiteunfold{\murec{X_i}{\phi_i}{\psi}} &\eqdef 
    \finiteunfold{\psi\{\subst{\murec{X_i}{\phi_i}{X_i}}{X_i}\}}\\
    \finiteunfold{\nurec{X_i}{\phi_i}{\psi}} &\eqdef 
    \finiteunfold{\psi\{\subst{\nurec{X_i}{\phi_i}{X_i}}{X_i}\}}\\
    \finiteunfold{\murec{X_i}{\phi_i}{\psi}} &\eqdef \atomprop \et \neg 
    \atomprop\\
    \finiteunfold{\nurec{X_i}{\phi_i}{\psi}} &\eqdef \atomprop \et \neg 
    \atomprop
  \end{align*}
\end{defn}

\begin{lem}
  \label{xml-calculus:lem:finite_unfold}
  Let \(\phi\) a cycle-free formula. If \(\focusedtreevar \in 
  \finterp{\phi}{V}\) then \(\focusedtreevar \in 
  \finterp{\finiteunfold{\phi}}{V}\).
\end{lem}

The reason why this lemma holds is the following. Given a tree satisfying \(\phi\), we deduce from the hypothesis that \(\phi\) is cycle free the fact that every node of the tree will be tested a finite number of times against every subformula of \(\phi\). As the tree and the number of subformulas are finite, the satisfaction derivation is finite hence only a finite number of unfolding is necessary to prove that the tree satisfies the formula, which is what the lemma states. As least and greatest fixpoints coincide when only a finite number of unfolding is required, this is sufficient to show that they collapse. Note that this would not hold if infinite trees were allowed: the formula \(\mu X. \emod{1} X\) is cycle free, but its interpretation is empty, whereas the interpretation of \(\nu X. \emod{1} X\) includes every tree with an infinite branch of \(\emod{1}\) children.

We now illustrate why formulas need to be cycle free for the fixpoints to collapse. Consider the formula \(\mu X. \emod{1} \emod{\overline{1}} X\). Its interpretation is empty. The interpretation of \(\nu X. \emod{1} \emod{\overline{1}} X\) however contains every focused tree that has one \(\emod{1}\) child.

  \iffull\begin{mypfo}

  The result is a consequence of the fact that a sub-formula is never 
  confronted twice to the same node of the focused tree as there is no cycle 
  in the formula.  It is thus possible to annotate occurrences of \(\nu\) and 
  \(\mu\) with the direction the formula is exploring for each variable, as in 
  Fig. \ref{xml-calculus:fig:cyclefree}, and prove the result by induction on the size of 
  focused tree in this direction.
  \iffull

  More precisely, each variable in every \(\mu\) and \(\nu\) of the initial 
  formula is given a unique identifier.
  
  The induction principle relies on the \emph{longest path} of a focused tree.  
  Given a tree and a direction (which may be \(\cfnil\)), we define the 
  longest path as the longest cycle-free path that starts in the initial 
  direction.
  
  We then prove the property that a tree \(\focusedtreevar\) belongs to the 
  finite unfolding of \(\phi\) by induction on the lexical order of:
  \begin{enumerate}
    \item the number of fixpoints not yet annotated;
    \item the max of the lengths of the longest path for a given unique 
      identifier according to the direction for this identifier;
    \item the size of the formula.
  \end{enumerate}

  The interesting case is an annotated formula recursion \(\phi = 
  \murec{X_i}{\phi_i}{\psi}\).  This formula may only have been produced by an 
  expansion.  As the formula is cycle-free, at least one modality has been 
  encountered since the expansion for each identifier associated with the 
  \(X_i\), and these modalities are compatible with the previous directions 
  (if they existed).  The longest path for each identifier is thus shorter 
  hence we have by induction that \(\focusedtreevar\) is in a finite expansion 
  of the expansion of \(\phi\).
  \fi
\end{mypfo}\fi

In the rest of the dissertation, only least fixpoints are considered. An important consequence of Lemma \ref{xml-calculus:lem:finite_unfold} is that the logic restricted in this way is closed under negation using De Morgan's dualities, extended to eventualities and fixpoints as follows:
\begin{align*}
\neg \emod{a}\phi & \eqdef \neg \emod{a}\true \ou \emod{a}\neg \phi \\
\neg \murec{X_i}{\phi_i}{\psi} & \eqdef \murec{X_i}{\neg 
\phi_i\{\overline{\subst{X_i}{\neg X_i}}\}}{\neg 
\psi\{\overline{\subst{X_i}{\neg X_i}}\}}
\end{align*}

\section{Translations of XML Concepts}
\label{xml-calculus:xml-concepts}
The interpretation of XPath expressions as sets of focused trees is given:
 $\hiddennotation{\xeinterp{\cdot}{\cdot}}{XPath interpretation as focused tree sets}{n:xeinterp}$

\begin{align*}
\xeinterp{\cdot}{\cdot} & : \lxpath \rightarrow \powerset{\focusedtrees} \rightarrow \powerset{\focusedtrees} \\
 \xeinterp{/p}{\focusedtreesetvar} & \eqdef \xpinterp{p}{\rootfocusedtrees{\focusedtreesetvar}} \\
 \xeinterp{p}{\focusedtreesetvar} & \eqdef \xpinterp{p}{\{\focusedtree{\tree{\starttrue{\nodelabel}}{\treelistvar}}{ \contextvar} \in \focusedtreesetvar\}} \\
 \xeinterp{e_1 \shortmid e_2}{\focusedtreesetvar} & \eqdef \xeinterp{e_1}{\focusedtreesetvar} \cup \xeinterp{e_2}{\focusedtreesetvar}\\
 \xeinterp{e_1 \cap e_2}{\focusedtreesetvar} &  \eqdef \xeinterp{e_1}{\focusedtreesetvar} \cap \xeinterp{e_2}{\focusedtreesetvar}
\end{align*}
\begin{align*}
 \xpinterp{\cdot}{\cdot} & : \dom{Path} \rightarrow \powerset{\focusedtrees} \rightarrow \powerset{\focusedtrees} \\
 \xpinterp{p_1/p_2}{\focusedtreesetvar} &  \eqdef \setof{\focusedtreevar'}{\focusedtreevar' \in \xpinterp{p_2}{\left(\xpinterp{p_1}{\focusedtreesetvar}\right)}}\\
 \xpinterp{\qualif{p}{q}}{\focusedtreesetvar} &  \eqdef  \setof{\focusedtreevar}{\focusedtreevar \in \xpinterp{p}{\focusedtreesetvar} \et \xqinterp{q}{\focusedtreevar}} \\ 
 \xpinterp{\step{\axisvar}{\nodelabel}}{\focusedtreesetvar} &  \eqdef \setof{\focusedtreevar}{\focusedtreevar \in \xainterp{\axisvar}{\focusedtreesetvar} \et \fname{(\focusedtreevar)}=\nodelabel}  \\ 
 \xpinterp{\step{\axisvar}{*}}{\focusedtreesetvar} &  \eqdef \setof{\focusedtreevar}{\focusedtreevar \in \xainterp{\axisvar}{\focusedtreesetvar}}
\end{align*}
\begin{align*}
 \xqinterp{\cdot}{\cdot} & : \dom{Qualif} \rightarrow \focusedtrees \rightarrow \{\text{true}, \text{false} \} \\
 \xqinterp{q_1 \op{and} q_2}{\focusedtreevar} & \eqdef \xqinterp{q_1}{\focusedtreevar} \et  \xqinterp{q_2}{\focusedtreevar} \\
 \xqinterp{q_1 \op{or} q_2}{\focusedtreevar} & \eqdef \xqinterp{q_1}{\focusedtreevar} \ou  \xqinterp{q_2}{\focusedtreevar} \\
 \xqinterp{\op{not}~q}{\focusedtreevar}  & \eqdef \neg~\xqinterp{q}{\focusedtreevar}\\
 \xqinterp{p}{\focusedtreevar} & \eqdef  \xpinterp{p}{\{\focusedtreevar\}} \neq \emptyset \\ \\          
\end{align*}

\begin{align*} 
  \xainterp{$\cdot$}{\cdot} & : \dom{Axis} \rightarrow \powerset{\focusedtrees} \rightarrow \powerset{\focusedtrees} \\
 \xainterp{\axis{self}}{\focusedtreesetvar} & \eqdef \focusedtreesetvar \\
 \xainterp{\axis{child}}{\focusedtreesetvar} & \eqdef 
 \fchildrenfocusedtrees{\focusedtreesetvar} \cup 
 \xainterp{\axis{following-sibling}}{\fchildrenfocusedtrees{\focusedtreesetvar}} \\
 \xainterp{\axis{following-sibling}}{\focusedtreesetvar} & \eqdef 
 \nsiblingsfocusedtrees{\focusedtreesetvar} \cup 
 \xainterp{\axis{following-sibling}}{\nsiblingsfocusedtrees{\focusedtreesetvar}}\\
 \xainterp{\axis{preceding-sibling}}{\focusedtreesetvar} & \eqdef 
 \psiblingsfocusedtrees{\focusedtreesetvar} \cup 
 \xainterp{\axis{preceding-sibling}}{\psiblingsfocusedtrees{\focusedtreesetvar}}\\ 
 \xainterp{\axis{parent}}{\focusedtreesetvar} & \eqdef 
 \parentfocusedtree{\focusedtreesetvar} \\
 \xainterp{\axis{descendant}}{\focusedtreesetvar} & \eqdef \xainterp{\axis{child}}{\focusedtreesetvar} \cup \xainterp{\axis{descendant}}{(\xainterp{\axis{child}}{\focusedtreesetvar})}\\
 \xainterp{\axis{descendant-or-self}}{\focusedtreesetvar} &  \eqdef \focusedtreesetvar \cup \xainterp{\axis{descendant}}{\focusedtreesetvar} \\
 \xainterp{\axis{ancestor}}{\focusedtreesetvar} & \eqdef 
 \xainterp{\axis{parent}}{\focusedtreesetvar} \cup 
 \xainterp{\axis{ancestor}}{(\xainterp{\axis{parent}}{\focusedtreesetvar})}\\
 \xainterp{\axis{ancestor-or-self}}{\focusedtreesetvar} & \eqdef \focusedtreesetvar \cup \xainterp{\axis{ancestor}}{\focusedtreesetvar} \\
 \xainterp{\axis{following}}{\focusedtreesetvar} & \eqdef 
 \xainterp{\axis{descendant-or-self}}{\left(\xainterp{\axis{following-sibling}}{(\xainterp{\axis{ancestor-or-self}}{\focusedtreesetvar})}\right)}\\
 \xainterp{\axis{preceding}}{\focusedtreesetvar} & \eqdef \xainterp{\axis{descendant-or-self}}{\left(\xainterp{\axis{preceding-sibling}}{(\xainterp{\axis{ancestor-or-self}}{\focusedtreesetvar})}\right)}
\end{align*}

\begin{align*}
  \fchildrenfocusedtrees{\focusedtreesetvar} & \eqdef \setof{\godown{\focusedtreevar}}{\focusedtreevar \in \focusedtreesetvar \et \isdefined{\godown{\focusedtreevar}}} \\
  \nsiblingsfocusedtrees{\focusedtreesetvar} & \eqdef \setof{\goright{\focusedtreevar}}{\focusedtreevar \in \focusedtreesetvar \et \isdefined{\goright{\focusedtreevar}}} \\
  \psiblingsfocusedtrees{\focusedtreesetvar} & \eqdef \setof{\goleft{\focusedtreevar}}{\focusedtreevar \in \focusedtreesetvar \et \isdefined{\goleft{\focusedtreevar}}} \\
\parentfocusedtree{\focusedtreesetvar} & \eqdef \{ \focusedtree{\tree{\startunk{\nodelabel}}{\revappend{\treelistvar_l}{\treelistcons{\treevar}{\treelistvar_r}}}}{\contextvar} \\ &\qquad \; | \; \focusedtree{\treevar}{\contextnode{\treelistvar_l}{\contextvar}{\startunk{\nodelabel}}{\treelistvar_r}} \in \focusedtreesetvar \}\\
  \revappend{\treelistnil}{\treelistvar_r} &\eqdef \treelistvar_r\\
  \revappend{\treelistcons{\treevar}{\treelistvar_l}}{\treelistvar_r} &\eqdef 
  \revappend{\treelistvar_l}{\treelistcons{\treevar}{\treelistvar_r}}
  \\
  \rootfocusedtrees{\focusedtreesetvar} & \eqdef \{\focusedtree{\tree{\starttrue{\nodelabel}}{\treelistvar}}{ \contexttop {\treelistvar}{\treelistvar}} \in \focusedtreesetvar\} \cup \rootfocusedtrees{\parentfocusedtree{\focusedtreesetvar}}
\end{align*}


\subsection{XPath Embedding}
\label{xml-calculus:sec:xpath}

An XPath expression can be translated into an equivalent formula in $\mulogic$ which performs navigation in focused trees in binary style, as presented in the  Section~\ref{analysis:xpath-embedding} of previous Chapter~\ref{analysis}. A stronger result can be proved:

\begin{prop}[Translation Correctness]
\label{xml-calculus:proposition2}
The following hold for an XPath expression $e$ and a \(\mulogic\) formula 
\(\phi\), with \(\psi = \mucalcE{e}{\phi}\):
\begin{enumerate}
\item $\finterp{\psi}{\emptyset}=\xeinterp{e}{\finterp{\phi}{\emptyset}}$
\item $\psi$ is cycle-free
\item the size of $\psi$ is linear in the size of $e$ and \(\phi\)
\end{enumerate}
\end{prop} 

\begin{figure}
\begin{center}
\begin{tikzpicture}[scale=0.65]
\draw (-1.1,5.2) node(bidon1) {};
\draw (8,6.01) node(bidon2) {};
\draw [white] (bidon1) -- (bidon2);

\draw (-0.5,6.03) node(example) {Translation of};
\draw (-1,5.2) node(inmulogic) {into $\mulogic$:};
\draw (3.9,6) node(xpathpart1) [text=purple]{$\step{following-sibling}{a}$};

\draw (3.8,5.2) node(xpathtrans1a) [text=purple]{$a$};
\draw(6.8,5.2) node(xpathtrans1b) [text=purple]{$\et \left(\mu Z . \emod{\overline{2}}\startatom \ou \emod{\overline{2}}Z\right)$};


\draw (8.5,6) node(xpathpart2) [text=violet]{$/\step{preceding-sibling}{b}$};

\draw (2,5.2) node(xpathtrans2) [text=violet]{$b \et \large{[} \mu Y. \emod{2}\large{(} $};
\draw (10.8,5.2) node(xpathtrans3) [text=violet]{$\large{)} \ou \emod{2}Y \large{]}$};

\draw (0,0) node(bidon1) {};
\draw (5,5.5) node(bidon2) {};
\draw [white] (bidon1) -- (bidon2);

\draw (0,4) node(x) [ball color=white,circle,text=black] {\begin{small}$\startatom$\end{small}};

\draw (1,3) node(c1) [ball color=white,circle,text=black]{b};
\draw [thick, ->, orange] (x) -- (c1);

\draw (2,2) node(c2) [ball color=purple,circle,text=black]{a};
\draw [dashed, thick, ->, orange] (c1) -- (c2);

\draw (3,1) node(c3) [ball color=white,circle,text=black]{c};

\draw [thick, ->, orange] (c2) -- (c3);

\draw (4,0) node(c4) [ball color=purple,circle,text=black]{a};

\draw [thick, ->, orange] (c3) -- (c4);

\draw (1,3) node(c1selected) [ball color=violet,circle,text=black]{b};
\end{tikzpicture}
\end{center}
\caption{Example of Back and Forth XPath Navigation Translation.}\label{xml-calculus:fig:xpathtransexample}
\end{figure}
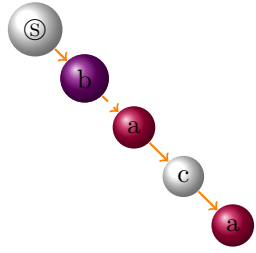

  \begin{mypfo}
 The proof uses a structural induction that ``peels off'' the compositional 
 layers of each set of rules over focused trees. The cycle-free part follows 
 from the fact that translated fixpoint formulas are closed and there is no 
 nesting of modalities with converse programs between a fixpoint variable and 
 its binder. Each XPath navigation step is cycle-free, and their composition yields a proper nesting of fixpoint formulas which is also cycle-free. Figure~\ref{xml-calculus:fig:xpathtransexample} illustrates this on an typical example. Finally, formal translations do not duplicate any subformula of arbitrary length.
 \end{mypfo}

\subsection{Embedding Regular Tree Languages}
\label{xml-calculus:xml-types}
%
%
The straightforward isomorphism between unranked and binary regular tree types (presented in Section~\ref{foundations:binary-tree-types} of Chapter~\ref{foundations}) is used. The translation from binary regular tree types into $\mulogic$ is given by the function $\Ttomu{\cdot}$ as follows:
$\hiddennotation{\Ttomu{\cdot}}{Translation of tree types into $\mulogic$}{n:ttomulogic}$
\begin{align*}
\Ttomu{\cdot} & : \lbtt \rightarrow \mulogic \\
\Ttomu{\emptyset} & \eqdef \atomprop \et \neg \atomprop \\
\Ttomu{\epsilon} & \eqdef \atomprop \et \neg \atomprop \\
\Ttomu{T_1 \tou T_2} & \eqdef \Ttomu{T_1} \ou \Ttomu{T_2} \\
\Ttomu{\nodelabel(X_1, X_2)} & \eqdef  \nodelabel \et \Tsucc{X_1}{1} \et \Tsucc{X_2}{2}\\
\Ttomu{\typebind{X}{T} T} & \eqdef \murec{X_i}{\Ttomu{T_i}}{\Ttomu{T}}
\end{align*}
where the formula \(\atomprop \et \neg \atomprop\) is used as ``false'', and the function $\Tsucc{\cdot}{\cdot}$ takes care of setting the type frontier:
$$\begin{array}{lll}
\Tsucc{X}{\alpha} &=& \left\{ \begin{array}{ll}\neg\emod{\alpha}\true \ou \emod{\alpha}X & \text{\emph{ if }} \nullable{X} \\ \emod{\alpha}X & \text{\emph{ if not }} \nullable{X}\end{array}\right.\\
\end{array}$$
according to the predicate $\nullable{\cdot}$ (defined in Section~\ref{analysis:xml-types-embedding} of previous chapter) which indicates whether a type contains the empty tree.

Note that the translation of a regular tree type uses only downward modalities 
since it describes the allowed subtrees at a given context. No additional 
restriction is imposed on the context from which the type definition starts.  
In particular, navigation is allowed in the upward direction so that type constraints for which only partial knowledge in a given 
direction is known can be supported. However, when the position of the root is known, conditions similar 
to those of absolute paths are added. This is particularly useful when a 
regular type is used by an XPath expression that starts its navigation at the 
root (\(/p\)) since the path will not go above the root of the type (by adding 
the restriction \(\phirootexpr\)).

On the other hand, if the type is compared with another type (typically to check 
inclusion of the result of an XPath expression in this type), then there is no 
restriction as to where the root of the type is (the translation does not 
impose the chosen node to be at the root). This is particularly useful since 
an XPath expression usually returns a set of nodes deep in the tree which 
may be compared to this partially defined type.

\mychapter{Satisfiability-Testing Algorithm}
\label{xml-calculus:sec:algo}


\section{Introduction}
This chapter presents the algorithm for deciding the logic introduced in previous chapter. It is shown sound and complete, and the time complexity boundary is proved. 
The combination of all these ingredients leads to the main result: a satisfiability algorithm for a logic for finite trees whose time complexity is a simple exponential of the size of a formula. 
With these proofs, a practically effective system for solving the 
satisfiability of a formula is described. The system has been experimented with some 
decision problems such as XPath containment, emptiness, overlap, and coverage, with or 
without type constraints. 

\paragraph{Chapter Outline}
Some preliminary notions are defined in Section~\ref{xml-calculus:preliminary-definitions}. The satisfiability algorithm is then introduced in Section \ref{xml-calculus:core-algorithm} and proven correct in Section \ref{xml-calculus:correctness}, with details of the implementation discussed in Section 
\ref{xml-calculus:sec:impl}.  Applications for type checking are described in Section 
\ref{xml-calculus:sec:experiments} along with some experimental results, before the approach outcome is discussed in \ref{xml-calculus:sec:outcome}.

\section{Preliminary Definitions} 
\label{xml-calculus:preliminary-definitions}
The \emph{unwinding} of a formula $\phi=(\murec{X_i}{\phi_i}{\psi})$, noted $\notation{\expand{\phi}}{Unwinding of a formula $\phi$}{n:unw}$, is defined as $\expand{\phi} \eqdef 
\psi\{\overline{\subst{\murec{X_i}{\phi_i}{X_i}}{X_i}}\}$ which denotes the 
formula $\psi$ in which every occurrence of a $X_i$ is replaced by 
$(\murec{X_i}{\phi_i}{X_i})$. 

The \emph{Fisher-Ladner closure} $\notation{\cl{\psi}}{Fisher-Ladner closure of $\psi$}{n:flclosure}$ of a formula $\psi$ is defined as 
the set of all subformulas of $\psi$ where fixpoint formulas are additionally 
unwound once. Specifically, the relation $\erel \subseteq 
\mulogic \times \mulogic$ is defined as the least relation that satisfies the 
following:
\begin{itemize}
\item $\phi_1 \et \phi_2 \erel \phi_1$, $\phi_1 \et \phi_2 \erel \phi_2$
\item $\phi_1 \ou \phi_2 \erel \phi_1$, $\phi_1 \ou \phi_2 \erel \phi_2$
\item $\emod{a}\phi' \erel \phi'$
\item $\murec{X_i}{\phi_i}{\psi} \erel \expand{\murec{X_i}{\phi_i}{\psi}}$
\end{itemize}
The closure $\cl{\psi}$ is the smallest set $S$ that contains $\psi$ and closed under the relation $\erel$, i.e. if $\phi_1 \in S$ and $\phi_1 \erel \phi_2$ then $\phi_2 \in S$.

$\notation{\ap{\psi}}{Set of atomic propositions for $\psi$}{n:setofatoms}$ denotes the set of atomic propositions used in $\psi$ along with 
an other name, \(\nodelabelother\), representing atomic propositions not 
occurring in \(\psi\). 

The extended closure is defined as $\notation{\extendedcl{\psi}}{Extended Fisher-Ladner closure of $\psi$}{n:extendedflclosure} = \cl{\psi} \cup \{ \neg \phi \mid \phi \in \cl{\psi}\}$.
Every formula $\phi \in \extendedcl{\psi}$ can be seen as a boolean 
combination of formulas of a set called the Lean of $\psi$, inspired from 
\cite{vardi-jancl06}. This set is noted $\notation{\lean{\psi}}{Lean of $\psi$}{n:lean}$ and defined as follows:
\begin{multline*}
\lean{\psi} = \setof{\emod{a} \true}{ a \in \domProg} \cup  \ap{\psi}\\ \cup 
\{\startatom\} \cup  \setof{\emod{a} \phi}{\emod{a} \phi \in  \cl{\psi}}
\end{multline*}

\label{xml-calculus:lean}

A $\psi$-\emph{type} (or simply a ``\emph{type}'') (Hintikka set in the temporal logic literature) is a set $t \subseteq \lean{\psi}$ such that: 
\begin{itemize}
\item $\forall \emod{a}\phi \in \lean{\psi}, \emod{a}\phi \in t \ourimplies  
  \emod{a}\true \in t$ (modal consistency);
\item $\emod{\overline{1}}\true \notin t \ou \emod{\overline{2}}\true \notin 
  t$ (a tree node cannot be both a first child and a second child);
\item exactly one atomic proposition $\atomprop \in t$ (XML labeling); the 
  function $\atomin{t}$ is used to return the atomic proposition of a type $t$;
\item \(\startatom\) may belong to \(t\).
\end{itemize}
 $\notation{\types{\psi}}{Set of $\psi$-types}{n:psitypes}$ denotes the set of $\psi$-types. For a \(\psi\)-type \(t\), the 
\emph{complement} of \(t\) is the set \(\lean{\psi} \setminus t\).

A type determines a truth assignment of every formula in $\extendedcl{\psi}$ 
with the relation $\notation{\extendedin}{Truth assignment relation}{n:extendedin}$ defined in Figure \ref{xml-calculus:fig:truth}.
\begin{figure}
\begin{mathpar}
  \inferrule{ }{\inbase{\true}{t}{\emptyset}{\emptyset}} \and
  \inferrule{ \phi \in \lean{\psi} \\ \phi \in t }{ 
  \inbase{\phi}{t}{\{\phi\}}{\emptyset} } \and
  \inferrule{ \inbase{\phi_1}{t}{T_1}{F_1} \\ \inbase{\phi_2}{t}{T_2}{F_2} }{ 
  \inbase{\phi_1 \et \phi_2}{t}{T_1 \cup T_2}{F_1 \cup F_2} } \and
  \inferrule{ \inbase{\phi_1}{t}{T_1}{F_1}}{ \inbase{\phi_1 \ou 
  \phi_2}{t}{T_1}{F_1} } \and
  \inferrule{ \inbase{\phi_2}{t}{T_2}{F_2}}{ \inbase{\phi_1 \ou 
  \phi_2}{t}{T_2}{F_2} } \and
  \inferrule{ \notinbase{\phi}{t}{T}{F} }{ \inbase{\neg \phi}{t}{T}{F} } \and
  \inferrule{ \inbase{\expand{\murec{X_i}{\phi_i}{\psi}}}{t}{T}{F} }{ 
  \inbase{\murec{X_i}{\phi_i}{\psi}}{t}{T}{F} } \and
  \inferrule{ \phi \in \lean{\psi} \\ \phi \not\in t }{ 
  \notinbase{\phi}{t}{\emptyset}{\{\phi\}} } \and
  \inferrule{ \notinbase{\phi_1}{t}{T_1}{F_1} \\ 
  \notinbase{\phi_2}{t}{T_2}{F_2} }{ \notinbase{\phi_1 \ou \phi_2}{t}{T_1 \cup 
  T_2}{F_1 \cup F_2} } \and
  \inferrule{ \notinbase{\phi_1}{t}{T_1}{F_1}}{ \notinbase{\phi_1 \et 
  \phi_2}{t}{T_1}{F_1} } \and
  \inferrule{ \notinbase{\phi_2}{t}{T_2}{F_2}}{ \notinbase{\phi_1 \et 
  \phi_2}{t}{T_2}{F_2} } \and
  \inferrule{ \inbase{\phi}{t}{T}{F} }{ \notinbase{\neg \phi}{t}{T}{F} } \and
  \inferrule{ \notinbase{\expand{\murec{X_i}{\phi_i}{\psi}}}{t}{T}{F} }{ 
  \notinbase{\murec{X_i}{\phi_i}{\psi}}{t}{T}{F} }
\end{mathpar}
\caption{Truth Assignment of a Formula}
\label{xml-calculus:fig:truth}
\end{figure}

Note that such derivations are finite because the number of naked 
\(\murec{X_i}{\phi_i}{\psi}\) (that do not occur under modalities) strictly 
decreases after each expansion.

The notation \(\phi \extendedin t\) is often used if there are some \(T,F\) such that 
\(\inbase{\phi}{t}{T}{F}\). A formula $\phi$ is true at a type $t$ 
iff $\phi \extendedin t$. 

The the truth status of a formula is now related to the truth assignment of its \(\psi\)-types.

\begin{prop}
  If \(\inbase{\phi}{t}{T}{F}\), then \(T \subseteq t\), \(F \subseteq  
  \lean{\psi}\setminus t\), and \(\bigwedge_{\psi \in T} \psi \et 
  \bigwedge_{\psi \in F} \neg \psi \implies \phi\).
  If \(\notinbase{\phi}{t}{T}{F}\), then \(T \subseteq t\), \(F 
  \subseteq  \lean{\psi}\setminus t\), and \(\bigwedge_{\psi \in T} \psi \et 
  \bigwedge_{\psi \in F} \neg \psi \implies \neg \phi\).
\end{prop}

\begin{mypfo}
  Immediate by induction on the derivations.
\end{mypfo}

A compatibility relation is now defined between types. This relation establishes which formulas must hold in a type in order for it to be a witness for a modal formula.

$\hiddennotation{\Delta_a(\cdot,\cdot)}{Compatibility relation for two $\psi$-types}{n:delta}$
\begin{defn}[Compatibility relation]:  
  Two types \(t,t'\) are \emph{compatible} under \(a \in \domFProg\), written 
  $\Delta_a(t,t')$, iff
\begin{align*}
\forall \emod{a}\phi \in \lean{\psi}, \emod{a}\phi \in t &\ourequiv \phi 
\extendedin t' \\
  \forall \emod{\overline{a}}\phi \in \lean{\psi}, \emod{\overline{a}}\phi \in 
  t' &\ourequiv \phi \extendedin t
\end{align*}
\end{defn}

\section{The Algorithm}
\label{xml-calculus:core-algorithm}





The algorithm works on sets of triples of the form $(t, w_1, w_2)$ where $t$ 
is a type, and $w_1$ and $w_2$ are sets of types which represent all possible 
witnesses for $t$ according to relations \(\Delta_1\) and \(\Delta_2\).

The algorithm proceeds in a bottom-up approach, repeatedly adding new triples until a satisfying model is found (i.e. a triple whose first component is a type implying the formula), or until no more triple can be added. Each iteration of the algorithm builds types representing deeper trees (in the \(1\) and \(2\) direction) with pending backward modalities that will be fulfilled at later iterations. Types with no backward modalities are satisfiable, and if such a type implies the formula being tested, then it is satisfiable. The main iteration is as follows:
$$\begin{array}{l}
X \leftarrow \emptyset \\
\instr{repeat} \\
\quad X' \leftarrow X \\
\quad X \leftarrow \func{Upd}(X') \\
\quad \instr{if} \func{FinalCheck}(\psi,X) ~\instr{then} \\
\quad \quad \instr{return} \text{``}\psi \text{ is satisfiable}\text{''} \\
\instr{until} X=X' \\
\instr{return} \text{``}\psi \text{ is unsatisfiable}\text{''} \\
\end{array}$$
where $X \subseteq \types{\psi} \times \powerset{\types{\psi}} \times \powerset{\types{\psi}}$ and the operations $\func{Upd}(\cdot)$ and $\func{FinalCheck}(\cdot)$ are defined on Figure~\ref{xml-calculus:fig:algo-func}.

\begin{figure*}[h]
$\begin{array}{lll}
  \func{Upd}(X) \eqdef X & \cup ~ \multilinesetof{(t,\func{w}_1(t,\unmarked{X}) ,\func{w}_2(t,\unmarked{X}))}{ & \startatom \notin t \subseteq \types{\psi}\\ 
    && \et \emod{1}\true \in t \ourimplies \func{w}_1(t,\unmarked{X})  \neq \emptyset \\ 
    && \et \emod{2}\true \in t \ourimplies \func{w}_2(t,\unmarked{X})  \neq \emptyset} \\
  & \cup ~ \multilinesetof{\marked{(t,\func{w}_1(t,\unmarked{X}) ,\func{w}_2(t,\unmarked{X}))}}{ & \startatom \in t \subseteq \types{\psi} \\
    && \et \emod{1}\true \in t \ourimplies \func{w}_1(t,\unmarked{X})  \neq \emptyset \\
    && \et \emod{2}\true \in t \ourimplies \func{w}_2(t,\unmarked{X})  \neq \emptyset} \\
  & \cup ~ \multilinesetof{\marked{(t,\func{w}_1(t,\marked{X}) ,\func{w}_2(t,\unmarked{X}))}}{& \startatom \notin t \subseteq \types{\psi} \\
    && \et \emod{1}\true \in t \ourimplies \func{w}_1(t,\marked{X})  \neq \emptyset \\
    && \et \emod{2}\true \in t \ourimplies \func{w}_2(t,\unmarked{X})  \neq \emptyset} \\
  & \cup ~ \multilinesetof{\marked{(t,\func{w}_1(t,\unmarked{X}) ,\func{w}_2(t,\marked{X}))}}{& \startatom \notin t \subseteq \types{\psi} \\
    && \et \emod{1}\true \in t \ourimplies \func{w}_1(t,\unmarked{X})  \neq \emptyset \\
    && \et \emod{2}\true \in t \ourimplies \func{w}_2(t,\marked{X})  \neq \emptyset}
\end{array}$

\begin{align*}
\func{w}_a(t,X) & \eqdef \setof{\tripletype{x}}{x \in X \et \emod{\overline{a}}\true \in 
\tripletype{x} \et \Delta_a(t,\tripletype{x})}\\
\func{FinalCheck}(\psi, X) & \eqdef \exists x \in \marked{X}, 
\descofsatisfies{x}{\psi} \et \forall a \in \domBProg, \emod{a}\true \notin 
\tripletype{x}  \\
\descofsatisfies{(t,w_1,w_2)}{\psi} & \eqdef \psi \extendedin t
\ou \exists x',  \descofsatisfies{x'}{\psi} \et (x' \in w_1 \ou x' \in w_2) \\
\end{align*}
\begin{align*}
\marked{X} &\eqdef \setof{x \in X}{x = \marked{(\allcases,\allcases,\allcases)}}\\
\unmarked{X}&\eqdef \setof{x \in X}{x = (\allcases,\allcases,\allcases)}\\
\tripletype{(t,w_1,w_2)}  & \eqdef t
\end{align*}
\caption{Operations used by the Algorithm.}\label{xml-calculus:fig:algo-func}
\end{figure*}

$\notation{X^i}{Set of triples after $i$ iterations}{n:setoftriples}$ and $\notation{T^i}{Set of $\psi$-types after $i$ iterations}{n:setofpsitypes}$ respectively denote the set of triples and the set of types after $i$ 
iterations: $T^i=\setof{\tripletype{x}}{x\in X^i}$. Note that $T^{i+1}$ is the 
set of types for which at least one witness belongs to $T^i$. 

\subsection{Example Run of the Algorithm}

\begin{figure*}
\begin{tikzpicture}

\draw (0,0)   node [text=white](bidon1) {3) $T^0=\emptyset$};
\draw (0,3.4) node (bidon2) {};
\draw [color=white](bidon1)--(bidon2);
\draw (0,0) node(t0) {3) $T^0=\emptyset$};

\draw (0,1) node(t1) {4) $T^1=~$}

      (0.5,1) node(t1left)  {$\{$}

      (1,1) node(t1s1) [ball color=white,circle,text=black] {$\sigma$}
      (2,1) node(t1s3) [ball color=white,circle,text=black]{$\sigma$}
      (3,1) node(t1s4) [ball color=white,circle,text=black]{$\sigma$}

      (4,1) node(t1a1) [ball color=white,circle,text=black]{$a$}
      (5,1) node(t1a3) [ball color=white,circle,text=black]{$a$}
      (6,1) node(t1a4) [ball color=white,circle,text=black]{$a$}

      (7,1) node(t1b1) [ball color=white,circle,text=black]{\begin{small}$b$\end{small}}
      (8,1) node(t1b3) [ball color=white,circle,text=black]{\begin{small}$b$\end{small}}
      (9,1) node(t1b4) [ball color=white,circle,text=black]{\begin{small}$b$\end{small}}

      (10,1) node(t1right)  {$\}$};

\draw (1.5,2) node(f2) {}
      (3.5,2) node(f3) {}
      
      (4.5,2) node(f5) {}
      (6.5,2) node(f6) {}
      
      (7.5,2) node(f8) {}
      (9.5,2) node(f9) {};

\draw [thick, ->, orange,dashed]
      (f2) -- (t1s3);   
\draw [thick, ->, cyan,dashed]
      (f3) -- (t1s4) ;
      
\draw [thick, ->, orange,dashed]      
      (f5) -- (t1a3);
\draw [thick, ->, cyan,dashed]
      (f6) -- (t1a4);

\draw [thick, ->, orange,dashed]
      (f8) -- (t1b3);
\draw [thick, ->, cyan,dashed]
      (f9) -- (t1b4);
\draw (4,3) node(t2c2) {5) Does};
\draw (7.4,3) node(t2d2) {belong to $T^2$ ?};
\draw (5,3) node(t22) [ball color=white,circle,text=black]{$a$};
\draw (5.7,3) node(t2b2) [text=cyan]{$\emod{1}\phi$};
\draw (4.5,2) node(t2w2) {};
\draw [thick, ->, cyan, dashed]
      (t22) -- (t2w2);
\draw (5.7,3) node(t2bred2) [text=red]{$\emod{1}\phi$};
\draw (3,1) node(dnw12) [ball color=red,circle,text=black]{$\sigma$};
\draw (6,1) node(dnw22) [ball color=red,circle,text=black]{$a$};
\draw (9,1) node(w12) [ball color=green,circle,text=black]{\begin{small}$b$\end{small}};
\draw (9.8,3) node(answer2) [text=green]{Yes! Witness:};
\draw (8.4,2.15) node(t2c2translated) {7) Does};
\draw (9.6,2.15) node(t22translated) [ball color=white,circle,text=black]{$a$};
\draw (10.5,2.15) node(secondq2) {satisfy};
\draw (10.6,1.8) node(secondq2b) {$\psi ?$};
\draw (9.6,2.15) node(t22g) [ball color=green,circle,text=black]{$a$};
\draw (10.6,1.4) node(answer2) [text=green]{Yes!};
\draw [thick, ->, cyan]
      (f9) -- (t1b4);
\draw (0.3,2) node(t2) {6) $\left|T^2\right|=81$};
\draw (1.1,2.6) node(t3)[text=green] {8) $\rightarrow$ return satisfiable!};
%
%
\end{tikzpicture}

$\begin{array}{l}\\
2)~ \lean{\psi}=\left\{\emod{1}\true,  ~ \emod{\overline{1}}\true,~ \emod{2}\true, ~ \emod{\overline{2}}\true, ~ \sigma, ~ a, ~ b, ~ \emod{1} \phi, ~ \emod{2}\phi \right\} \\ \\
1)~ \psi = a \et \emod{1}\phi$ with $\phi = \mu X . (b \et \startatom) \ou \emod{2}X \equiv \expand{\phi}=(b \et \startatom) \ou \emod{2}\phi
\end{array}$

\caption{Run of the Algorithm for Checking Emptiness of \texttt{self$::$b/parent$::$a}}\label{fig:examplerun}
\end{figure*}
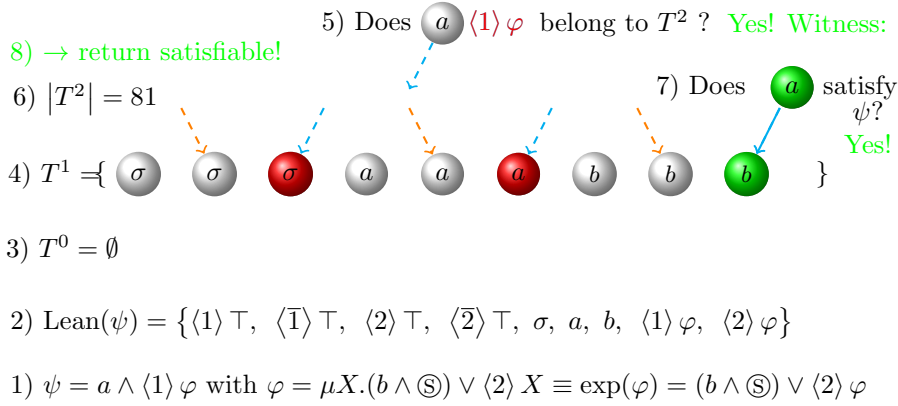

Figure~\ref{fig:examplerun} illustrates a run of the algorithm for checking the non-emptiness of the simple XPath expression $e=\texttt{self$::$b/parent$::$a}$. This expression is first compiled into the logic as explained in section~\ref{xml-calculus:sec:xpath}. The resulting formula $\psi = \mucalcE{e}{\true}$ is shown on Figure~\ref{fig:examplerun} (step 1). As a second step, $\lean{\psi}$ is computed. Then the fixpoint computation starts: the set of types $T^1$ contains all possible leaves (step 3). For each type in $T^2 \setminus T^1$, a witness must be found in $T^1$. The algorithm notably finds a witness for a particular $\psi$-type $t$ such that $a \et \emod{1}\phi \in t$ (step 5). 
$T^2$ finally contains 81 $\psi$-types (step 6). $t$ happens to satisfy the initial formula $\psi$ (step 7), therefore the algorithm stops just after computing $T^2$ (step 8) because the structure built by connecting $t$ and its witness (as drawn on Figure~\ref{fig:examplerun}) is a finite tree which contains a node on which $\psi$ is satisfied. Thus \texttt{self$::$b/parent$::$a} is satisfiable.

\section{Correctness and Complexity}
\label{xml-calculus:correctness}
In this section the correctness of the satisfiability testing 
algorithm, is proved, and it is shown that its time complexity is 
$2^{O(\setcardinal{\lean{\psi}})}$.

\begin{thm}[Correctness] \label{xml-calculus:theorem1}
The algorithm decides satisfiability of $\mulogic$ formulas over finite focused trees.
\end{thm}

\paragraph{Termination}
For $\psi \in \mulogic$, since $\cl{\psi}$ is a finite set, $\lean{\psi}$ and $\powerset{\lean{\psi}}$ are also finite. Furthermore, $\func{Upd}(\cdot)$ is monotonic and each $X^i$ is included in the finite set $\types{\psi} \times \powerset{\types{\psi}} \times \powerset{\types{\psi}}$, therefore the algorithm terminates. To finish the proof, it thus suffices to prove soundness and completeness.

\paragraph{Preliminary Definitions for Soundness}

First, a notion of partial satisfiability is introduced for a formula. In this partial satisfiability notion, 
backward modalities are only checked up to a given level. A formula $\phi$ is 
partially satisfied iff $\fpinterp{\phi}{V}{0} \neq \emptyset$ as defined in 
Figure \ref{xml-calculus:fig:fpinterp}.

$\hiddennotation{\fpinterp{\cdot}{\cdot}{\cdot}}{Partial satisfiability of a formula}{n:partialsatisf}$

\begin{figure}
\begin{align*}
\fpinterp{\true}{V}{n} &\eqdef \focusedtrees &
\fpinterp{X}{V}{n} &\eqdef V(X)\\
\fpinterp{\phi \ou \psi}{V}{n} &\eqdef \fpinterp{\phi}{V}{n} \cup 
\fpinterp{\psi}{V}{n} &
\fpinterp{p}{V}{n} &\eqdef \setof{\focusedtreevar}{\fname{(\focusedtreevar)} = 
p}\\
\fpinterp{\phi \et \psi}{V}{n} &\eqdef \fpinterp{\phi}{V}{n} \cap 
\fpinterp{\psi}{V}{n}&
\fpinterp{\neg p}{V}{n} &\eqdef 
\setof{\focusedtreevar}{\fname{(\focusedtreevar)} \neq p} \\
\fpinterp{\emod{\overline{1}} \phi}{V}{0} &\eqdef \focusedtrees &
\fpinterp{\startatom}{V}{n} &\eqdef \setof{\focusedtreevar}{\focusedtreevar 
= \focusedtree{\tree{\starttrue{\nodelabel}}{\treelistvar}}{\contextvar}} \\
\fpinterp{\emod{\overline{2}} \phi}{V}{0} &\eqdef \focusedtrees &
\fpinterp{\neg \startatom}{V}{n} &\eqdef 
\setof{\focusedtreevar}{\focusedtreevar = 
\focusedtree{\tree{\nodelabel}{\treelistvar}}{\contextvar}}
\end{align*}
\begin{align*}
\fpinterp{\emod{\overline{1}} \phi}{V}{n>0} &\eqdef 
\setof{\focusedtreevar\emod{1}}{\focusedtreevar \in \fpinterp{\phi}{V}{n-1} 
\land \isdefined{\focusedtreevar\emod{1}}}\\
\fpinterp{\emod{\overline{2}} \phi}{V}{n>0} &\eqdef 
\setof{\focusedtreevar\emod{2}}{\focusedtreevar \in \fpinterp{\phi}{V}{n-1} 
\land \isdefined{\focusedtreevar\emod{2}}}\\
\fpinterp{\emod{1} \phi}{V}{n} &\eqdef \setof{\focusedtreevar\emod{\overline 
1}}{\focusedtreevar \in \fpinterp{\phi}{V}{n+1} \land 
\isdefined{\focusedtreevar\emod{\overline 1}}}\\
\fpinterp{\emod{2} \phi}{V}{n} &\eqdef \setof{\focusedtreevar\emod{\overline 
2}}{\focusedtreevar \in \fpinterp{\phi}{V}{n+1} \land 
\isdefined{\focusedtreevar\emod{\overline 2}}}\\
\fpinterp{\neg \emod{a} \true}{V}{n} &\eqdef 
\setof{\focusedtreevar}{\isundefined{\focusedtreevar\emod{a}}}\\
\fpinterp{\murec{X_i}{\phi_i}{\psi}}{V}{n} &\eqdef \text{let \(T_i = 
\left(\bigcap \setof{\overline{T_i} \subseteq 
\overline{\focusedtrees}}{\fpinterp{\overline{\phi_i}}{V[\overline{T_i/X_i}]}{n} 
\subseteq \overline{T_i}}\right)_i\)}
\\
&\qquad \text{ in } \fpinterp{\psi}{V[\overline{T_i/X_i}]}{n}
\end{align*}
\caption{Partial Satisfiability}
\label{xml-calculus:fig:fpinterp}
\end{figure}

For a type $t$, $\notation{\typeformula{t}}{Most constrained formula for a $\psi$-type $t$}{n:mostcontrained}$ denotes the most constrained formula, where 
atoms are taken from \(\lean{\psi}\). In the following, \(\startatomopt\) 
stands for \(\startatom\) if \(\startatom \in t\), and for \(\neg 
\startatom\) otherwise.
\begin{equation*}
\typeformula{t}=\atomin{t} \et \bigwedge _{\atomprop \in \atomprops, \atomprop \notin t} \neg 
\atomprop \et \startatomopt \et \bigwedge_{\emod{a}\phi \in t} \emod{a}\phi \et 
\bigwedge _{\emod{a}\phi \notin t} \neg\emod{a}\phi
\end{equation*}

A notion of \emph{paths} is now introduced. Paths written \(\notation{\pathvar}{Path (concatenation of modalities)}{n:path}\) are 
concatenations of modalities: the empty path is written \(\emptypath\), and 
path concatenation is written \(\conspath{\pathvar}{a}\).

Every path may be given a \emph{depth}:
\begin{align*}
\depth{\emptypath} & \eqdef 0 \\
\depth{\conspath{\pathvar}{a}} & \eqdef \depth{\pathvar}+1 \quad \text{if } a 
\in \{1,2\} \\
\depth{\conspath{\pathvar}{a}} & \eqdef \depth{\pathvar}-1 \quad \text{if } a 
\in \{\overline{1},\overline{2}\}
\end{align*}

A forward path is a path that only mentions forward modalities.

A tree of types \(\treetypevar\) is defined as a tree whose nodes are types, 
\(\treetypenode{\treetypevar} = t\), with at most two children, 
\(\treetypedown{\treetypevar}\) and \(\treetyperight{\treetypevar}\). The 
navigation in tree of types is trivially extended to forward paths. A tree of 
types is \emph{consistent} iff for every forward path \(\pathvar\) and for 
every child \(a\) of \(\treetypevar\emod{\pathvar}\), the following holds: 
\(\treetypenode{\treetypevar\emod{\pathvar}} = t\), 
\(\treetypenode{\treetypevar\emod{\pathvar a}} = t'\) implies \(\emod{a}\true 
\in t\),
\(\emod{\overline{a}}\true \in t'\), and \(\Delta_a(t,t')\).

Given a consistent tree of types \(\treetypevar\), a dependency 
graph is now defined. In this graph, nodes are pairs of a forward path \(\pathvar\) and a formula in 
\(t = \treetypenode{\treetypevar\emod{\pathvar}}\) or the negation of a 
formula in the complement \(t\).  The directed edges of the graph are 
modalities consistent with the tree.  For every \( (\pathvar,\phi) \) in the 
nodes the following edges are built:
\begin{itemize}
  \item \(\phi \in \ap{\psi} \cup \neg \ap{\psi} \cup \{\startatom, \neg 
    \startatom, \emod{a}\true, \neg \emod{a}\true\}\): no edge
  \item \(\pathvar = \emptypath, \phi = \emod{\overline{a}}\phi'\) with \(a 
    \in \{1,2\}\): no edge
  \item \(\pathvar = \pathvar' a, \phi = \emod{a'}\phi'\):
    let \(t = \treetypenode{\treetypevar\emod{\pathvar}}\).
    Let first consider the case where \(a' \in \{1,2\}\) and let \(t' = 
    \treetypenode{\treetypevar\emod{\pathvar a'}}\). As \(\treetypevar\) is
    consistent, \(\phi' \extendedin t'\) hence there are \(T,F\) such 
    that \(\inbase{\phi'}{t'}{T}{F}\) with \(T\) a subset of \(t'\), and \(F\) 
    a subset of the complement of \(t'\). For every \(\phi_T \in T\) an 
    edge \(a'\) is added to \( (\pathvar a', \phi_T)\), and for every \(\phi_F \in F\) 
    an edge \(a'\) is added to \( (\pathvar a', \neg \phi_F)\).
    Consider now the case where \(a' \in \{\overline{1},\overline{2}\}\) 
    and first show that \(a' = \overline{a}\). As \(\treetypevar\) is 
    consistent, \(\emod{\overline{a}} \true\) in \(t\). Moreover, as 
    \(t\) is a tree type, it must contain \(\emod{a'} \true\). As \(a'\) is a 
    backward modality, it must be equal to \(\overline{a}\) as at most one may 
    be present. Hence \(\pathvar' a a' = \pathvar'\) holds. Let \(t' = 
    \treetypenode{\treetypevar\emod{\pathvar'}}\). By consistency, 
    \(\phi' \extendedin t'\), hence \(\inbase{\phi'}{t'}{T}{F}\) and  
    edges are added as in the previous case: to \( (\pathvar', \phi_T) \) and to \( 
    (\pathvar', \neg \phi_F)\).
  \item \(\pathvar = \pathvar' a, \phi = \neg \emod{a'}\phi'\):
    let \(t = \treetypenode{\treetypevar\emod{\pathvar}}\). If 
    \(\emod{a'}\true\) is not in \(t\) then no edge is added. Otherwise, one
    proceeds as in the previous case. For downward modalities, let \(t' = 
    \treetypenode{\treetypevar\emod{\pathvar a'}}\) and compute
    \(\notinbase{\phi'}{t'}{T}{F}\) which is known to hold by consistency. Edges are then added to \( (\pathvar a', \phi_T)\) and to \( (\pathvar a', \neg 
    \phi_F)\) as before. For upward modalities, as \(\emod{a'}\true\) holds
    in \(t\), one must have \(a' = \overline{a}\) and let \(t' = 
    \treetypenode{\treetypevar\emod{\pathvar'}}\).  
    \(\notinbase{\phi'}{t'}{T}{F}\) is computed and edges are added to \( (\pathvar', 
    \phi_T) \) and to \( (\pathvar', \neg \phi_F)\) as before.
\end{itemize}

\begin{lem}
The dependency graph of a consistent tree of types of a cycle-free formula is 
cycle free.
\end{lem}

\begin{mypfo}
  The proof proceeds by induction on the depth of the cycle, relying on the 
  fact that the dependency graph is consistent with the tree structure (i.e.  
  if a \(1\) edge reaches a node, no \(\overline{2}\) edge may leave this 
  node). The induction case is trivial: if there is a cycle of depth \(n\), 
  there must be a cycle of depth \(n-1\), a contradiction.

  The base case is for a cycle of depth \(1\). One case is described, where the 
  cycle is \( (\pathvar, \emod{1}\phi) \longrightarrow^1 (\pathvar 1, 
  \emod{\overline{1}}\psi) \longrightarrow^{\overline{1}} (\pathvar, 
  \emod{1}\phi) \). As \(\phi\) must be a subformula of \(\psi\) and \(\psi\) 
  a subformula of \(\phi\), they are both recursive formula. An analysis of 
  the shape of \(\phi\), based on the derivations \(\inbase{\phi}{t}{T}{F}\) 
  and \(\inbase{\psi}{t'}{T'}{F'}\) with \(\emod{1}\psi \in T\) and 
  \(\emod{\overline{1}}\phi \in T'\) then shows that \(\phi\) is not a 
  cycle-free formula, a contradiction.
\end{mypfo}

\begin{lem}[Soundness]
Let \(T\) be the result set of the algorithm. For any type $t \in T$ and any 
\(\phi\) such that \(\phi \extendedin t\), then 
\(\fpinterp{\phi}{\emptyset}{0} \neq \emptyset\).
\end{lem}

\begin{mypfo}
 
The proof proceeds by induction on the number of steps of the algorithm. For 
every \(t\) in \(T^n\) and every witness tree \(\treetypevar\) rooted at \(t\) 
built from \(X^n\), one can show that \(\treetypevar\) is a consistent tree type 
and one can build a focused tree \(\focusedtreevar\) that is rooted (i.e. of the 
shape 
\(\focusedtree{\tree{\startunk{\nodelabel}}{\treelistvar}}{\contexttop{\treelistnil}{\treelistvar'}}\)).
The tree \(\focusedtreevar\) is
in the partial interpretation of \(\typeformula{t}\): 
\(\focusedtreevar\emod{\pathvar} \in 
\fpinterp{\typeformula{\treetypenode{\treetypevar\emod{\pathvar}}}}{\emptyset}{\depth{\pathvar}}\) 
for any path \(\pathvar\) whose depth is \(0\) or more, and  
\(\focusedtreevar\) contains the context marker only if \(\startatom\) 
occurs in \(\treetypevar\). Then one shows that for all \(\phi \extendedin t\), 
\(\focusedtreevar \in \fpinterp{\phi}{\emptyset}{0}\) holds.

The base case is trivial by the shape of \(t\): it may only contain backward 
modalities (trivially satisfied at level \(0\)), one atomic proposition, and 
one context proposition. Moreover there is only one tree of witnesses to 
consider, the tree whose only node is \(t\). If the atomic proposition is 
\(\atomprop\), then the focused tree returned is either 
\(\focusedtree{\tree{\starttrue{\atomprop}}{\treelistnil}}{\contexttop{\treelistnil}{\treelistnil}}\) 
or 
\(\focusedtree{\tree{\atomprop}{\treelistnil}}{\contexttop{\treelistnil}{\treelistnil}}\) 
depending on the context proposition.

In the inductive case, every witness types for both downward 
modalities, \(t_1\) and \(t_2\) are considered. For each of them, every tree type 
\(\treetypevar_1\) and \(\treetypevar_2\) are considered and a tree type rooted at 
\(t\) is built which is consistent by definition of the algorithm. By induction,  
 \(\focusedtreevar_1\) and \(\focusedtreevar_2\) such that 
\(\focusedtreevar_1\emod{\pathvar} \in 
\fpinterp{\typeformula{\treetypenode{\treetypevar\emod{1 
\pathvar}}}}{\emptyset}{\depth{\pathvar}}\) and
\(\focusedtreevar_2\emod{\pathvar} \in 
\fpinterp{\typeformula{\treetypenode{\treetypevar\emod{2 
\pathvar}}}}{\emptyset}{\depth{\pathvar}}\) for any path \(\pathvar\) whose 
depth is \(0\) or more. If either \(\treetypevar_1\) or \(\treetypevar_2\) 
contains \(\startatom\), then \(\focusedtreevar_1\) or \(\focusedtreevar_2\) 
contains the context marker by induction. Moreover, by definition of the 
algorithm, it is the case for only one of them and \(\startatom\) is not in 
\(t\).

Let \(\focusedtreevar_1\) be 
\(\focusedtree{\tree{\startunk{\nodelabel_1}}{tl_1}}{\contexttop{\treelistnil}{tr_1}}\) 
and \(\focusedtreevar_2\) be
\(\focusedtree{\tree{\startunk{\nodelabel_2}}{tl_2}}{\contexttop{\treelistnil}{tr_2}}\).
Let \(\focusedtreevar = 
\focusedtree{\tree{\startunk{\atomin{t}}}{\treelistcons{\tree{\startunk{\nodelabel_1}}{tl_1}}{tr_1}}}{\contexttop{\treelistnil}{\treelistcons{\tree{\startunk{\nodelabel_2}}{tl_2}}{tr_2}}}\) 
where \(\startunk{\atomin{t}}\) is \(\starttrue{\atomin{t}}\) if 
\(\startatom \in t\), and \(\atomin{t}\) otherwise. Note that 
\(\focusedtreevar\) contains exactly one context marker iff \(\startatom \in 
\treetypevar\).

Next, one shows that \(\focusedtreevar_1\emod{\pathvar} \in 
\fpinterp{\typeformula{\treetypenode{\treetypevar\emod{1 
\pathvar}}}}{\emptyset}{\depth{\pathvar}}\) implies \(\focusedtreevar\emod{1 
\pathvar} \in \fpinterp{\typeformula{\treetypenode{\treetypevar\emod{1 
\pathvar}}}}{\emptyset}{\depth{\pathvar}}\), and the same for the other 
modality, by induction on the depth of the path, remarking that every backward 
modality at level \(0\) is trivially satisfied.

Then one proceeds to show that $\focusedtreevar$ satisfies $\typeformula{t}$ at 
level $0$. To do so, a further induction on the dependency tree is needed. Let 
$\pathvar$ be a path of the dependency tree and $\psi$ be a formula at that 
path in the dependency tree, one shows that $\focusedtreevar\emod{\pathvar} \in 
\fpinterp{\psi}{V}{\depth{\pathvar}}$. To do so, one relies on 
$\focusedtreevar\emod{\pathvar} \in \fpinterp{\psi}{V}{\depth{\pathvar}-1}$ if 
\(\depth{\pathvar} \neq 0\). In the base case at depth \(0\), the result is by 
construction as the formula is either a backward modality or an atomic 
formula. In the base case at another depth, the case is immediate by induction 
as the formula has to be an atomic formula whose interpretation does not 
depend on the depth. In the induction case, one concludes by the inductive 
hypothesis and by definition of partial satisfiability.

The proof is concluded by noticing that the final selected type has no backward 
modality, hence \(\fpinterp{\typeformula{t}}{0}{\emptyset} = 
\finterp{\typeformula{t}}{\emptyset}\).

\end{mypfo}

\begin{lem}[Completeness]
For a cycle-free closed formula $\phi \in \mulogic$, if 
$\finterp{\phi}{\emptyset} \neq \emptyset$ then the algorithm terminates   
with a set of triples \(X\) such that \(\func{FinalCheck}(\phi,X)\).
\end{lem}

\begin{mypfo}
  Let \(\focusedtreevar \in \finterp{\phi}{\emptyset}\) be a smallest focused 
  tree validating the formula such that the names occurring in 
  \(\focusedtreevar\) are either also occurring in \(\phi\) or are a single 
  other name \(\nodelabelother\).  By Lemma~\ref{xml-calculus:lem:finite_unfold}, there is 
  a finite unfolding of \(\phi\) such that \(\focusedtreevar\) belongs to its 
  interpretation.  Hence there is a finite satisfiability derivation, defined 
  in Figure \ref{xml-calculus:fig:treeval}, of \(\treeval{\focusedtreevar}{\emptypath}{\phi}\).

$\hiddennotation{\treeval{\cdot}{\pathvar}{\cdot}}{Satisfiability relation}{n:satisfrel}$

  \begin{figure}
  \begin{mathpar}
    \inferrule{ }{ \treeval{\focusedtreevar}{\pathvar}{\true} } \and
    \inferrule{ \fname{(\focusedtreevar)} = \atomprop }{ 
    \treeval{\focusedtreevar}{\pathvar}{\atomprop} } \and
    \inferrule{ \fname{(\focusedtreevar)} \neq \atomprop }{ 
    \treeval{\focusedtreevar}{\pathvar}{\neg \atomprop} } \and
    \inferrule{ }{ 
    \treeval{\focusedtree{\tree{\starttrue{\atomprop}}{\treelistvar}}{\contextvar}}{\pathvar}{\startatom} 
    } \and
    \inferrule{ }{ 
    \treeval{\focusedtree{\tree{\atomprop}{\treelistvar}}{\contextvar}}{\pathvar}{\neg 
    \startatom} } \and
    \inferrule{ \treeval{\focusedtreevar}{\pathvar}{\phi} }{ 
    \treeval{\focusedtreevar}{\pathvar}{\phi \ou \psi} } \and
    \inferrule{ \treeval{\focusedtreevar}{\pathvar}{\psi} }{ 
    \treeval{\focusedtreevar}{\pathvar}{\phi \ou \psi} } \and
    \inferrule{ \treeval{\focusedtreevar}{\pathvar}{\phi} \\ 
    \treeval{\focusedtreevar}{\pathvar}{\psi} }{ 
    \treeval{\focusedtreevar}{\pathvar}{\phi \et \psi} } \and
    \inferrule{ \treeval{\godown{\focusedtreevar}}{\pathvar 1}{\phi} }{ 
    \treeval{\focusedtreevar}{\pathvar}{\emod{\hasleftsucc}\phi} } \and
    \inferrule{ \treeval{\goright{\focusedtreevar}}{\pathvar 2}{\phi} }{ 
    \treeval{\focusedtreevar}{\pathvar}{\emod{\hasrightsucc}\phi} } \and
    \inferrule{ \treeval{\goup{\focusedtreevar}}{\pathvar \overline{1}}{\phi} 
    }{ \treeval{\focusedtreevar}{\pathvar}{\emod{\isleftsucc}\phi} } \and
    \inferrule{ \treeval{\goleft{\focusedtreevar}}{\pathvar 
    \overline{2}}{\phi} }{ 
    \treeval{\focusedtreevar}{\pathvar}{\emod{\isrightsucc}\phi} } \and
    \inferrule{ \isundefined{\focusedtreevar\emod{a}} }{ 
    \treeval{\focusedtreevar}{\pathvar}{\neg \emod{a} \true} } \and
    \inferrule{ 
    \treeval{\focusedtreevar}{\pathvar}{\expand{\murec{X_i}{\phi_i}{\psi}}} }{ 
    \treeval{\focusedtreevar}{\pathvar}{\murec{X_i}{\phi_i}{\psi}} }
  \end{mathpar}
  \caption{Satisfiability Relation}
  \label{xml-calculus:fig:treeval}
\end{figure}

  In the satisfiability derivation, paths are assumed to be normalized (\(1 
  \overline{1} = \emptypath\)). Hence every path is a concatenation of a 
  (possibly empty) backward path \(\pathvar_b\) followed by a forward path 
  \(\pathvar_f\).

  This derivation has the following property, immediate by induction: let 
  \(\focusedtreevar\) the initial focused tree, then 
  \(\treeval{\focusedtreevar'}{\pathvar}{\phi}\) implies \(\focusedtreevar' = 
  \focusedtreevar\emod{\pathvar}\). Hence if 
  \(\treeval{\focusedtreevar_1}{\pathvar}{\phi_1}\) and 
  \(\treeval{\focusedtreevar_2}{\pathvar}{\phi_2}\), then \(\focusedtreevar_1 
  = \focusedtreevar_2\).

  Next, one uses the satisfiability derivation to construct a run of the 
  algorithm that concludes that \(\phi\) is satisfiable. One first associates 
  each path to a type, which one then saturates (adding formulas that are true 
  even though the satisfiability relation does not mention them at that path).  
  One next shows that every formula at a path in the satisfiability relation is 
  implied by the type at that path, and that types are consistent according to 
  the \(\Delta_a(t,t')\) relation. One then concludes that the types are created 
  by a run of the algorithm by induction on the paths.

  More precisely, let first describe how \(t_\pathvar\) is built. Let 
  \(\Phi_\pathvar\) the set of formulas at path \(\pathvar\). One first adds 
  every formula of \(\Phi_\pathvar\) that is in \(\lean{\phi}\), then one 
  completes this set to yield a correct type: if \(\emod{a}\psi \in 
  \Phi_\pathvar\) then one adds \(\emod{a}\true\); for every modality \(a\) for 
  which \(\focusedtreevar\emod{a}\) is defined one adds \(\emod{a}\true\); if 
  there is no atomic proposition in \(\Phi_\pathvar\) then one adds 
  \(\fname{(\focusedtreevar\emod{\pathvar})}\); finally if 
  \(\focusedtreevar\emod{\pathvar}\) has the context marker one adds 
  \(\startatom\).

  One next saturates the types. For every path \(t_\pathvar\) if \(t_{\pathvar 
  a}\) exists, if \(\emod{a} \psi \in \lean{\phi}\), and if \(\psi \extendedin 
  t_{\pathvar a}\) then one adds \(\emod{a} \psi\) to \(t_\pathvar\). This 
  procedure is repeated until it does not change any type. Termination is a 
  consequence of the finite size of the lean and of the number of paths. The 
  resulting types are satisfiable as they are before saturation (since a 
  focused tree satisfies them) and each formula added during saturation is 
  first checked to be implied by the type.
  
  One next shows (*): for any given path \(\pathvar\), if \(\phi_\pathvar \in 
  \Phi_\pathvar\) then \(\phi_\pathvar \extendedin t_\pathvar\), by induction 
  on the satisfiability derivation. Base cases with no negation are immediate 
  by definition of \(t_\pathvar\) as these are formulas of the lean. For base 
  cases with negation, one relies on the fact that 
  \(\focusedtreevar\emod{\pathvar}\) satisfies the formula, hence one cannot 
  for instance have \(\atomprop\) and \(\neg \atomprop\) in \(\Phi_\pathvar\).  
  If \(\neg \emod{a} \true \in \Phi_\pathvar\) then one cannot also have 
  \(\emod{a} \psi \in \Phi_\pathvar\) as \(\pathvar a\) is not a valid path, 
  hence \(\emod{a} \true\) is not in \(t_\pathvar\) thus \(\neg \emod{a} \true 
  \extendedin t_\pathvar\).  The inductive cases of this induction 
  (disjunction, conjunction, recursion) are immediate as they correspond to 
  the definition of \(\cdot \extendedin \cdot\).

  One next shows that for every type \(t_\pathvar\) and \(t_{\pathvar a}\) where 
  \(a\) is a forward modality, \(\emod{\overline{a}} \true \in 
  t_{\pathvar a}\) and \(\Delta_a(t_\pathvar,t_{\pathvar a})\) hold. (Note that, by 
  path normalization, the types considered may be 
  \(t_{\overline{1}\overline{2}}\) and \(t_{\overline{1}}\) for modality 
  \(2\).) The first condition is immediate by construction of \(t_{\pathvar 
  a}\) as \(\focusedtreevar\emod{\pathvar a}\) is defined. For the second 
  condition, let \(\emod{a} \psi \in t_\pathvar\). If \(\emod{a} \psi \in 
  \Phi_\pathvar\), then it occurs in the satisfiability derivation with an 
  hypothesis \(\treeval{\focusedtreevar_{\pathvar a}}{\pathvar a}{\psi}\). In 
  this case  \(\psi \extendedin t_{\pathvar a}\) holds by (*). If \(\emod{a} 
  \psi \notin \Phi_\pathvar\) then it was added during saturation and the 
  result is immediate by construction. Conversely, if \(\psi \extendedin 
  t_{\pathvar a}\) then by saturation \(\emod{a} \psi \in 
  t_\pathvar\). The case \(\emod{\overline{a}} \psi \in 
  t_{\pathvar a}\) is now considered. The proof goes exactly as before, distinguishing the case 
  where the formula is in \(\Phi_{\pathvar a}\) and the case where it was 
  added by saturation.

  One now shows that there is a run of the algorithm that produces these types.  
  The proof proceeds by induction on the paths in the downward direction: if 
  \(t_{\pathvar a}\) has been proven for a partial run for \(a \in 
  \domFProg\), then \(t_\pathvar\) is proven for the next step of the 
  algorithm. Moreover, one shows that \( (t_\pathvar, \{t_{\pathvar 1}\}, 
  \{t_{\pathvar 2}\}) \) is marked iff a forward subtree of 
  \(\focusedtreevar\emod{\pathvar}\) contains the context mark. The base case 
  is for paths with no descendants, hence no witness is required. The 
  algorithm then adds \( (t_\pathvar, \emptyset, \emptyset) \) to its set of 
  types, with a mark iff \(\startatom \in t_\pathvar\), iff 
  \(\focusedtreevar\emod{\pathvar}\) is marked.

  The inductive case is now considered. By induction, a partial run of the 
  algorithm returns \(t_{\pathvar 1}\) and/or \(t_{\pathvar 2}\). One first 
  shows that \(t_\pathvar\) is returned in the next step of the algorithm, 
  taking these two types as witnesses.  One first remarks that if either witness 
  is marked then the other is not and the mark is not at 
  \(\focusedtreevar\emod{\pathvar}\), since there is only one context mark in 
  \(\focusedtreevar\), and if the mark is at 
  \(\focusedtreevar\emod{\pathvar}\), then neither witness is marked. For each 
  child \(a \in \domFProg\),  \(\Delta_a(t_\pathvar, t_{\pathvar a})\) 
  and \(\emod{\overline{a}} \true \in t_{\pathvar a}\), hence the triple \( 
  (t_\pathvar, W_1, W_2) \) with \(t_{\pathvar 1} \in W_1\) and \(t_{\pathvar 
  2} \in W_2\) is added by the algorithm.

  One may now conclude. At the end of the induction, the last path considered, 
  \(\pathvar_0\), has no predecessor, hence it is the longest backward only 
  path. Since \(\focusedtreevar\emod{\pathvar_0}\) is the root of the tree, \(\emod{\overline{1}} \true \notin t_{\pathvar_0}\) and \(
  \emod{\overline{2}} \true \notin t_{\pathvar_0}\). Moreover, as the context 
  mark is somewhere in \(\focusedtreevar\), it is in a forward subtree of 
  \(\focusedtreevar\emod{\pathvar_0}\), hence the final type is marked.  
  Finally, \(t_\emptypath\) is in the witness tree of the final type, and 
  since \(\treeval{\focusedtreevar}{\emptypath}{\phi}\), \(\phi 
  \extendedin t_\emptypath\).
\end{mypfo}

\begin{lem}[Complexity] For a formula $\psi \in \mulogic$ the 
  satisfiability problem $\finterp{\psi}{\emptyset} \neq\emptyset$ is 
  decidable in time $2^{O(n)}$ where $n=\setcardinal{\lean{\psi}}$.
\end{lem}

\begin{mypfo}
$\setcardinal{\types{\psi}}$ is bounded by $\setcardinal{\powerset{\lean{\psi}}}$ 
which is $2^{O(n)}$. During each iteration, the algorithm adds at least one 
new type (otherwise it terminates), thus it performs at most $2^{O(n)}$ 
iterations. What is done at each iteration is now detailed. For each type that 
may be added (there are \(2^{O(n)}\) of them), there are two traversals of the 
set of types at the previous step to collect witnesses. Hence there are \(2 * 
2^{O(n)} * 2^{O(n)} = 2^{O(n)}\) witness tests at each iteration. Each witness 
test involves a membership test and a \(\Delta_a\) test. In the implementation 
these are precomputed: for every formula \(\emod{a}\phi\) in the lean, the 
subsets \( (T,F) \) of the lean that must be true and false respectively for 
\(\phi\) to be true are precomputed, so testing \(\phi \extendedin t\) are 
simple inclusion and disjunction tests. The $\func{FinalCheck}$ condition test 
at most $2^{O(n)}$ $\psi$-types and each test takes at most \(2^{O(n)}\) 
(testing the formulas containing \(\startatom\) against \(\psi\)).  
Therefore, the worst case global time complexity of the algorithm does not 
exceed $2^{O(n)}$.
\end{mypfo}

\section{Implementation Techniques}
\label{xml-calculus:sec:impl}

This section describes the main techniques used in the complete implementation \cite{solver-implementation} of the $\mulogic$ decision procedure. 

\subsection{Implicit Representation of Sets of $\psi$-Types}
The implementation relies on a symbolic representation and manipulation of sets of types using Binary Decision Diagrams (BDDs) \cite{bryant86}. BDDs provide a canonical representation of boolean functions. Experience has shown that this representation is very compact for very large boolean functions. Their effectiveness is notably well known in the area of formal verification of systems \cite{clarke-book99}.

First, one may observe that the implementation can avoid keeping track of every possible witnesses of each $\psi$-type. In fact, for a formula $\phi$, one can test $\finterp{\phi}{\emptyset}\neq \emptyset$ by testing the satisfiability of the (linear-size) ``plunging'' formula $\psi=\mu X. \phi \ou \emod{1}X \ou \emod{2}X$ at the root of focused trees. That is, checking $\fpinterp{\psi}{\emptyset}{0} \neq \emptyset$ while ensuring there is no unfulfilled upward eventuality at top level $0$. One advantage of proceeding this way is that the implementation only need to deal with a current set of $\psi$-types at each step.


A bit-vector representation of $\psi$-types is now introduced. Types are complete in the sense that either a subformula or its negation must belong to a type. It is thus possible for a formula $\phi \in \lean{\psi}$ to be represented using a single BDD variable. For $\lean{\psi}=\{ \phi_1, ..., \phi_m\}$, a subset $t\subseteq \lean{\psi}$ is represented by a vector $\notation{\vect{t}}{Bit-vector representation of $t$}{n:bitvector}=\left<t_1,...,t_m \right> \in \{0,1\}^m$ such that $\phi_i \in t$ iff $t_i=1$. A BDD with $m$ variables is then used to represent a set of such bit vectors. 

For a program $a\in \domFProg$, some auxiliary predicates on a vector $\vect{t}$ are defined:
\begin{itemize}
\item $\needwitness{a}{\vect{t}}$ is read ``$\vect{t}$ is a parent for program $a$'' and is true iff the bit for $\emod{a}\true$ is true in $\vect{t}$
\item $\iswitness{a}{\vect{t}}$ is read ``$\vect{t}$ is a child for program $a$'' and is true iff the bit for $\emod{\overline{a}}\true$ is true in $\vect{t}$
\end{itemize}

For a set $T \subseteq \powerset{\lean{\psi}}$, its corresponding characteristic function is denoted $\notation{\charact{T}}{Characteristic function of a set $T$}{n:charactfun}$. 
Encoding $\charact{\types{\psi}}$ is straightforward with the previous definitions.

The equivalent of $\extendedin$ is defined on the bit vector representation:
$$\status{\phi}{\vect{t}} \eqdef \left\{ 
\begin{array}{ll}
 t_i &\text{ if } \phi \in \lean{\psi} \\
 \status{\phi'}{\vect{t}} \et \status{\phi''}{\vect{t}} &\text{ if } \phi = \phi' \et \phi'' \\
 \status{\phi'}{\vect{t}} \ou \status{\phi''}{\vect{t}} &\text{ if } \phi = \phi' \ou \phi'' \\
 \neg \status{\phi'}{\vect{t}} &\text{ if } \phi = \neg \phi' \\
 \status{\expand{\phi}}{\vect{t}} &\text{ if } \phi = \murec{X_i}{\phi_i}{\psi}
\end{array} \right.$$

$a \implication b$ and $a \equivalence b$ respectively denote the implication and equivalence of two boolean formulas $a$ and $b$ over vector bits.
The BDD of the relation $\Delta_a$ for $a \in \domFProg$ can now be constructed. This BDD relates all pairs $(\vect{x}, \vect{y})$ that are consistent w.r.t the program $a$, i.e., such that $\vect{y}$ supports all of $\vect{x}$'s $\emod{a}\phi$ formulas, and vice-versa $\vect{x}$ supports all of $\vect{y}$'s $\emod{\overline{a}}\phi$ formulas:

$$\Delta_a(\vect{x}, \vect{y}) \eqdef  \bigwedge_{1 \leq i \leq m} \left\{
\begin{array}{l} 
x_i \equivalence \status{\phi}{\vect{y}} \quad \text{if } \phi_i = \emod{a}\phi \\
y_i \equivalence \status{\phi}{\vect{x}} \quad \text{if } \phi_i = \emod{\overline{a}}\phi  \\
\true \quad \text{otherwise }
\end{array} \right.$$

For $a \in \domFProg$, the set of witnessed vectors is defined:
$$\charact{\func{Wit}_a(T)}(\vect{x}) \eqdef  \needwitness{a}{\vect{x}} \implication \existsvect{\vect{y}}{h(\vect{y}) \et \Delta_a(\vect{x}, \vect{y})}$$
where $h(\vect{y}) = \charact{T}(\vect{y}) \et \iswitness{a}{\vect{y}}$. 

Then, the BDD of the fixpoint computation is initially set to the false constant, and the main function $\func{Upd}(\cdot)$ is implemented  as:
$$\charact{\func{Upd}(T)}(\vect{x}) \eqdef \charact{T}(\vect{x}) \ou \left( \charact{\types{\psi}}(\vect{x}) \et \bigwedge_{a \in \domFProg} \charact{\func{Wit}_a(T)}(\vect{x}) \right)$$

Finally, the solver can be implemented as iterations over the sets $\charact{\func{Upd}(T)}$ until a fixpoint is reached. The final satisfiability condition consists in checking whether $\psi$ is present in a $\psi$-type of this fixpoint with no unfulfilled upward eventuality:
$$\existsvect{\vect{t}}{\charact{T}(\vect{t}) \et \bigwedge_{a \in \domFProg} \neg \iswitness{a}{\vect{t}} \et \status{\psi}{\vect{t}}}$$

\subsection{Satisfying Model Reconstruction}

The implementation keeps a copy of each intermediate set of types computed by the algorithm, so that whenever a formula is satisfiable, a minimal satisfying model can be extracted.
The top-down (re)construction of a satisfying model starts from a root (a $\psi$-type for which the final satisfiability condition holds), and repeatedly attempts to find successors.
In order to minimize model size, only required left and right branches are 
built. Furthermore, for minimizing the maximal depth of the model, left and 
right successors of a node are successively searched in the intermediate sets 
of types, in the order they were computed by the algorithm. For readability 
purposes, the extracted satisfying model can be enriched by annotating the 
context mark $\startatom$ from which XPath evaluation started and a target 
node selected by the XPath expression. The annotated model is then provided to 
the user in XML unranked tree syntax.

\subsection{Conjunctive Partitioning and Early Quantification}

\iffull

The BDD-based implementation involves computations of \emph{relational products} of the form:  
\begin{equation}
\existsvect{\vect{y}}{ h(\vect{y}) \et \Delta_a(\vect{x}, \vect{y}) }
\label{xml-calculus:relational-product}
\end{equation}
It is well-known that such a computation may be quite time and space consuming, because the BDD corresponding to the relation $\Delta_a$ may be quite large.
 
One famous optimization technique consists in using \emph{conjunctive partioning} \cite{clarke-book99} and \emph{early quantification}  \cite{vardi-jancl06}. The idea is to compute the relational product without ever building the full BDD of the relation $\Delta_a$. This is possible by taking advantage of the form of $\Delta_a$ along with properties of existential quantification. By definition, $\Delta_a$ is a conjunction of $n$ equivalences relating $\vect{x}$ and $\vect{y}$ where $n$ is the number of $\emod{b}\phi$ formulas in $\lean{\psi}$ where $\phi\neq \true$ and $b \in \{a,\overline{a}\}$:
$$\Delta_a(\vect{x}, \vect{y}) = \bigwedge_{i=1}^{n}  R_i(\vect{x},\vect{y})$$

If a variable $y_k$ does not occur in the clauses $R_{i+1},...,R_{n}$ then the relational product (\ref{xml-calculus:relational-product}) can be rewritten as:
%
%
$$\bigexists{y_1,...,y_{k-1},y_{k+1},...,y_m}{\existsvect{y_k}{h(\vect{y}) \et \bigwedge_{1 \leq j \leq i} R_j(\vect{x}, \vect{y})} \et \bigwedge_{i+1 \leq l \leq n} R_l(\vect{x}, \vect{y})}$$


This allows to apply existential quantification on intermediate BDDs and thus to compose smaller BDDs. Of course, there are many ways to compose the $R_i(\vect{x},\vect{y})$. Let $\rho$ be a permutation of $\{0,...,n-1\}$ which determines the order in which the partitions $R_i(\vect{x},\vect{y})$ are combined. For each $i$, let $D_i$ be the set of variables $y_k$ with $k \in \{1,...,m\}$ that $R_i(\vect{x},\vect{y})$ depends on. $E_i$ is defined as the set of variables contained in $D_{\rho(i)}$ that are not contained in $D_{\rho(j)}$ for any $j$ larger than $i$:
$$E_i=D_{\rho(i)} \setminus \bigcup_{j=i+1}^{n-1}D_{\rho(j)}$$
The $E_i$ are pairwise disjoint and their union contains all the variables. The relational product (\ref{xml-calculus:relational-product}) can be computed by starting from:
$$h_1(\vect{x}, \vect{y}) = \bigexists{y_k \in E_0}{h(\vect{y}) \et R_{\rho(0)}(\vect{x}, \vect{y})}$$
and successively computing $h_{p+1}$ defined as follows:
$$h_{p+1}(\vect{x}, \vect{y}) = \left\{ 
\begin{array}{cl} 
\bigexists{y_k \in E_{p}}{h_{p}(\vect{x},\vect{y}) \et R_{\rho(p)}(\vect{x}, \vect{y})} & \text{if } E_p \neq \emptyset \\
\\
h_{p}(\vect{x},\vect{y}) \et R_{\rho(p)}(\vect{x}, \vect{y})  & \text{if } E_p = \emptyset
\end{array}
\right.$$
until reaching $h_n$ which is the result of the relational product.
The ordering $\rho$ determines how early in the computation variables can be quantified out. This directly impact the sizes of BDDs constructed and therefore the global efficiency of the decision procedure. It is thus important to choose $\rho$ carefully. The overall goal is to minimize the size of the largest BDD created during the elimination process. A heuristic taken from \cite{clarke-book99} is used. It seems to provide a good approximation as in practice it yields the best observed performance. It defines the cost of eliminating a variable $y_k$ as the sum of the sizes of all the $D_i$ containing $y_k$: 
$$\sum_{1 \leq i \leq n, y_k \in D_i} \left| D_i \right|$$
The ordering $\rho$ on the relations $R_i$ is then defined in such a way that variables can be eliminated in the order given by a greedy algorithm which repeatedly eliminates the variable of minimum cost.

\else
The BDD-based implementation involves computations of \emph{relational products} of the form
$\existsvectnesp{\vect{y}}{ h(\vect{y}) \et \Delta_a(\vect{x}, \vect{y}) }$. Such a computation may be quite time and space consuming, because the BDD corresponding to the relation $\Delta_a$ may be quite large.
A well-known optimization technique is that of \emph{conjunctive partitioning} combined with \emph{early quantification} \cite{clarke-book99,vardi-jancl06}.
The idea is to compute the relational product while avoiding the construction of the monolithic BDD of the relation $\Delta_a(\vect{x}, \vect{y})$. This is possible by considering that $\Delta_a(\vect{x}, \vect{y})$ is a conjunction of conditions. Computing the relational product thus consists in evaluating a quantified boolean formula of the form
$\exists y_1...\exists y_n (c_1 \et ... \et c_m)$ where the $c_i$s are boolean formulas. If a variable $y_j$ does not occur in the clauses $c_{i+1},..., c_m$ then the formula can be rewritten as $\exists y_1... \exists y_{j-1} \exists y_{j+1}... \exists y_n (\left[ \exists y_j (c_1 \et ... \et c_i) \right] \et c_{i+1} \et ... \et c_m)$.
This allows to apply existential quantification on intermediate BDDs and thus 
to compose smaller BDDs \cite{clarke-book99}.
\fi

\subsection{BDD Variable Ordering}

The cost of BDD operations is very sensitive to variable ordering. Finding the optimal variable ordering is known to be NP-complete \cite{hojati-iccd96}. However, several heuristics are known to perform relatively well in practice \cite{clarke-book99}. Choosing a good initial order of $\lean{\psi}$ formulas does significantly improve performance. To this end, preserving locality of the initial problem happens to be essential. Experience has shown that the variable order determined by the breadth-first traversal of the formula $\psi$ to solve, which keeps sister subformulas in close proximity, yields better results in practice.

\section{Typing Applications and Experimental Results}
\label{xml-calculus:sec:experiments}
\label{xml-calculus:xml-decision-pb}
For XPath expressions $e_1,..., e_n \in \lxpath$, the decision problems presented in Section~\ref{analysis:xml-decision-pb} can be generalized in the presence of several XML type expressions $T_1, ..., T_n$ and formulated as follows: 
\begin{itemize}
\item XPath containment: $\mucalcE{e_1}{(\startatom \et \Ttomu{T_1})} \et \neg 
  \mucalcE{e_2}{(\startatom \et \Ttomu{T_2})}$  (if the formula is unsatisfiable then all nodes 
  selected by $e_1$ under type constraint $T_1$ are selected by $e_2$ under 
  type constraint $T_2$)
\item XPath emptiness: $\mucalcE{e_1}{(\startatom \et \Ttomu{T_1})}$
\item XPath overlap: $\mucalcE{e_1}{(\startatom \et \Ttomu{T_1})} \et \mucalcE{e_2}{(\startatom \et \Ttomu{T_2})}$ 
\item XPath coverage: $\mucalcE{e_1}{(\startatom \et \Ttomu{T_1})} \et \bigwedge_{2\leq i \leq n} \neg  \mucalcE{e_i}{(\startatom \et \Ttomu{T_i})}$ 
\end{itemize}

The advantage of generalizing all the previous problem formulations with distinct types $T_1$ and $T_2$ is particularly useful for applications where types evolve. For instance, it is common that a file format of some company (described by an XML schema for instance) evolves over time. In this case, transformations that operated on the old document type must be updated to operate on the new type. Analysing XPath queries of a transformation (written in XSLT for instance) under different type constraints (the old one and the new one) can be used for helping the programmer to identify and understand the consequences of the evolution of the document type.

The system can also be used to check basic subtyping: $\Ttomu{T_1} \et \neg \Ttomu{T_2}$. However, since XPath (and therefore reverse navigation) is not used in that case, algorithms specialized for this restricted case such as the ones proposed in \cite{hosoya-toit03} or in \cite{tozawa-ciaa03} may perform better on practical instances.

Additionally, two decision problems are of special interest for XML static type checking:
\begin{itemize}
\item Static type checking of an annotated XPath query: 
  $\mucalcE{e_1}{(\startatom \et \Ttomu{T_1})} \et \neg {\Ttomu{T_2}}$ (if the formula is 
  unsatisfiable then all nodes selected by $e_1$ under type constraint $T_1$ 
  are included in the type $T_2$.)
\item XPath equivalence under type constraints, checked by $\mucalcE{e_1}{(\startatom \et \Ttomu{T_1})} 
  \et \neg \mucalcE{e_2}{(\startatom \et \Ttomu{T_2})}$ and $\neg \mucalcE{e_1}{(\startatom \et \Ttomu{T_1})} 
  \et \mucalcE{e_2}{(\startatom \et \Ttomu{T_2})}$ (This test can be used to check that the 
  nodes selected after a modification of a type $T_1$ by $T_2$ and an XPath 
  expression $e_1$ by $e_2$ are the same, typically when an input type changes 
  and the corresponding XPath query has to change as well.)
\end{itemize}


\subsection{Experimental Results}

Extensive tests of the implementation \cite{solver-implementation} have been carried out\footnote{Experiments have been conducted with a Java implementation running on a Pentium 4, 3 Ghz, with 512Mb of RAM with Windows XP.}. This section gathers a few of them. All times reported correspond to the actual running time (in milliseconds) of the $\mulogic$ satisfiability solver without the extra (negligible) time spent for parsing XPath and translating into $\mulogic$.

First, an XPath benchmark \cite{xpathmark} is used. Its goal is to cover XPath features by gathering a significant variety of XPath expressions met in real-world applications. In this first test series, types are not yet considered, and the focus is only given to the XPath containment problem, since its logical formulation (presented in Section~\ref{analysis:xml-decision-pb}) is the most complex (as it requires the logic to be closed under negation). 
This first test series consists in finding the relation holding for each pair of queries from the benchmark. This means checking the containment of each query of the benchmark against all the others. $q_i \subseteq q_j$ denotes that the query $q_i$ is contained in the query $q_j$. Comparisons of two queries $q_i$ and $q_j$ may yield to three different results:
\begin{enumerate}
\item $q_i \subseteq q_j$ and $q_j \subseteq q_i$, the queries are semantically equivalent, which is denoted by $q_i \equivalent q_j$
\item  $q_i \subseteq q_j$ but $q_j \not \subseteq q_i$, denoted by $q_i \contained q_j$ or alternatively by $q_j \contains q_i$
\item $q_i \not \subseteq q_j$ and $q_j \not \subseteq q_i$, queries are not related, denoted by $q_i \norelation q_j$
\end{enumerate}
Queries are presented on Figure~\ref{xml-calculus:fig:xpathmark-queries} (where ``//'' is used as a shorthand for ``/\step{\axis{descendant-or-self}}{*}/''). Corresponding results together with running times of the decision procedure are summarized on Table~\ref{table:xpathmark-results}. Obtained results show that all tests are solved in several milliseconds. These first results suggest that several XPath expressions used in real-world scenarios can be efficiently handled in practice.

\begin{figure}
\begin{small}
\centering
$\begin{array}{ll}
	   q_1 & /\name{site}/\name{regions}/\text{*}/\name{item}  \\
		 q_2 & /\name{site}/\name{auctions}/\name{auction}/\name{annotation}/\name{description}/\name{parlist}/\name{listitem}/\name{text}/\name{keyword}  \\
		 q_3 & //\name{keyword}  \\
		 q_4 & /\step{descendant-or-self}{\name{listitem}}/\step{descendant-or-self}{\name{keyword}}  \\
		 q_5 & /\name{site}/\name{regions}/\text{*}/\qualif{\name{item}}{\step{parent}{\name{namerica}} \op{or} \step{parent}{\name{samerica}}}  \\
		 q_6 & //\name{keyword}/\step{ancestor}{\name{listitem}} \\
		 q_7 & //\name{keyword}/\step{ancestor-or-self}{\name{mail}}  \\
		 q_8 &  /\name{site}/\name{regions}/\name{namerica}/\name{item} \shortmid /\name{site}/\name{regions}/\name{samerica}/\name{item}  \\	
		 q_9 &  /\name{site}/\name{people}/\qualif{\name{person}}{\name{address} \op{and} (\name{phone} \op{or} \name{homepage})}  \\
\end{array}$
\end{small}
\caption{Queries Taken from the XPathmark Benchmark.}\label{xml-calculus:fig:xpathmark-queries}
\end{figure}

\begin{table}
\centering
$\begin{array}{lll}
\begin{array}{|c|c|c|}
\hline 
\multirow{2}{*}{\text{~~Relation~~}} & \multicolumn{2}{|c|}{\text{~~Time (ms)~~}} \\  
 \cline{2-3} 
 & \subseteq & \supseteq  \\
\hline 
\hline 
q_1 \norelation q_2  &  ~~~17~~~ & ~~~21~~~  \\
q_1 \norelation q_3  & 13  &  20 \\
q_1 \norelation q_4  & 12  & 16 \\
q_1 \contains q_5    & 14  & 9 \\
q_1 \norelation q_6  & 21  & 17 \\
q_1 \norelation q_7  & 13  & 11 \\
q_1 \contains q_8    & 8  & 13 \\
q_1 \norelation q_9  & 14  & 17 \\
q_2 \contained q_3   & 32  & 35 \\
q_2 \contained q_4   & 33   & 38 \\
q_2 \norelation q_5  & 24  & 22 \\
q_2 \norelation q_6  & 21  & 38 \\
q_2 \norelation q_7  & 30 & 31 \\
q_2 \norelation q_8  & 22  & 23 \\
q_2 \norelation q_9  & 35  & 37 \\
q_3 \contains q_4    &  14 & 23 \\
q_3 \norelation q_5  &  7 & 9 \\
q_3 \norelation q_6  &  5 & 8 \\
\hline
\end{array}
& \quad & 
\begin{array}{|c|c|c|}
\hline 
\multirow{2}{*}{\text{~~Relation~~}} & \multicolumn{2}{|c|}{\text{ ~~Time (ms)~~}} \\  
 \cline{2-3} 
 & \subseteq & \supseteq \\
\hline 
\hline 
q_3 \norelation q_7  &  ~~~13~~~ & ~~~11~~~  \\
q_3 \norelation q_8  &  16 & 4  \\
q_3 \norelation q_9  &  13 & 16  \\
q_4 \norelation q_5  &  22 & 14  \\
q_4 \norelation q_6  &  5 & 12  \\
q_4 \norelation q_7  &  22 & 11 \\
q_4 \norelation q_8  &  13 & 17  \\
q_4 \norelation q_9  &  15 & 17 \\
q_5 \norelation q_6  &  10 & 10  \\
q_5 \norelation q_7  &  13 & 8  \\
q_5 \equivalent q_8  &  9 & 14  \\
q_5 \norelation q_9  &  17 & 21  \\
q_6 \norelation q_7  &  21 & 22  \\
q_6 \norelation q_8  &  17 & 17 \\
q_6 \norelation q_9  &  13 & 19  \\
q_7 \norelation q_8  &  22 & 19 \\
q_7 \norelation q_9  &  14 & 17  \\
q_8 \norelation q_9  &  9 & 11  \\
\hline
\end{array}
\end{array}$
\caption{Results for Comparisons of Benchmark Queries.}\label{table:xpathmark-results}
\end{table}

As a second test series, several expressions found in research papers on the containment of XPath expressions are compared. Figure~\ref{xml-calculus:fig:research-expr} presents the collected expressions. Figure~\ref{xml-calculus:fig:research-expr} also shows the  obtained results. The first containment instance of Figure~\ref{xml-calculus:fig:research-expr} was first formulated in \cite{suciu-miklau-jacm04} as an example for which the proposed tree pattern homomorphism technique is incomplete. The third example was not solvable in acceptable time and space bounds using the technique based on WS2S presented in Chapter~\ref{containment}. For this instance, the $\mulogic$ technique is orders of magnitude faster, and yields acceptable memory footprints. These results suggest that the system is reasonably able to handle containment instances which are difficult or impossible to solve using other techniques.

\begin{figure}
\centering
$\begin{array}{ll}
\begin{small}
\begin{array}{ll}
e_1 &  /\name{a}[.//\name{b}[\name{c}/\text{*}//\name{d}]/\name{b}[\name{c}//\name{d}]/\name{b}[\name{c}/\name{d}]] \\
e_2 &  /\name{a}[.//\name{b}[\name{c}/\text{*}//\name{d}]/\name{b}[\name{c}/\name{d}]]  \\ \\ 

e_3 & \name{a}[\name{b}]/\text{*}/\name{d}/\text{*}/\name{g} \\
e_4 & \name{a}[\name{b}]/(\name{b} \shortmid \name{c})/\name{d}/(\name{e}|\name{f})/\name{g} \\
e_5 & (\name{a}[\name{b}]/\name{b}/\name{d}/\name{e}/\name{g}) \shortmid (\name{a}/\name{b}/\name{d}/\name{f}/\name{g}) \\  \\   

e_6 & \name{a}/\name{b}/\name{s}//\name{c}/\name{b}/\name{s}/\name{c}//\name{d} \\
e_7 & \name{a}//\name{b}/\text{*}/\name{c}//\text{*}/\name{d} \\  \\

e_8 & \name{a}[\name{b}/\name{e}][\name{b}/\name{f}][\name{c}] \\
e_9 & \name{a}[\name{b}/\name{e}][\name{b}/\name{f}] \\ \\ 

e_{10} & /\qualif{\step{descendant}{\name{editor}}}{\step{parent}{\name{journal}}} \\
e_{11} & /\step{descendant-or-self}{\name{journal}}/\name{editor} 
\end{array}
\end{small}
&
\begin{array}{|c|c|c|c|c|}
\hline 
\multirow{2}{*}{\text{Relation}} & \multicolumn{2}{|c|}{\text{~Time (ms)~}} \\ 
 \cline{2-3} 
 & \subseteq & \supseteq  \\
 \hline 
\hline 
e_1 \contained e_2  &  ~323~ & ~~248~~   \\
e_3 \contains e_4  &  18 & 25  \\
e_3 \contains e_5  &  23 & 17  \\
e_4 \contains e_5  &  24 & 25 \\
e_6 \contained e_7  & 37 &  30 \\
e_8 \contained e_9  &  8 & 9  \\
e_{10} \equivalent e_{11}  &  17 & 14  \\
\hline 
\end{array}

\end{array}$
\caption{Results for Instances Found in Research Papers.} \label{xml-calculus:fig:research-expr}
\end{figure}

Figure~\ref{xml-calculus:fig:research-expr-following-preceding} presents the results of a third test series including examples with intersection, and axes such as ``following'' and ``preceding'', which are not illustrated in the previous series. 

\begin{figure}
\centering
$\begin{array}{cc}
\begin{small}
\begin{array}{ll}
e_{12} & \name{a}/\name{b}//\name{c}/\step{\axis{following-sibling}}{\name{d}}/\name{e} \\
e_{13} & \name{a}//\name{d}[\step{\axis{preceding-sibling}}{\name{c}}]/\name{e} \\
e_{14} & //\name{a}//\name{b}//\name{c}/\step{\axis{following-sibling}}{\name{d}}/\name{e} \\
e_{15} &  //\name{b}[\step{\axis{ancestor}}{\name{a}}]//\text{*}[\step{\axis{preceding-sibling}}{\name{c}}]/\name{e} \\
e_{16} &  /\name{b}[\step{\axis{preceding}}{\name{a}}]//\step{\axis{following}}{\name{c}} \\
e_{17} &  /\name{a}/\name{b}//\step{\axis{following}}{\name{c}} \\
e_{18} &  \name{a}/\name{b}[//\name{c}]/\step{\axis{following}}{\name{d}}/\name{e} \\
e_{19} &  \name{a}//\name{d}[\step{\axis{preceding}}{\name{c}}]/\name{e} \\
e_{20} &  \name{a}/\name{b}//\name{d}[\step{\axis{preceding-sibling}}{\name{c}}]/\name{e}  \\
e_{21} &  \name{a}/\name{c}/\step{\axis{following}}{\name{d}}/\name{e} \\
e_{22} &  \name{a}/\name{d}[\step{\axis{preceding}}{\name{c}}]/\name{e} \\
e_{23} &  \name{a}/\name{b}[//\name{c}]/\step{\axis{following}}{\name{d}}/\name{e} \cap \name{a}/\name{d}[\step{\axis{preceding}}{\name{c}}]/\name{e} \\
e_{24} &  \name{a}/\name{c}/\step{\axis{following}}{\name{d}}/\name{e} \cap \name{a}/\name{d}[\step{\axis{preceding}}{\name{c}}]/\name{e} \\
\end{array}
\end{small}
&
\begin{array}{|l|c|c|}
\hline 
\multirow{2}{*}{\text{Relation}} & \multicolumn{2}{|c|}{\text{~Time (ms)~}} \\  
 \cline{2-3} 
 & ~\subseteq~ & \supseteq  \\
\hline 
 \hline                 
e_{12} \contained e_{13}   & 23  & 17    \\
e_{14} \contained e_{15}   & 12  & 23  \\
e_{16} \contained e_{17}   & 18  & 22 \\
e_{18} \contained e_{19}   & 17  & 15 \\
e_{20} \equiv e_{12}       & 23  & 24 \\
e_{21} \norelation e_{22}  & 15  & 19  \\
e_{23}  \contained e_{21}   &  22 &  19 \\
e_{24}  \norelation e_{18}   & 16  & 11  \\
\hline
\end{array}
\end{array}$
\caption{Results for Instances with Horizontal Navigation.} \label{xml-calculus:fig:research-expr-following-preceding}
\end{figure}

In the fourth test series, several XPath expressions (shown on Figure~\ref{xml-calculus:xpath-for-use-with-dtds}) are used in the presence of two real-world XML types: the DTDs of the SMIL \cite{smil} and XHTML \cite{xhtml} W3C recommendations. Table~\ref{xml-calculus:table:tested-dtds} gives the size of each DTD by presenting the number of symbols used (alphabet size) and the number of grammar production rules (type variables) in the unranked and binary representations.
Several decision problems and their results are presented on Table~\ref{xml-calculus:table:results-xpath-dtd}. For example, the emptiness test for $p_9$ shows that the official XHTML DTD does not syntactically prohibit the nesting of anchors. Obtained results suggest that deciding XPath problems remains practically feasible, especially for static analysis purposes where such operations are performed at compile-time.


\begin{figure}
\centering
\begin{small}
$\begin{array}{ll}
p_5 & \text{switch/layout} \\
p_6 & \text{smil/head//layout} \\
p_7 &\text{smil/head//layout[ancestor::switch]} \\ 
p_8 & \text{*//switch[ancestor::head]/descendant::seq//audio[preceding-sibling::video]} \\ \\

p_9 & \text{descendant::a[ancestor::a]} \\
p_{10} &\text{/descendant::*}  \\
p_{11} &\text{html/(head} \shortmid \text{body)} \\
p_{12} &\text{html/head/descendant::*} \\
p_{13} &\text{html/body/descendant::*} \\
p_{14} &\text{//img} \\
p_{15} &\text{//img[not *]}
\end{array}$
\end{small}
\caption{Queries Used in the Presence of DTDs.}\label{xml-calculus:xpath-for-use-with-dtds}
\end{figure}

\begin{table}
\centering
$\begin{array}{|l|c|c|c|}
\hline
\text{DTD} & \text{Labels} & \text{Tree Type Variables} \\
\hline
\text{SMIL 1.0 \cite{smil}}  &  19 & 29 \text{ unranked}, 11\text{ binary}\\
\text{XHTML 1.0 \cite{xhtml}} & 77 & 104 \text{ unranked},  325 \text{ binary} \\
\hline
\end{array}$
\caption{Types Used in Experiments.} \label{xml-calculus:table:tested-dtds}
\end{table}

\begin{table}
\centering
$\begin{array}{|c|c|c|c|c|}
\hline
\text{Question} & \text{Instance} & \text{DTD} & \text{Answer} & \text{Time (ms)}\\
\hline
 \text{Non-Emptiness}  & p_5 & \text{SMIL} & \text{yes}& 56\\
  \text{Overlap}  & p_5 \cap p_6 \neq \emptyset & \text{SMIL} &\text{no} & 75\\
   \text{Containment}   & p_6 \subseteq p_7 &  \text{SMIL} &\text{no}  & 81\\
  \text{Non-Emptiness}  & p_8 & \text{SMIL} & \text{yes}& 94\\
 \text{Non-Emptiness}  & p_9 & \text{XHTML} &\text{yes} &  2530\\  
 \text{Coverage}  & p_{10} \subseteq p_{11} \cup p_{12} \cup p_{13} & \text{XHTML} & \text{yes} & 2723\\
  \text{Containment}  & p_{14} \subseteq p_{15} & \text{XHTML} & \text{yes} &  2 937\\
\hline
\end{array}$
\caption{Results in the Presence of DTDs.}\label{xml-calculus:table:results-xpath-dtd}
\end{table}

An additional benefit of the technique is that it automatically outputs a satisfying XML document, which can easily be enriched with XPath context and target information. For instance, the solver trace for the emptiness test for $p_8$ is given below:
\begin{small}
\begin{verbatim}
Checking emptiness of 
*//switch[ancestor::head]/descendant::seq//audio[preceding-sibling::video]
in the presence of 'smil.dtd'.
Parsing XPath [249 ms].
Compilation of XPath to Tree Logic Formulas [15 ms].
Input DTD read from 'sampleDTDs/smil.dtd'.
Start symbol is $smil

Converted DTD into BTT [140 ms].
CFT: 29 type variables and 19 terminals.
BTT: 11 type variables and 17 terminals.

Translated BTT into Tree Logic [16 ms].

Computing Relevant Closure
Computed Relevant Closure [46 ms].
Computed Lean [0 ms].
The Lean has size 53. It contains 35 eventualities and 18 symbols.
Fixpoint Computation Initialized [31 ms].
Computing Fixpoint......[94 ms].
Formula is satisfiable [171 ms].
A satisfying finite binary tree model was found [94 ms]:
smil(head(switch(seq(video(#, audio), layout), meta), #), #)
In XML syntax:
<smil context="true">
  <head>
    <switch>
      <seq>
        <video/>
        <audio target="true"/>
      </seq>
      <layout/>
    </switch>
    <meta/>
  </head>
</smil>

*//switch[ancestor::head]/descendant::seq//audio[preceding-sibling::video]
is satisfiable in presence of 'smil.dtd'
\end{verbatim}
\end{small}

These experiments shed a first light on the cost of solving XML decision problems in practice, and suggest that the presented system is already able to handle realistic scenarios.

\section{Outcome}
\label{xml-calculus:sec:outcome}

The essence of the obtained results lives in a sub-logic of the alternation free modal \(\mu\)-calculus with converse, with some syntactic restrictions on formulas, and where models are finite trees. As detailed in Chapter~\ref{xml-calculus:the-logic-for-xml}, the syntactic restrictions allow to prove that formulas of the logic are cycle-free. The cycle-free property is used to prove that the least and greatest fixpoint operators collapse in a single fixpoint operator. This provides closure under negation, which is the key property for solving the containment (a logical implication). Deep connections between this logic and XML decision problems can then be revealed: XPath expressions and regular tree type formulas conform to the $\mulogic$ syntactic restrictions. Furthermore, XPath expressions and regular tree languages can surprisingly be efficiently embedded since they are linear in the size of the corresponding formulas in the logic.

A sound and complete algorithm for testing the satisfiability of the logic is described in this chapter. Its upper bound time complexity is $2^{O(n)}$ w.r.t. the length $n$ of the given formula. The combination of all these ingredients yields the main result: sound and complete decision procedures, with the same upper bound complexity, for XML decision problems involving regular tree types and XPath queries. The implementation appears efficient in practice. A benefit of the approach is that the system can be effectively used in static analyzers for programming languages manipulating both XPath expressions and XML type annotations (input and output).

%


 \ifglobalcompil
  \else 
   \bibliographystyle{apalike}
   \bibliography{references}
   \end{document}
  \fi

\newif\ifglobalcompil\globalcompiltrue
  \ifglobalcompil
  \else 
    \input{Preamble} 
    \input{Markup}                       
    \newcommand{\chapterabstract}[1]{Chapter Abstract: #1}
    \begin{document} 
  \fi
  
\mychapter{Conclusion}
\label{conclusion}
\section{Summary of the Main Contributions}
The main contribution of this thesis is a new logic for finite trees, derived from the $\mu$-calculus. The logic is expressive enough to capture regular tree types along with multi-directional navigation in finite trees. It is decidable in single exponential time (specifically in $2^{O(n)}$ steps where $n$ is the size of the input formula defined as its number of atomic propositions and eventualities). This improves the best known computational complexity for finite trees. As such, this logic offers a new compromise between expressivity and complexity, specifically interesting in the context of XML. 

Another contribution of this thesis is to show how to linearly compile queries and regular tree types (including DTDs and XML Schemas) in the logic. 
The logic takes almost full XPath into account and supports the largest fragment that has been treated for static analysis. Another advantage is that the logic is a sublogic of an existing one (the $\mu$-calculus) thus facilitating known optimization techniques and warranting extensibility.

This solves the major decision problems needed in the static analysis of XML specifications. These problems involve containment, emptiness, equivalence, overlap, and coverage of XPath queries (in the presence or absence of regular tree types), static type-checking of an annotated XPath query, and XPath equivalence under type constraints.

Furthermore, implementation techniques that yield concrete design and effective algorithm implementation in practice are presented. The fully implemented system is already able to handle realistic scenarios.

\section{Perspectives}

There are a number of interesting and promising directions for further research that builds on the results and ideas developed in this dissertation. 


\subsection{Further Optimizations of the Logical Solver}
The worst-case complexity upper bound for deciding $\mulogic$ cannot be less than exponential time (since it can be used to decide FTA containment, or alternatively since it contains the CTL subsystem). Nevertheless, several techniques can be further developed for continuing to improve the performance of the implementation. One perspective is to use dynamic reordering of BDD variables whenever it can speed up the decision procedure. Another interesting direction of further research is to attempt to statically reduce Lean contents by exploiting peculiarities of particular problem instances such as locality. 


\subsection{Pushing the XPath Decidability Envelope Further}

One perspective of this thesis consists in extending the considered XPath fragment in order to support restricted data value comparisons (in the manner of \cite{segoufin-pods06}). Another direction for extending the fragment consists in integrating related work on counting \cite{dal-zilio-popl04,seidl-icalp04} to the logic. 


\subsection{Enhancing the Translation of Regular Tree Types}

Another perspective consists in considering a modification of the translation of types such that it 
imposes the context of a type to also follow the regular tree language 
definition (stating for instance that the parent of a given node may only be 
some specific other nodes). This would allow a yet more precise and interesting reporting on type-checking instances.

\subsection{Efficiently Supporting Attributes and Data Values}

Most theoretical work on XML and XPath models XML documents by finite labeled ordered trees, where the labels are taken from a finite alphabet.  Attributes and data values are usually ignored. This thesis makes the same abstractions, and thus still offers perspectives for supporting more XML features. There is a reason for each previous widespread abstractions. 

The difficulty for supporting XML attributes arises from the fact that they are unordered \cite{xml} which forces to consider mixed ordered and unordered contents in the underlying data model. There are several directions that can be followed for supporting constraints over mixed content while avoiding blow-ups caused by a naive modeling of unordered data on top of an ordered data model. Shuffle automata introduced in the 1970's provide a way to deal with an interleave operator \cite{jedrzejowicz-tcs01}. The work found in \cite{dal-zilio-aaecc06} introduces the Sheaves logic and a related new class of automata (sheaves automata) suited for ordered trees. The logic combines regularity and counting constraints, and provides an interleaving operator. The work found in \cite{murata-em03} proposes an automata rewriting technique for handling attribute-element constraints, which has been implemented in a validator for RELAX NG. The approach presented in this dissertation can easily be extended for supporting unordered XML attributes in an alternative manner, by observing that the algorithm proposed in Chapter~\ref{xml-calculus:sec:algo} is based on $\psi$-types. Since a $\psi$-type is simply a set of formulas, attributes could naturally be modeled by a new class of atomic propositions, with the same complexity.

The usual reason for ignoring data values comes from the fact that they quickly lead to languages whose static analysis is undecidable \cite{alon-jcss03,neven-icdt03,benedikt-pods05}. Nevertheless, there exists examples of decidable static reasoning tasks involving attribute values \cite{arenas-it05,buneman-is03,segoufin-pods06}. A perspective of this thesis is to extend the algorithm proposed in Chapter~\ref{xml-calculus:sec:algo} to deal with attribute values. This could help at identifying more precisely the upper-bound complexity of decision problems involving XPath with limited data value comparison, which has been observed to be between NEXPTIME and 3-NEXPTIME in the recent work found in \cite{segoufin-pods06}.

\subsection{Query Optimization}
Another perspective of this thesis is to take advantage of the static analysis of XPath expressions for optimization purposes.
This allows for instance to automatically detect contradictions and eliminate redundancies from XML queries at compile time, as preliminary investigated in \cite{geneves-doceng04}. One perspective is to extend this work with some trace-based semantics for XPath (in the manner of \cite{hartel-time05}) in order to capture optimality of a query w.r.t a given evaluation context. Then, the optimal query could be calculated by using the automatic comparison of queries described in this thesis.

\subsection{Query Evaluation via Model-Checking}
The linear translation of XPath into the $\mu$-calculus opens perspectives for query evaluation. A direction of future work consists in revisiting XPath evaluation (reduced to model-checking) based on existing techniques \cite{mateescu-tacas02,mateescu-scp03}.

\subsection{Application to the Static Analysis of Transformations}
Last but not least, a perspective of this thesis is to apply the presented XPath static analysis techniques to the type-checking of XML transformation languages. 
Results presented in this dissertation open the way to the construction of debuggers, compilers, and type-checkers for XSLT and XQuery. For example, the recent work found in \cite{moller-rr05} could benefit from using the exact algorithm of Chapter~\ref{xml-calculus:sec:algo} instead of their conservative approximation. The practical experiments reported in Chapter~\ref{xml-calculus:sec:algo} strengthen the hope for an effective analysis of this kind in the near future.


 \ifglobalcompil
  \else 
   \bibliographystyle{apalike}
   \bibliography{references}
   \end{document}
  \fi

\backmatter
\bibliographystyle{apalike}
\bibliography{references}

\begin{appendices}
\newif\ifglobalcompil\globalcompiltrue
  \ifglobalcompil
  \else 
    \input{Preamble} 
    \input{Markup}                       
    \newcommand{\chapterabstract}[1]{Chapter Abstract: #1}
    \begin{document} 
  \fi

\mychapter{Computational Complexity for Logical Satisfiability Dealt With in this Dissertation}
\label{appendix:complexity-classes}
\newcommand{\largeurmax}{9.8}

 \begin{tikzpicture}{-2cm}{}{}{}
    \pgfsetlinewidth{0.6pt}
    
    \pgfsetendarrow{\pgfarrowto};
    \pgfxyline(0,0.5)(0,12.5);
    \pgfclearendarrow;
	
	  \pgfxyline(-3,1)(-3,11.5);
	  \pgfxyline(\largeurmax,1)(\largeurmax,11.5);
	
	  \pgfxyline(-3,1)(\largeurmax,1);
	      \pgfsetdash{{0.4cm}{0.4cm}{0.4cm}{0.4cm}}{0cm}
	  \pgfxyline(-3,2.5)(\largeurmax,2.5);
	      \pgfsetdash{}{0pt}
    \pgfxyline(-3,4)(\largeurmax,4);
	  \pgfxyline(-3,5.5)(\largeurmax,5.5);
    \pgfxyline(-3,7)(\largeurmax,7);
    \pgfxyline(-3,8.5)(\largeurmax,8.5);
    \pgfxyline(-3,10)(\largeurmax,10);
    \pgfxyline(-3,11.5)(\largeurmax,11.5);
  
    \pgfputat{\pgfxy(-2.9,12)}{\pgfbox[left,base]{Undecidable}};
    \pgfputat{\pgfxy(-2.9,10.6)}{\pgfbox[left,base]{Decidable}};
    \pgfputat{\pgfxy(-2.9,9.2)}{\pgfbox[left,base]{Elementary}};
    \pgfputat{\pgfxy(-2.9,7.6)}{\pgfbox[left,base]{EXPSPACE}};
    \pgfputat{\pgfxy(-2.9,6.2)}{\pgfbox[left,base]{EXPTIME}};
    \pgfputat{\pgfxy(-2.9,4.6)}{\pgfbox[left,base]{PSPACE}};
    \pgfputat{\pgfxy(-2.9,3.2)}{\pgfbox[left,base]{NP}};
    \pgfputat{\pgfxy(-2.9,1.6)}{\pgfbox[left,base]{P (PTIME)}};
        

    \pgfcircle[fill]{\pgfxy(0,10.5)}{2pt};                                                               
    \pgfputat{\pgfxy(0.4,10.4)}{\pgfbox[left,base]{$2^{\left.2^{\cdot^{\cdot^{2^{O(n)}}}}\right\}k}$ WS2S \cite{meyer72} used in Chapter~\ref{containment}.}};
    
  
    \pgfcircle[fill]{\pgfxy(0,6.7)}{2pt};
    \pgfputat{\pgfxy(0.4,6.6)}{\pgfbox[left,base]{$2^{O(n^4\cdot \fun{log}n)}$ Full $\mu$-calculus \cite{gradel-book02}.}}

      \pgfcircle[fill]{\pgfxy(0,6.2)}{2pt};
      \pgfputat{\pgfxy(0.4,6.1)}{\pgfbox[left,base]{$2^{O(n\cdot \fun{log}n)}$ AFMC \cite{tozawa-tableaux05} used in Chapter~\ref{analysis}.}}

      \pgfcircle[fill]{\pgfxy(0,5.7)}{2pt};
      \pgfputat{\pgfxy(0.4,5.6)}{\pgfbox[left,base]{$2^{O(n)}$   $\mulogic$ logic proposed in Chapters~\ref{xml-calculus:the-logic-for-xml} and \ref{xml-calculus:sec:algo}.}}




    \end{tikzpicture}

\ifglobalcompil
 \else 
  \bibliographystyle{apalike}
  \bibliography{references}
  \end{document}
 \fi

\newif\ifglobalcompil\globalcompiltrue

  \ifglobalcompil
  \else 
    \input{Preamble}    
    \usepackage[latin1]{inputenc} 
    \usepackage[cyr]{aeguill}
    \usepackage[francais]{babel}

    \input{Markup}                       
    \newcommand{\chapterabstract}[1]{Chapter Abstract: #1}
    \begin{document} 
  \fi

\mychapter{R\'esum\'e \'etendu}
\label{apercu}
\section*{Motivation et objectifs}

Ce travail a \'et\'e initialement motiv\'e par le besoin d'analyseurs statiques efficaces pour les langages de manipulation de donn\'ees et de documents XML.
Ces langages de programmation utilisent des sch\'emas \cite{xml-schemas} et des requ\^etes XPath \cite{xpath} comme constructions de premi\`ere classe. Des exemples actuels de ces langages incluent la recommandation du W3C XSLT \cite{xslt} pour la transformation de documents XML, et la future recommandation XQuery \cite{xquery} pour l'interrogation de bases de donn\'ees XML. Equiper ces langages de syst\`emes d\'ecidables et efficaces pour la v\'erification statique de types a \'et\'e l'un des d\'efis de recherche majeurs de la derni\`ere d\'ecennie, qui a entre autres rassembl\'e les communaut\'es travaillant sur les langages de programmation, les bases de donn\'ees, les documents structur\'es, et l'informatique th\'eorique. Ce travail poursuit l'effort de recherche initi\'e dans les travaux d\'ecrits dans \cite{murata-pdp96, tozawa-doceng01, milo-jcss03, hosoya-toit03}.

Ce travail a abouti \`a la conception d'une logique d'arbre finis adapt\'ee \`a XML, et sa proc\'edure de d\'ecision, pr\'esent\'ees dans cette th\`ese. Le solveur logique a \'et\'e implant\'e au c\oe ur d'un syst\`eme pour l'analyse statique g\'en\'erale et le typage des sp\'ecifications XML. Le syst\`eme peut \^etre utilis\'e comme un composant d'analyseurs statiques pour les langages de programmation utilisant \`a la fois des expressions XPath et des types XML.

Cette th\`ese pr\'esente les investigations th\'eoriques qui ont conduit aux fondations de cette nouvelle logique d'arbres finis, avec les bases algorithmiques et les principes d'implantation sur lesquels le solveur logique repose. Ces d\'ecouvertes sont appliqu\'ees \`a la r\'esolution des probl\`emes de typage XML, qui sont traduits dans la logique. Les probl\`emes r\'esolus incluent le typage statique du langage XPath en pr\'esence de types r\'eguliers d'arbres.

\subsection*{Documents XML et sch\'emas}

\emph{Extensible Markup Language} (XML) \cite{xml} est un format de fichier texte pour la repr\'esentation de structures arborescentes sous une forme standard.

La structure compl\`ete d'un document XML, si on s'abstrait des d\'etails d'importance moindre, est un arbre d'arit\'e variable, dans lequel les n\oe uds (aussi appel\'es \emph{\'el\'ements} dans le jargon XML) sont \'etiquett\'es, les feuilles de l'arbre sont des n\oe uds textes, et l'ordre entre les n\oe uds enfants est important. XML peut \^etre vu comme une syntaxe concr\`ete pour la description de telles structures en utilisant des balises. Un exemple de document XML suit:

\label{apercu:well-formed-sample-xml-doc}
\begin{verbatim}
<plante>
  <categorie>Vasculaire</categorie>
  <tissu>
    <nom>Phloeme</nom>
    <def>Le phloeme est un tissu vivant servant au transport 
         dans toutes les parties de la plante.</def>
    <note>Dans les arbres, c'est une partie de l'ecorce.</note>
  </tissu>
</plante>
\end{verbatim}

Un \'el\'ement est d\'ecrit par une paire compos\'ee d'une balise ouvrante $<...>$ et d'une balise fermante $</...>$, entre lesquelles le contenu de l'\'el\'ement est ins\'er\'e. Dans l'exemple pr\'ec\'edent ``\texttt{plante}'', ``\texttt{categorie}'', ``\texttt{tissu}'', ``\texttt{nom}'', ``\texttt{def}'', et ``\texttt{note}'' sont des \'etiquettes (\emph{noms d' \'el\'ement} dans le jargon XML).



\begin{figure}[h]
\centering

\begin{tikzpicture}
\tikzstyle{every node}=[ball color=white,circle,text=black]
\draw (1,1) node(root) {\begin{small}plante\end{small}};
\draw (-1,-0.5) node(n1) {\begin{tiny}categorie\end{tiny}};
\draw (3,-0.5) node(n2) {\begin{small}tissu\end{small}};

\draw (1,-2) node(n3) {\begin{small}nom\end{small}};
\draw (3,-2) node(n4) {\begin{small}\;def\;\end{small}};
\draw (5,-2) node(n5) {\begin{small}note\end{small}};

\tikzstyle{every node}=[text=black]

\draw (-1,-2) node(n6) {Vasculaire};
\draw (1,-3.5) node(n7) {Phlo\`eme};
\draw (3,-3.5) node(n8) {Le (...)};
\draw (5,-3.5) node(n9) {Dans (...)};

\draw [thick, ->, black] (root) -- (n1);
\draw [thick, ->, black] (root) -- (n2);

\draw [thick, ->, black] (n1) -- (n6);

\draw [thick, ->, black] (n2) -- (n3);
\draw [thick, ->, black] (n2) -- (n4);
\draw [thick, ->, black] (n2) -- (n5);

\draw [thick, ->, black] (n3) -- (n7);
\draw [thick, ->, black] (n4) -- (n8);
\draw [thick, ->, black] (n5) -- (n9);

\end{tikzpicture}
\caption{Exemple: arbre d'un document bien-form\'e.}\label{apercu:fig:tree-sample}
\end{figure}
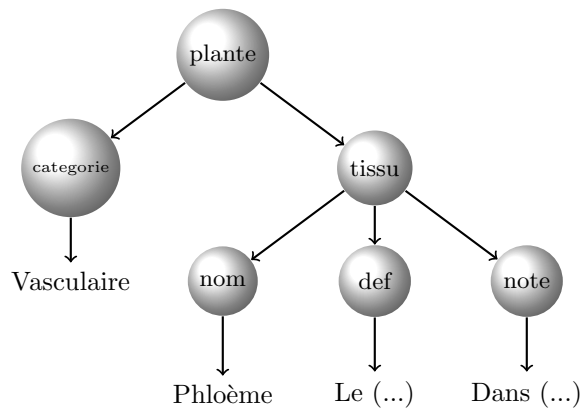

La sp\'ecification XML ne d\'efinit pas a priori l'ensemble des \'etiquettes permises dans un document XML, et ne d\'efinit pas non plus de s\'emantique pour les \'etiquettes. Seules des conditions de bonne formation sont d\'efinies pour s'assurer que les \'el\'ements sont bien imbriqu\'es, ce qui permet de consid\'erer les documents XML comme les arbres. Par exemple, la Figure~\ref{apercu:fig:tree-sample} donne une repr\'esentation plus visuelle du pr\'ec\'edent document XML bien form\'e.

L'ensemble des \'etiquettes qui apparaissent dans un document XML est d\'etermin\'e par des \emph{sch\'emas} qui peuvent \^etre librement d\'efinis par les utilisateurs. Un \emph{sch\'ema} (aussi appel\'e un \emph{type XML}) est une description des contraintes sur la structure des documents, comme les \'etiquettes permises et leur possible structure d'imbrication. Un sch\'ema d\'efinit ainsi une classe de documents XML. Deux niveaux de correction peuvent donc \^etre distingu\'es pour les documents XML:


\begin{itemize}
\item le caract\`ere \emph{bien-form\'e} qui s'applique aux documents qui v\'erifient la condition n\'ecessaire et suffisante (d\'efinie par la sp\'ecification XML) pour \^etre interpr\'et\'es comme des arbres;
\item la \emph{validit\'e} qui s'applique aux documents qui v\'erifient les contraintes additionnelles d\'ecrites par un sch\'ema donn\'e.
\end{itemize}

La validit\'e d'un document implique son caract\`ere bien-form\'e puisque un sch\'ema d\'ecrit des contraintes sur l'arbre et non sur la repr\'esentation textuelle du document XML.

Chaque application peut d\'efinir son propre format de donn\'ees en d\'efinissant des sch\'emas, \`a un plus haut niveau d'abstraction (structures arborescentes). De ce fait, XML est souvent appel\'e un m\'etalangage ou un ``format pour les formats de donn\'ees''.


S\'eparer les deux niveaux de correction permet aux applications de partager des outils logiciels g\'en\'eriques pour manipuler des documents bien form\'es (analyseurs syntaxiques, \'editeurs, requ\^etes, outils d'interrogation et de transformation...). Ces outils implantent tous les m\^emes conventions d\'efinies par la sp\'ecification XML (comme la fa\c con d'inclure des commentaires, des fragments externes, des caract\`eres sp\'eciaux...). XML rend ainsi possible un premier niveau de traitement pour un document XML d\`es lors qu'il est bien-form\'e, sans faire l'hypoth\`ese additionnelle beaucoup plus forte qu'il est valide par rapport \`a un certain sch\'ema. Cette g\'en\'ericit\'e est l'une des forces de XML. En cons\'equence, l'adoption de XML s'est faite \`a une vitesse et une ampleur in\'egal\'ee. De nombreux sch\'emas ont \'et\'e d\'efinis et sont actuellement largement utilis\'es en pratique, par exemple: XHTML (la version XML de HTML), SVG (pour le graphisme vectoriel), SMIL (pour la synchronisation des documents multim\'edias), MatML (pour les formules math\'ematiques), SOAP (pour l'appel de proc\'edure \`a distance), XBRL et FIX (pour les informations financi\`eres et les transactions de valeurs), SMD (pour la musique), X3D (pour la mod\'elisation 3D), et CML (pour les structures chimiques).

\subsection*{XPath}

\label{apercu:xpath-intro}


XPath \cite{xpath, xpath20} a \'et\'e introduit par le W3C comme le langage de requ\^etes standard pour s\'electionner et r\'ecup\'erer de l'information dans les documents XML. Il permet de naviguer dans les arbres XML et de retourner un ensemble de n\oe uds v\'erifiant certaines conditions. En tant que tel, XPath forme l'essence de l'acc\`es aux donn\'ees XML.

\begin{figure*}
\centering
  \includegraphics[width=10cm, keepaspectratio=true]{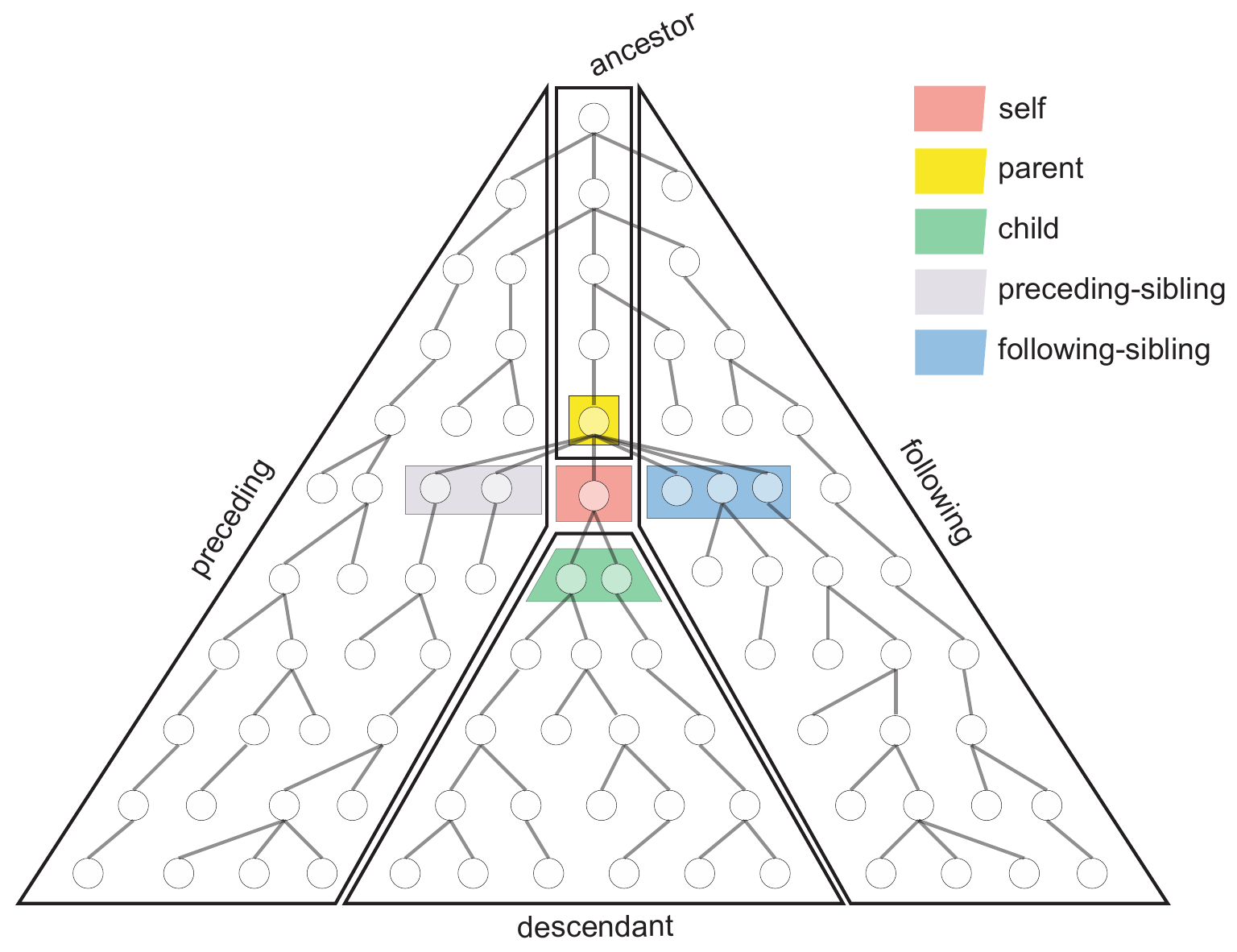}
   \caption{Partition des axes depuis un n\oe ud de contexte.}\label{apercu:fig:xpath-axes}
\end{figure*}

Dans leur forme la plus simple, les expressions XPath ressemblent \`a des ``chemins de navigation dans les r\'epertoires''. Par exemple, l'expression XPath
\xpath{/\name{livre}/\name{chapitre}/\name{section}}
navigue \`a partir de la racine d'un document (d\'esign\'ee par le ``/'' en t\^ete) \`a travers les n\oe uds ``livre'' au premier niveau, vers leurs n\oe uds enfants ``chapitre'', jusqu'\`a leurs n\oe uds enfants nomm\'es ``section''. Le r\'esultat de l'\'evaluation de l'expression compl\`ete est l'ensemble de tous les n\oe uds ``section'' qui peuvent \^etre atteints de cette mani\`ere. De plus, \`a chaque \'etape de la navigation, les n\oe uds s\'electionn\'es peuvent \^etre filtr\'es avec des qualifieurs. Un qualifieur est une expression bool\'eenne entre crochets qui peut tester l'existence ou l'absence de chemins. Si on formule par exemple la requ\^ete suivante :
\xpath{  /\name{livre}/\name{chapitre}/\qualif{\name{section}}{\name{citation}}}
alors le r\'esultat est constitu\'e de \emph{tous} les \'el\'ements ``section'' qui ont au moins un \'el\'ement fils nomm\'e ``citation''. La situation devient plus int\'eressante lorsque les capacit\'es de navigation de XPath selon d'autres ``axes'' que l'axe ``child'' sont utilis\'ees. En effet, l'expression XPath pr\'ec\'edente est un raccourci pour:
\xpath{/\axis{child::}\name{livre}/\axis{child::}\name{chapitre}/\qualif{\axis{child::}\name{section}}{\axis{child::}\name{citation}}}
qui fait apparaitre de mani\`ere explicite que chaque \'etape de navigation utilise l'axe ``child'' contenant tous les n\oe uds enfants des n\oe uds s\'electionn\'es lors de l'\'etape pr\'ec\'edente. Si on formule la requ\^ete suivante :
\xpath{/\axis{child::}\name{livre}/\qualif{\axis{descendant::}\name{*}}{\axis{child::}\name{citation}}}
alors la derni\`ere \'etape s\'electionne les n\oe uds de n'importe quel nom qui sont parmi les descendants de l'\'el\'ement ``livre'' et qui ont un sous-\'el\'ement nomm\'e ``citation''. Il est aussi possible d'utiliser des axes comme ``\axis{preceding-sibling}'' pour naviguer vers les pr\'ec\'edents n\oe uds fils du m\^eme parent, ou ``\axis{ancestor}'' pour naviguer r\'ecursivement vers le haut (cf. Figure~\ref{apercu:fig:xpath-axes}).  \emph{L'ordre du document} est d\'efini comme l'ordre dans lequel les n\oe uds sont visit\'es par un parcours en profondeur d'abord de l'arbre. Les axes qui effectuent de la navigation dans l'ordre inverse de l'ordre du document sont appel\'es ``axes inverses''.

 
Les exemples pr\'ec\'edents illustrent tous des expressions XPath absolues puisqu'elles commencent toutes avec un ``/'' qui se r\'ef\`ere \`a la racine. La s\'emantique d'une expression \emph{relative} (sans le ``/'' en t\^ete) est d\'efinie par rapport \`a un \emph{n\oe ud de contexte} dans l'arbre. Le \emph{n\oe ud de contexte} d\'esigne simplement le n\oe ud de l'arbre depuis lequel la navigation d\'ebute. A partir d'un n\oe ud de contexte quelconque dans un arbre, tous les autres n\oe uds peuvent \^etre facilement atteints: les axes XPath forment une partition de l'arbre. La Figure~\ref{apercu:fig:xpath-axes} illustre cela sur un exemple. Plus de d\'etails informels sur le langage XPath complet peuvent \^etre trouv\'es dans la sp\'ecification du W3C \cite{xpath}.

XPath est de plus en plus populaire du fait de son expressivit\'e et de sa syntaxe compacte. Ces deux avantages ont conf\'er\'e \`a XPath un r\^ole central dans d'autres sp\'ecifications cl\'es et applications XML. Il est utilis\'e dans XQuery \cite{xquery} comme le langage c\oe ur pour formuler des requ\^etes; dans XSLT \cite{xslt} pour la s\'election des n\oe uds dans les transformations; dans XML Schema \cite{xml-schemas} pour d\'efinir les cl\'es; dans XLink \cite{xlink} et XPointer \cite{xpointer} pour r\'ef\'erencer des parties de donn\'ees XML. XPath est aussi utilis\'e dans de nombreuses applications comme les langages de mise \`a jour \cite{xmlupdates} et de contr\^ole d'acc\`es \cite{xml-access-control}.

\subsection*{V\'erification statique de type}


Les applications XML utilisent la plupart du temps les sch\'emas pour effectuer de la validation (aussi appel\'ee \emph{v\'erification dynamique de type}). La validation consiste en l'utilisation d'un validateur de sch\'ema qui analyse un document XML particulier par rapport \`a un certain sch\'ema dans le but de s'assurer que le document est bien conforme aux attentes de l'application.


En pratique cependant, les documents XML sont souvent g\'en\'er\'es dynamiquement par un certain programme. Typiquement, les programmes qui manipulent du XML acc\`edent tout d'abord aux donn\'ees (se conformant possiblement \`a un certain sch\'ema) avec des expressions XPath, et construisent et retournent ensuite un document XML r\'esultat qui se conforme aux exigeances d'un autre sch\'ema.

Une approche ambitieuse est la \emph{v\'erification statique de type} pour ces programmes, qui consiste \`a s'assurer au moment de la compilation, que le code traitant les donn\'ees XML ne peut pas produire de document non valide. Un v\'erificateur statique de type analyse un programme, possiblement en pr\'esence des sch\'emas qui d\'ecrivent ses entr\'ees et sorties (si ceux-ci s'av\`erent disponibles). La difficult\'e du probl\`eme est fonction du langage dans lequel le programme et les sch\'emas sont exprim\'es.


Les langages de sch\'emas ont fait l'objet de nombreuses \'etudes et sont maintenant bien compris comme des sous-ensembles des langages r\'eguliers d'arbres \cite{murata-toit05}.
Cependant, bien que de nombreuses tentatives aient \'et\'e faites pour mieux comprendre les techniques de typage statique, en particulier \`a travers la conception de langages de programmation sp\'ecifiques au domaine \cite{hosoya-toit03}, aucune approche est effectivement capable de supporter XPath, qui demeure n\'eanmoins l'essence de la navigation et de l'acc\`es aux donn\'ees XML.

\subsection*{D\'efis de recherche}



Les limitations des approches existantes sont justifi\'ees par la difficult\'e de l'analyse statique de XPath. Il est connu que l'analyse statique du langage XPath complet est ind\'ecidable. L'importance et l'ampleur des applications motivent n\'eanmoins des questions de recherche: quel est le plus gros fragment de XPath dont l'analyse statique est d\'ecidable ? Quels fragments peuvent \^etre efficacement d\'ecid\'es en pratique ? Comment d\'eterminer si une expression XPath est satisfaisable sur l'un des arbres XML d\'efinis par un sch\'ema donn\'e ? Comment savoir si deux requ\^etes vont toujours donner le m\^eme r\'esultat lorsqu'elles sont \'evalu\'ees sur un document valide par rapport \`a un certain sch\'ema ? Est ce que le r\'esultat d'une expression XPath sur un document valide se conforme toujours aux exigeances d'un autre sch\'ema ? Existe-t-il un algorithme capable de r\'epondre \`a ces questions d'une mani\`ere efficace de telle sorte qu'il soit utilisable en pratique ?

Une source de difficult\'e pour un tel algorithme est qu'il doit v\'erifier des propri\'et\'es sur une quantification possiblement infinie sur un ensemble d'arbres. Une vari\'et\'e d'autres facteurs contribuent de plus \`a sa complexit\'e comme les op\'erateurs permis dans les requ\^etes XPath et leur composition (cf. Chapitre~\ref{foundations:queries}). Une cons\'equence de ces difficult\'es est que de telles questions de recherche sont toujours ouvertes.

\section*{Aper\c cu de cette th\`ese}


Cette th\`ese part de l'id\'ee que deux probl\`emes doivent \^etre r\'esolus pour pouvoir r\'epondre \`a des probl\`emes de d\'ecision dans le monde XML. Tout d'abord, identifier une logique appropri\'ee avec une expressivit\'e suffisante pour supporter \`a la fois les langages d'arbres r\'eguliers et la navigation et la s\'emantique de s\'election de n\oe uds \`a la XPath. Deuxi\`emement, r\'esoudre efficacement le probl\`eme de la satisfaisabilit\'e de cette logique qui permet de d\'eterminer si une formule donn\'ee de la logique admet un document XML qui la satisfait.

\subsection*{Principales contributions}

La contribution principale de cette th\`ese est une nouvelle logique pour les arbres finis, d\'eriv\'ee du $\mu$-calcul. La logique est suffisamment expressive pour capturer les langages r\'eguliers d'arbres et la navigation multi-directionelle dans les arbres finis. Elle est d\'ecidable en temps simplement exponentiel (plus pr\'ecis\'ement en $2^{O(n)}$ \'etapes o\`u $n$ est la taille de la formule dont le statut de v\'erit\'e est d\'etermin\'e d\'efinie comme le nombre de propositions atomiques et d'\'eventualit\'es qu'elle comporte). Cela am\'eliore la meilleure complexit\'e computationnelle connue pour les arbres finis. En tant que telle, cette logique offre un nouveau compromis entre expressivit\'e et complexit\'e, sp\'ecifiquement int\'eressant dans le contexte de XML).



Une autre contribution de cette th\`ese est de montrer comment traduire lin\'eairement les requ\^etes et les types r\'eguliers d'arbres (incluant les DTDs et les XML Schemas) dans la logique. La logique prend en compte XPath dans sa quasi globalit\'e, et supporte le plus gros fragment qui a \'et\'e trait\'e pour l'analyse statique. Un autre avantage est que la logique est une sous-logique d'une existante (le $\mu$-calcul) ce qui facilite l'application de techniques d'optimisation connues et l'extensibilit\'e.

Cela r\'esout les probl\`emes de d\'ecision majeurs rencontr\'es dans l'analyse statique des langages manipulant des structures XML. Ces probl\`emes englobent l'inclusion, la satisfaisabilit\'e, l'\'equivalence, le recouvrement, la couverture des requ\^etes XPath (en pr\'esence ou absence de types r\'eguliers d'arbres), le typage statique d'une requ\^ete XPath annot\'ee, et l'\'equivalence des requ\^etes sous contraintes de type.

De plus, des techniques d'implantation sont pr\'esent\'ees, qui conduisent \`a la r\'ealisation concr\`ete et \`a l'implantation d'algorithmes efficaces en pratique. Le syst\`eme enti\`erement implant\'e est d\'ej\`a capable de traiter des sc\'enarios r\'ealistes.

\subsection*{Applications} 


La principale application de ce travail est une nouvelle classe d'analyseurs statiques de programmes manipulant des donn\'ees et des documents XML. Cette classe d'analyseurs utilise directement les r\'esultats d\'ecrits dans cette th\`ese, qui r\'esolvent les probl\`emes de d\'ecision auxquels ils sont confront\'es. La r\'esolution de chaque probl\`eme particulier offre des applications importantes.


Par exemple, le probl\`eme le plus fondamental pour un langage de requ\^ete est la satisfaisabilit\'e: une expression retourne-t-elle toujours un r\'esultat vide ? La satisfaisabilit\'e de XPath est importante pour l'optimisation des langages h\^otes de XPath: par exemple, si on peut savoir au moment de la compilation qu'une requ\^ete est insatisfaisable, alors tous les calculs qui en d\'ependent peuvent \^etre \'evit\'es.

Un autre probl\`eme fondamental est le probl\`eme de l'\'equivalence: deux requ\^etes retournent-elles toujours les m\^emes r\'esultats ? Savoir d\'eterminer l'\'equivalence entre deux requ\^etes est utile pour la reformulation et l'optimisation de la requ\^ete elle-m\^eme \cite{geneves-doceng04}, qui vise \`a s'assurer de propri\'et\'es op\'erationnelles tout en pr\'eservant la s\'emantique de la requ\^ete \cite{abiteboul-Jcss99,pierce-dbpl05}.

Le probl\`eme le plus critique pour le typage statique des transformations XML est l'inclusion de requ\^etes XPath: est ce que, pour tout arbre, le r\'esultat d'une requ\^ete particuli\`ere est inclus dans le r\'esultat d'une autre ? Ce probl\`eme se pose pour l'analyse du flot de contr\^ole de XSLT \cite{moller-rr05}. Savoir d\'eterminer l'inclusion est aussi utile pour v\'erifier les contraintes d'int\'egrit\'es \cite{xml-schemas}, et pour v\'erifier la politique et les droits d'acc\`es dans les applications de s\'ecurit\'e XML \cite{xml-access-control}.

D'autres probl\`emes de d\'ecision utiles dans les applications incluent par exemple la couverture mutuelle des requ\^etes (deux expressions peuvent-elles s\'electionner des n\oe uds communs ?) et la couverture (les n\oe uds s\'electionn\'es par une requ\^ete sont-ils toujours contenus dans l'union des r\'esultats s\'electionn\'es par d'autres requ\^etes ?). Ces probl\`emes sont par exemple utiles pour d\'etecter statiquement les erreurs de programmation.


Cette th\`ese r\'esout ces probl\`emes de d\'ecision, en pr\'esence ou en l'absence de contraintes de types XML comme les DTDs \cite{xml} ou les XML Schemas \cite{xml-schemas}. Cela permet de s'assurer de propri\'et\'es locales ou globales importantes (comme le bon typage ou des optimisations) au moment de la compilation, pour un traitement plus s\^ur et plus efficace des donn\'ees XML. Les r\'esultats pr\'esent\'es dans cette th\`ese ouvrent notamment des perspectives prometteuses concernant l'analyse statique des transformations XML.

\subsection*{Organisation de la th\`ese}

Cette th\`ese est divis\'ee en trois parties.
La premi\`ere partie est d\'edi\'ee \`a l'\'etat de l'art et pr\'esente les techniques de pointe existantes et les travaux de recherche reli\'es. A cette fin, le chapitre~\ref{foundations} introduit quelques fondations th\'eoriques et formalismes utilis\'es dans la suite de cette th\`ese, tout en introduisant progressivement les travaux reli\'es au fur et \`a mesure que leurs concepts sous-jacents ont \'et\'e pr\'esent\'es.


Dans une seconde partie, les chapitres~\ref{containment} et~\ref{analysis} conduisent des investigations pr\'eliminaires avec des logiques connues dans le cadre de XML. Plus pr\'ecis\'ement, le chapitre~\ref{containment} \'etudie dans quelle mesure la logique monadique du second ordre peut \^etre utilis\'ee en pratique, en d\'epit de sa grande complexit\'e, pour r\'esoudre des probl\`emes d'analyse statique comme l'inclusion des requ\^etes XPath. Une proc\'edure de d\'ecision correcte pour l'inclusion est propos\'ee. Le chapitre~\ref{analysis} introduit le $\mu$-calcul sans alternance comme un puissant remplacement pour la logique monadique du second ordre, et \'etudie son usage pour raisonner sur les arbres XML. Les probl\`emes de d\'ecision mettant en jeu les requ\^etes XPath et les types r\'eguliers sont r\'eduits \`a la satisfaisabilit\'e de cette logique sur des structures de Kripke g\'en\'erales.


Gr\^ace aux le\c cons tir\'ees des investigations pr\'ec\'edemment men\'ees, la troisi\`eme partie de cette th\`ese pr\'esente la contribution finale.
Le chapitre~\ref{xml-calculus:the-logic-for-xml} propose une logique d'arbres finis sp\'ecifiquement con\c cue pour XML. Le chapitre~\ref{xml-calculus:sec:algo} propose un algorithme pour tester la satisfaisabilit\'e de la logique, ainsi que les techniques pour son implantation. Des exp\'erimentations sont men\'ees avec une implantation compl\`ete du syst\`eme, qui s'av\`ere efficace sur plusieurs sc\'enarios r\'ealistes. Enfin, le chapitre~\ref{conclusion} conclut cette th\`ese et donne de nouvelles perspectives.


 \ifglobalcompil
  \else 
   \bibliographystyle{apalike}
   \bibliography{references}
   \end{document}
  \fi
\end{appendices}

\end{document}